\documentclass[aps,prb,longbibliography,twocolumn,amsmath,amssymb,floatfix]{revtex4-2}
\usepackage[english]{babel}
\usepackage{blindtext}
\usepackage[utf8]{inputenc}
\usepackage{graphicx}
\usepackage{array} 
\usepackage{bm}
\usepackage{latexsym}
\usepackage{physics}
\usepackage{amsfonts}
\usepackage{bbold}
\usepackage{cancel}
\usepackage{subfig}
\usepackage{scalerel}
\usepackage{appendix}
\usepackage[normalem]{ulem}
\usepackage{physics,braket}

\usepackage{ragged2e}
\usepackage{mathtools}
\usepackage[export]{adjustbox}
\newcolumntype{H}{>{\setbox0=\hbox\bgroup}c<{\egroup}@{}}

\usepackage[normalem]{ulem}

\usepackage[colorlinks = true,
            linkcolor = blue,
            urlcolor  = blue,
            citecolor = blue,
            anchorcolor = blue]{hyperref}
\usepackage[usenames,dvipsnames]{color}
\usepackage{tikz}
\usetikzlibrary{shapes.geometric, arrows}
\makeatletter
\newcommand*{\rom}[1]{\expandafter\@slowromancap\romannumeral #1@}
\newcommand{\vebm}[1]{
\ifcat\noexpand#1\relax
    {\bm #1}
\else
    {\bf #1}
\fi
}
\newsavebox{\@brx}
\newcommand{\llangle}[1][]{\savebox{\@brx}{\(\m@th{#1\langle}\)}%
  \mathopen{\copy\@brx\kern-0.5\wd\@brx\usebox{\@brx}}}
\newcommand{\rrangle}[1][]{\savebox{\@brx}{\(\m@th{#1\rangle}\)}%
  \mathclose{\copy\@brx\kern-0.5\wd\@brx\usebox{\@brx}}}
\makeatother
\begin{document}
\definecolor{myPurple}{RGB}{240, 180, 210}
\definecolor{myGreen}{RGB}{228, 240, 180}
\definecolor{myYellow}{RGB}{240, 240, 160}
\definecolor{myBlue}{RGB}{140, 200, 255}

\newcommand{\APS}[1]{\textcolor{cyan}{#1} }

\title{Exotic superconducting states in altermagnets}

\date{\today}

\author{Kirill Parshukov}
\email{k.parshukov@fkf.mpg.de}
\affiliation{Max Planck Institute for Solid State Research, Heisenbergstrasse 1, D-70569 Stuttgart, Germany}

\author{Andreas P. Schnyder}
\email{a.schnyder@fkf.mpg.de}
\affiliation{Max Planck Institute for Solid State Research, Heisenbergstrasse 1, D-70569 Stuttgart, Germany}

\begin{abstract}

The interplay between magnetism and superconductivity is one of the central topics of condensed matter physics, which has recently been put into new light by the discovery of altermagnets. Here, we study this interplay from a fundamental symmetry perspective using irreducible co-representations of the altermagnetic spin-point groups.
We construct and tabulate all symmetry-allowed pairing functions for altermagnets, 
which uncovers numerous exotic pairing states. 
We focus on three of them, namely: (i) a non-unitary ``\textit{{altermagnetic} superconductor}'' with different spatial anisotropies for the spin-up and spin-down condensates, (ii) a ``\textit{half-and-half metal-superconductor}" where only electrons with one of the two spin components form Cooper pairs, and (iii) a ``\textit{spin chiral superconductor}'' with spin-polarized chiral edge states.
Interestingly, the first of these three superconductors exhibits an unusual fractional ac Josephson current for only one of the two spin polarizations{, even though the two condensates are symmetry related}.
We present phenomenological Ginzburg-Landau theories for these unconventional superconductors and show that they correspond to stable minima of the free energies. 
We examine their topological properties, study the effects of small spin-orbit coupling, consider possible material examples, and investigate their topological responses.

\end{abstract}

\maketitle

\textit{Introduction}---
The interplay between magnetism and superconductivity has been and 
continues to be of great interest in condensed matter physics~\cite{ RevModPhys.56.755, RevModPhys.81.1551, doi:10.1143/JPSJ.81.011004,Keimer2015, RevModPhys.87.855, Qimiao2016, Aoki_2022,  Eremin2021}. On the one hand, magnetic fluctuations can serve as a pairing glue
for superconductivity.
On the other hand, the symmetry lowering of the magnetic order imposes 
restrictions on the possible pairing states.
For example, 
ferromagnetism breaks time-reversal symmetry and lifts the spin degeneracy of the Fermi surface, 
leading to unconventional 
spin-triplet pairing~\cite{RevModPhys.81.1551,aoki_ferro_SC_nature_2001,aoki_review_ferro_SC_JPSJ}.
In contrast, antiferromagnets exhibit spin-degenerate
Fermi surfaces, such that
conventional spin-singlet pairing is generally the dominant instability~\cite{RevModPhys.81.1551,RevModPhys.87.855}.

The recently discovered altermangets exhibit both
ferro- and antiferro-like characteristics~\cite{PhysRevX.12.031042, PhysRevX.12.040501, smejkal_nat_review_22}. 
Due to their unique symmetry properties, described by spin-point groups (SPGs), they show collinear-compensated magnetic order, while at the same time having spin-split electronic bands~\cite{
PhysRevX.12.040002, https://doi.org/10.1002/adfm.202409327, AFM_Spintronics_review2025, PhysRevLett.132.176702, PhysRevB.105.064430,PhysRevB.102.144441,doi:10.7566/JPSJ.88.123702, PhysRevB.111.174407, delre2024diracpointstopologicalphases, PhysRevB.111.094412, PhysRevB.111.224406, PhysRevLett.134.096703}, as recently confirmed in photoemission spectroscopy~\cite{krempasky_jungwirth_nature_24,Reimers2024,Yang2025}.
As a consequence,  
altermagnetic metals 
have fully spin-polarized Fermi surfaces, whereby spin-up and spin-down parts are related by a symmetry of the SPG.
Because of these unique features of both the magnetic order and the Fermi surface, one may
expect a distinctive novel type of interplay between magnetism and superconductivity in altermagnets~\cite{mazin2022notesaltermagnetismsuperconductivity,SC_AMreview2025, PhysRevB.108.184505, PhysRevB.108.224421, PhysRevB.109.134515, leraand2025phononmediatedspinpolarizedsuperconductivityaltermagnets, zora264790, PhysRevLett.131.076003, PhysRevB.110.L060508, PhysRevResearch.5.043171, PhysRevB.110.024503,PhysRevB.111.L100507, parthenios2025spinpairdensitywaves, PhysRevB.111.144508, chakraborty2024constraintssuperconductingpairingaltermagnets,PhysRevB.108.054511,PhysRevB.108.075425, feng2024superconductingorderparametersspin, wu2025intraunitcellsingletpairingmediated,sumita2025phasemodulatedsuperconductivityaltermagnetism, PhysRevB.111.064502, PhysRevB.111.184515,PhysRevLett.126.067001}. 
This begs the following questions: 
(i) What types of superconductivity can altermagnets support?
(ii) Does altermagnetism lead to exotic pairing states not present in
conventional ferro- or antiferromagnets? (iii) What types 
of novel devices and responses can be engineered using
superconducting altermagnets?

In this article, we answer 
these questions from a fundamental symmetry point of view. 
Using the irreducible co-representations~\cite{xiao2023spin,ren2023enumeration,jiang2023enumeration,chen2023spin,schiff2023spin} (coreps) of the SPGs~\cite{https://doi.org/10.1002/pssb.2220100206,doi:10.1098/rspa.1966.0211,10.1063/1.1708514,LITVIN1974538,Yang2021SymmetryIA},
we construct and tabulate all symmetry-allowed pairing states
of superconducting altermagnets. 
Importantly, due to negligible spin-orbit coupling (SOC),
 elements of the  SPGs act differently on the spin and lattice degrees of freedom, in contrast to regular magnetic point groups (MPGs).
Hence, the coreps of SPGs have a very different structure from those of MPGs, enabling the existence of novel types of pairing states in altermagnets.
To derive the basis functions 
of the superconducting orders, we employ
two complementary approaches, namely, 
{(1) constructing them from the irreducible representations of the unitary halving group using Dimmock indicators~\cite{Dimmock1963, bradley1972mathematical, dresselhaus2008group}, and (2) obtaining them directly from the coreps matrices via projector operators~\cite{Dimmock1963}.}
Both methods yield identical results and unveil a number of exotic pairing states, whose spin and spatial parts are 
interrelated in a highly nontrivial manner.

{In Ref.~\cite{feng2024superconductingorderparametersspin}, the authors presented a general methodology for constructing superconducting basis functions from SPG coreps. In particular, they discussed two- and four-dimensional channels in altermagnets. The present work provides a comprehensive tabulation of pairing basis functions for all SPGs describing collinear compensated magnets. We study topology and physical responses associated with several exotic states.
In particular, we analyze:}

(i) A spin-triplet superconductor 
whose spin-up and spin-down pairing functions are related 
by a (improper) rotation combined with a spin flip. 
Since this type of pairing exhibits an altermagnetic-like symmetry, we call it
an ``\emph{altermagnetic superconductor}".
The state represents the most striking and important result of our analysis, {because of its spin-polarized topological responses}
Its order parameter is {non-unitary} and its
spin-up and spin-down parts
have different spatial anisotropies, giving rise to exotic spin-polarized responses.
We consider, in particular, a spin-polarized fractional ac Josephson effect~\cite{10.1063/1.1789931}
and identify an organic material where this could be observed.
{Additional details of this state are provided in the Appendices.}

\begin{figure*}[t!]
    \centering
\includegraphics[width=0.95\textwidth]{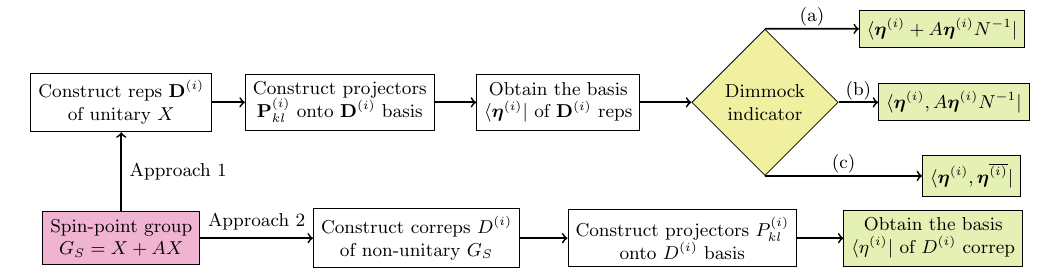}
    \caption{\justifying Diagram describing the two approaches employed to obtain the basis functions of the superconducting order parameters in altermagnets. With the approach 1 we construct first the basis functions for the unitary halving group. Approach 2 allows to construct the basis functions directly using the corep matrices. }
    \label{approaches}
\end{figure*}

(ii) A superconductig state, where only the spin-up electrons form Cooper pairs, while the spin-down electrons remain metallic, which we term a ``\emph{half-and-half metal-superconductor}". 
The superconducting part of this state is
similar to the A$_1$ phase of $^3$He~\cite{PhysRevLett.30.81} and ferromagnetic superconductors~\cite{PhysRevLett.86.850},
while its normal part is a half-metal~\cite{half_metal_PRL_83}.

(iii) A number of chiral pairing states, which break antiunitary symmetries of the normal-state altermagnet, which we call
``\emph{spin chiral superconductors}".
These occur for SPGs with correps of dimension four or six,
such that the antiunitary symmetry can be broken in each spin subspace by a complex superposition of two (or three) basis states.

For the above three examples,
we discuss their topological classifications 
and present phenomenological Ginzburg-Landau (GL) theories~\cite{RevModPhys.63.239},
describing second order phase transition from the altermagnetically ordered state into the superconducting state.
We show that these unconventional superconductors
arise as stable minima of the GL free-energy functional.

\textit{Spin-point group basis functions} ---
We begin by considering the SPGs that describe the symmetries of compensated collinear magnets without SOC.
In the absence of SOC the symmetry actions $(s||g)$ on spin $s$ and lattice $g$ are decoupled. Collinear magnetic order exhibits an SO(2) spin rotation symmetry and a combined symmetry 
{
$(2_{\perp} \mathcal{T}||E)$ consisting of a unitary $\pi$ spin rotation about an axis perpendicular to the ordering direction followed by time reversal.}
The magnetic moments are compensated due to the $(2_{\perp}||g)$ symmetry, where $g$ is an operation that relates magnetic sublattices. If $g$ is the space inversion or translation, the resulting phase is antiferromagneic. If $g$ is an (improper) rotation the group corresponds to the altermagnetic order. 

SPG $G_S$ combines a unitary halving {group} $X$ and an antiunitary part $ X(\mathcal{T}2_{\perp} || E)$
\begin{align} 
    G_S &= X +  X(\mathcal{T}2_{\perp} || E),\\
    X &= X_{1/2} + (2_{\perp} || g) X_{1/2}, \label{halvinggroup} \\
    X_{1/2} &= \text{SO(2)} \times H.
\end{align}
The unitary group $X$ is further split into two cosets, $X_{1/2}$ -- the sublattice symmetry group, and 
$(2_{\perp} || g) X_{1/2}$, whose elements flip the spin relating spin sublattices. $X_{1/2}$ consists of a normal subgroup $H$, which is the symmetry group of each magnetic site, and the SO(2) spin rotational group. With the group decomposition we can construct its irreducible co-representations~\cite{schiff2023spin}.

To construct coreps basis functions of a group with anti-unitary elements $G_S = X + XA$ we will discuss two approaches [see Fig.~\ref{approaches}]. The first approach is based on the derivation of the basis functions for the unitary 
group $X$ irreducible representaions (reps) 
{$\mathbf{\Gamma}^{(i)}$. The basis functions are obtained using the projector operator}
\begin{equation}\label{projector}
    \mathbf{P}_{kl}^{(i)} = \frac{l_i}{|X|}\sum_{R\in X}\mathbf{D}^{(i)*}(R)_{kl}R,
\end{equation}
where $l_i$ is the dimension of $\mathbf{\Gamma}^{(i)}$ reps with matrices $\mathbf{D}^{(i)}$, $R$ represents the action of the group element. For the continuous SO(2) spin-rotational symmetry, the summation is {replaced} by the integration with the measure $\tfrac{1}{2\pi} d\phi$. The projector transforms $l^\text{th}$ basis vector to the $k^\text{th}$. To determine the basis, we act with $\mathbf{P}_{ll}^{(i)}$ onto an arbitrary function. The $\mathbf{D}^{(i)}$ matrices were not tabulated in Ref.~\cite{schiff2023spin}, we repeat the procedure used by the authors and obtain reps for unitary halving groups of all considered SPGs (see Supplemental Material~\cite{sm_note} and references \cite{aroyo2006bilbao, 10.3389/fphy.2018.00086, GTPack, Karavaev}). 

The {coreps} of the $G_S$ group can be obtained from the unitary group $X$ reps. Generally there are three possibilities according to Dimmock rules~\cite{Dimmock1963}
\begin{align}
 &(a) \ \bra{\boldsymbol{\eta}^{(i)} + A\boldsymbol{\eta}^{(i)} N^{-1}}, \ \mathbf{\Gamma}^{(i)} \stackrel{N}{\sim} \overline{\mathbf{\Gamma}^{(i)}}, \ {NN^* = \mathbf{D}(A^2)}, \notag\\
 &(b) \ \bra{\boldsymbol{\eta}^{(i)}, A\boldsymbol{\eta}^{(i)} N^{-1}}, \ \mathbf{\Gamma}^{(i)}\stackrel{N}{\sim} \overline{\mathbf{\Gamma}^{(i)}}, \ {NN^* = -\mathbf{D}(A^2)}, \notag \\
&(c) \ \bra{\boldsymbol{\eta}^{(i)}, \boldsymbol{\eta}^{\overline{(i)}}}, \ \mathbf{\Gamma}^{(i)}{\not\sim} \overline{\mathbf{\Gamma}^{(i)}}, \label{DimmockClasses}
\end{align}
where $\bra{\boldsymbol{\eta}^{(i)}}$ is the row of the basis functions of the rep {$\mathbf{\Gamma}^{(i)}$}. In the first two cases, the irreducible representation $\overline{\mathbf{\Gamma}^{(i)}}$ with matrices $\overline{\mathbf{D}^{(i)}}(R) = \mathbf{D}^{(i)*}(A^{-1}RA)$ is equivalent to $\mathbf{\Gamma}^{(i)}$ {via a} unitary transformation matrix $N$, {i.e. $ \mathbf{D}(R) = N \mathbf{D}^*(A^{-1}RA) N^{-1}$}. In the later case the reps are not equivalent. 
In the cases (b,c) the dimesionality of the irreducible representation is increased.

There is another approach to construct the corep basis functions for a group with non-unitary elements. Instead of constructing the basis functions for the unitary halving group reps, we introduce the corep projector operator
\begin{equation}
    P_{kl}^{(i)} = \frac{l_i}{|G_S|}\sum_{R\in X}[D^{(i)*}(R)_{kl}R + D^{(i)*}(RA)_{kl}RA],
\end{equation}
where 
$D$ are the coreps matrices~\cite{sm_note}.

We apply these approaches to construct the superconducting basis states.
The Cooper pair wave function for a one-band SC
can be expressed as a linear combination of basis functions 
\begin{align}
\begin{split}
    \Delta_{\sigma,\sigma'}(\mathbf{k}) &= \sum_{i} {\upsilon_i} \Delta_{\sigma, \sigma'}^{(i)}(\mathbf{k}),\\ \label{order_decomp}
    \Delta_{\sigma, \sigma'}^{(i)}(\mathbf{k}) &= i[ \psi_i(\mathbf{k}) +    (d_i(\mathbf{k}) \cdot \hat{\tau})] \tau^y_{\sigma,\sigma'},
\end{split}
\end{align}
where $\psi_i(\mathbf{k})$ is a scalar spin-singlet part,  
$d_i(\mathbf{k})$ is a three-dimensional (3D) vector corresponding to the triplet pairing, $\hat{\tau}$ is the vector of Pauli matrices.
With the discussed approaches we obtained coreps basis functions for all SPGs $G_S$ describing collinear compensated magnetic phases~\cite{sm_note}. 
The unique coreps of the SPG have even dimensionality with spin-polarized triplet basis states $d_{\pm}(\mathbf{k}) = (1, \pm i, 0)^T \gamma_{\pm}(\mathbf{k})$, related by the altermagnetic symmetry $(2_{\perp} || g)$. 
In this expression $\gamma_{\pm}(\mathbf{k})$ is an odd function of $\mathbf{k}$, such that $\gamma_{-}(\mathbf{k}) = g\gamma_{+}(\mathbf{k})$. $\text{D}_{\pm} = (1, \pm i, 0)^T$ are the eigenvectors of the SO(2) spin-rotations with eigenvalues $e^{\pm i \phi}$. These states correspond to the pairing $\Delta_{\uparrow\uparrow/\downarrow\downarrow}$, {see the Appendices for more details on how to construction the basis functions. 
} 
{The exotic superconducting phenomena uncovered in this work occur exclusively in the 
spin-triplet $S=\pm 1$ sector. Even though spin-singlet states can be unconventional (e.g., d-wave), they do not show altermagnetic spin–space locking, because the $S=0$ order is invariant under spin transformations.}

\textit{Half-and-half
metal-superconductor and altermagnetic superconductor} ---
To obtain stable superconducting states we minimize the GL free energy. The functional respects the same symmetries as the system's Hamiltonian. 
For the two-dimensional coreps describing $S=\pm 1$ states
the general free energy expansion up to the fourth order takes the form
\begin{align}\label{2_dim_free_energy}
\begin{split}
    f[{\upsilon_1, \upsilon_2}] &= \alpha (|\upsilon_1|^2 +|\upsilon_2|^2) + \beta_1 (|\upsilon_1|^2 + |\upsilon_2|^2)^2 \\ &+ \beta_2 |\upsilon_1|^2|\upsilon_2|^2,
\end{split}
\end{align}
where ${\upsilon}_i$ determine the expression $\Delta(\mathbf{k}) = {\upsilon}_1 \Delta_{\text{D}_-}(\mathbf{k})+{\upsilon}_2 \Delta_{\text{D}_+}(\mathbf{k})$, with the basis vectors of the coreps $\Delta_{\text{D}_\pm}(\mathbf{k})$.
The spin-rotational SO(2) and the charge U(1) symmetries do not allow to have the term ${\upsilon}_1 {\upsilon}^*_2 + {\upsilon}_2 {\upsilon}^*_1$. Below the phase transition we fix $\beta_1 > 0$ and $\alpha < 0$. 
As an example we consider the $\text{B}_{+1}\text{B}_{-1}$ corep of SPG $^{\overline{1}}$m$^{\overline{1}}$m$^1$2 (see Appendices).
The basis functions are given by
$\text{D}_+ \left(r_x k_x+ r_y k_y\right)   \text{and} \
 \text{D}_- \left(-r_x k_x+ r_y k_y\right)$ with real coefficients $r_x, r_y \in \mathbb{R}$.

\begin{figure}[t]
    \centering
    \includegraphics[width=0.9\linewidth]{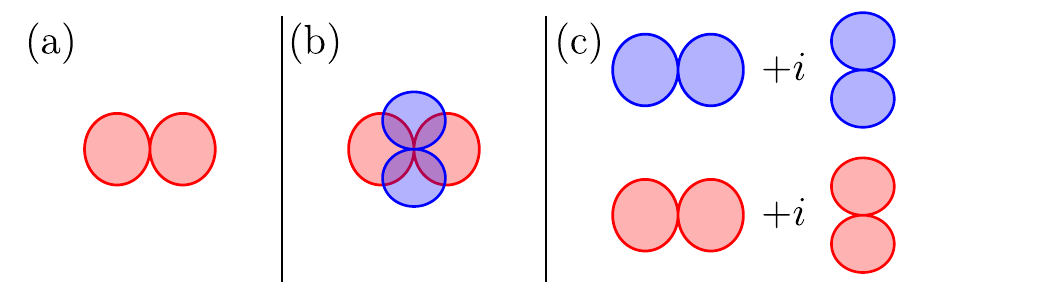}
    \caption{\justifying Three types of exotic superconducting states in altermagnets. (a) Half-and-half
metal-superconductor with the gap only in one of the spin subspaces. (b) Altermagnetic superconductor with different spatial anisotropy
for $S = 1$ and $S = -1$ parts of the pairing potential, related by the altermagnetic symmetry $(2_{\perp}||g)$. (c) Spin chiral state from the four-dimensional corep, the relative phase breaks antiunitary 
$(\mathcal{T}2_{\perp}||E)$ symmetry in each of the spin subspaces.}
    \label{fig:exotic_SC}
\end{figure}

First, we consider $\beta_2 > 0$. The stable superconducting state is $|{\upsilon}_1| = 0, |{\upsilon}_2| \neq 0$ or $|{\upsilon}_2| = 0, |{\upsilon}_1| \neq 0$. In this case, the superconducting gap opens only for one of the spin subspaces, and leads to half-and-half
metal-superconductor [see Fig.~\ref{fig:exotic_SC}(a)]. The triplet spin-polarized pairing $\Delta_{\uparrow\uparrow/\downarrow\downarrow} \neq 0$ is non-unitary, and breaks the altermagnetic symmetry $(2_{\perp}||g)$.
{A similar triplet state was discussed in the context of $^3$He in high magnetic fields~\cite{PhysRevLett.30.81}, corresponding to the A$_1$ phase. }
In our case, the exotic state appears in the compensated magnet without SOC and with spin-split bands.

If $\beta_2 <0$ there is a minimum at $|{\upsilon}_1| = |{\upsilon}_2| \neq 0$. Spin up and down subspaces are still decoupled, and the state preserves the altermagnetic symmetry $(2_{\perp}||g)$. The two spin-polarized pairing potentials with $S=\pm1 $ are related by the $(2_{\perp}||g)$ leading to the altermagnetic superconducting state [see Fig.~\ref{fig:exotic_SC}(b)]. We note that the state is not a unitary state.
The different anisotropies for the two spin-polarized pairings $\text{D}_+ \left(r_x k_x+ r_y k_y\right)   \text{and} \
 \text{D}_- \left(-r_x k_x+ r_y k_y\right)$,  with $S=\pm1$, lead to the non-unitary state $d(\mathbf{k}) = \text{D}_+ \left(r_x k_x+ r_y k_y\right) + e^{i\xi}
 \text{D}_- \left(-r_x k_x+ r_y k_y\right)$. The pair correlation is different for up and down-spins, i.e., $i d(\mathbf{k}) \cross d(\mathbf{k})^* = 8 r_x r_yk_x k_y\text{D}_{\text{z}}$.
 The state is a unique state of the altermagnetic phase and 
 {cannot arise in conventional (anti-)ferromagnets}~\footnote{{In an antiferromagnet, the state that combines two spin-polarized condensates related by the $(2_{\perp}||I)$ symmetry takes the form $d_{\text{AFM}}(\mathbf{k}) = \text{D}_+ \gamma (\mathbf{k}) + \eta
 \text{D}_- \gamma (-\mathbf{k})  =  (\text{D}_+ -\eta \text{D}_- )\gamma (\mathbf{k}) $, with $|\eta| = 1$.  The state is unitary and describes two overlapping spin-polarized condensates,
    $i d_{\text{AFM}}(\mathbf{k}) \times d_{\text{AFM}}(\mathbf{k})^* =0$.}}.

The minimum $|{\upsilon}_1| = |{\upsilon}_2| \neq 0$ has a large degeneracy, as the relative phase can take any value. 
The degeneracy is lifted in the presence of small SOC {(see Appendices)}.
In this case, we can include an additional quadratic term in the free energy expansion $\alpha_{\text{SOC}}({\upsilon}_1 {\upsilon}^*_2 + {\upsilon}_2 {\upsilon}^*_1)$.
If $\alpha_{\text{SOC}} < 0$ the minimum is at ${\upsilon}_1={\upsilon}_2$. When $\alpha_{\text{SOC}} > 0$ it is at ${\upsilon}_1=-{\upsilon}_2$. 

The obtained superconducting states do not break the antiunitary $(\mathcal{T}2_{\perp}||E)$ symmetry. Together with the particle-hole symmetry, this leads to a chiral (sublattice) symmetry in each of the spin subspaces (see Appendices and SM~\cite{sm_note} and Ref.~\cite{PhysRevB.78.195125}). The chiral symmetry quantizes 1D winding number. In 2D superconductors the symmetry protects spin-polarized point nodes at zero energy in the quasiparticle excitations bands. 
At the boundary of the superconductor we expect to observe spin-polarized zero-energy modes, i.e., Majorana modes (see Fig.~\ref{fig:TB_SC})~\cite{Majorana_ZM_review2024}.
In 3D materials the symmetry protects spin-polarized line-nodes.

\textit{Spin chiral superconductor} --- {Four-dimensional superconducting channels 
basis functions of SPGs were previously discussed in Ref.~\cite{feng2024superconductingorderparametersspin}.}
{The presence of higher-dimensional coreps, i.e., four and six-dimensional coreps, is unique to SPGs that describe collinear compensated magnets.
For the four-dimensional coreps each of the spin subspaces has two basis vectors
with an undetermined relative phase
that can break $(\mathcal{T}2_{\perp}||E)$, leading to chiral superconductivity. 
As an example, we}
consider the altermagnetic SPG ${}^{\bar{1}}6/{}^{\bar{1}}m{}^1m{}^{\bar{1}}m$ describing symmetries of CrSb~\cite{Reimers2024,Yang2025}. {There is a four-dimensional corep $\text{E}^{\text{u}}_{+1}\text{E}^{\text{u}}_{-1}$ with four basis vectors 
$\frac{-i}{\sqrt2}\text{D}_- (k_x+i k_y)$,
$\frac{-i}{\sqrt2}\text{D}_- (k_x-i k_y)$,
$\frac{-i}{\sqrt2}\text{D}_+(k_x+i k_y)$,
$\frac{-i}{\sqrt2}\text{D}_+(k_x-i k_y)$
 (see Appendices).
We introduce corresponding order parameters ${\upsilon}_1, {\upsilon}_2, {\upsilon}_3$ and ${\upsilon}_4$ [see Eq.~(\ref{order_decomp})].} 
The GL free energy has five quartic terms
\begin{align}
    f_4 &= \beta_1 (|{\upsilon}_1|^2 + |{\upsilon}_2|^2 + |{\upsilon}_3|^2 + |{\upsilon}_4|^2)^2 \label{f4_6mmm}\\ 
    +  \beta_2 (|{\upsilon}_1|^2 |{\upsilon}_2|^2 &+ |{\upsilon}_3|^2 |{\upsilon}_4|^2) + \beta_3 (|{\upsilon}_1|^2 |{\upsilon}_3|^2 + |{\upsilon}_2|^2 |{\upsilon}_4|^2) \notag\\ 
    +  \beta_4 (|{\upsilon}_2|^2 |{\upsilon}_3|^2 &+ |{\upsilon}_1|^2 |{\upsilon}_4|^2) + 2 \beta_5 |{\upsilon}_1||{\upsilon}_2||{\upsilon}_3||{\upsilon}_4| \cos{\theta},\notag
\end{align}
where $\theta = \theta_1 -\theta_2 -\theta_3+\theta_4$ is given by the phases of ${\upsilon}_n = |{\upsilon}_n|e^{i\theta_n}$.
There are a number of stable minima depending on the choice of the parameters, however, there is the continuous degeneracy in the relative phase. We consider one of the solutions
$|\upsilon_1|=|\upsilon_3|\ne0$,
$\upsilon_2=\upsilon_4=0$, and its opposite-chirality partner
$|\upsilon_2|=|\upsilon_4|\ne0$,
$\upsilon_1=\upsilon_3=0$ (for the parameter conditions, see
Appendices).
 The first state is $\text{D}_- (k_x + ik_y) + {e^{i\xi}}\text{D}_+(k_x+ik_y)$, 
[see Fig.~\ref{fig:exotic_SC}(c)].
The relative phase in $k_x\pm ik_y$ breaks the antiunitary symmetry $(\mathcal{T}2_{\perp}||E)$, while the relative spin phase $\xi$ remains arbitrary.

In a 2D superconductor, the gapped state is described by a Chern number~\cite{Volovik1997}. In 3D material, there can be point nodes with topological charges. In Table~\ref{tab:topolo_class_main} we summarize the topological classification of the discussed superconducting states~\cite{sm_note}. 

\begin{table}[t]
    \centering
        \renewcommand{\arraystretch}{1.5}
    \begin{tabular}{c|c|c|c|c}
         Class  & $(\mathcal{T}2_{\perp}||E)$ & 1 & 2 & 3  \\
         \hline
         {D}  & 0 & {$\mathbb{Z}_2$} & $\mathbb{Z}$&0\\
         \hline
         {BDI} & \checkmark  & $\mathbb{Z}$ & 0&{0}\\
    \end{tabular}
    \caption{\justifying Classification of topological phases of superconducting states in altermagnets. The invariants are determined for each of the spin-subspaces. If the anti-unitary symmetry $(\mathcal{T}2_{\perp}||E)$ is broken by a complex order parameter, the two-dimensional invariant is the Chern number for each spin-subspace. {When} $(\mathcal{T}2_{\perp}||E)$ {is preserved}, the invariant is one-dimensional winding number.}
    \label{tab:topolo_class_main}
\end{table}

\textit{Material example}---
$\kappa$-(BEDT-TTF)$_2$Cu[N(CN)$_2$]Br is a member of  quasi-2D organic compound family $\kappa$-(BEDT-TTF)$_2$X. The compound becomes altermagnetic at low temperatures, and goes through the superconducting phase transition under applied pressure. The phase separation region appears between superconducting and compensated collinear order~\cite{PhysRevLett.89.017003,doi:10.1143/JPSJ.74.2351,PhysRevB.76.054509}.

The magnetic phase has the symmetries of the 2D spin-space group $^1$2$^{\overline{1}}$g$^{\overline{1}}$g~\cite{PhysRevB.111.224406}. We show the 2D unitcell, band structure and the Fermi surface of the four-site tight-binding model~\cite{PhysRevB.89.045102} in Fig.~\ref{fig:TB_normal} {(see Appendices)}. 
The material has glide mirror symmetries that flip the spin $(2_{\perp}||M_{x/y}(\tfrac{1}{2},\tfrac{1}{2}))$ and the two-fold rotational symmetry $(E||2_z)$. The corep table for the SPG $^{\overline{1}}$m$^{\overline{1}}$m$^1$2 can be found in the Appendices.
For the SPG we obtained the basis functions describing singlet s-wave $\psi(\mathbf{k}) = 1$, d-wave $\psi(\mathbf{k}) = k_x k_y$, transforming as $\text{A}_+$ and $\text{A}_-$ coreps respectively. Basis states for the triplet $S = 0$ states are $i\text{D}_{\text{z}} k_x$ and $i\text{D}_{\text{z}} k_y$ corresponding to $\text{B}_+$ and $\text{B}_-$ coreps, {$\text{D}_{\text{z}}=(0,0,1)^T$}. The unique corep $\text{A}_{+1}\text{A}_{-1}$ is not compatible with superconducting order, as it describes $S\neq0$ states that are even under space inversion (here $(E||2_z)$ rotation). The other unique corep $\text{B}_{+1}\text{B}_{-1}$ gives rise to spin-polarized triplet basis states $\text{D}_+ \left(r_x k_x+ r_y k_y\right)   \text{and} \
 \text{D}_- \left(-r_x k_x+ r_y k_y\right)$ with real coefficients $r_x, r_y$. The two states are related by the altermagnetic symmetry $(2_{\perp}||M_{x})$.

\begin{figure}[b]
\centering
    \includegraphics[width=\linewidth]{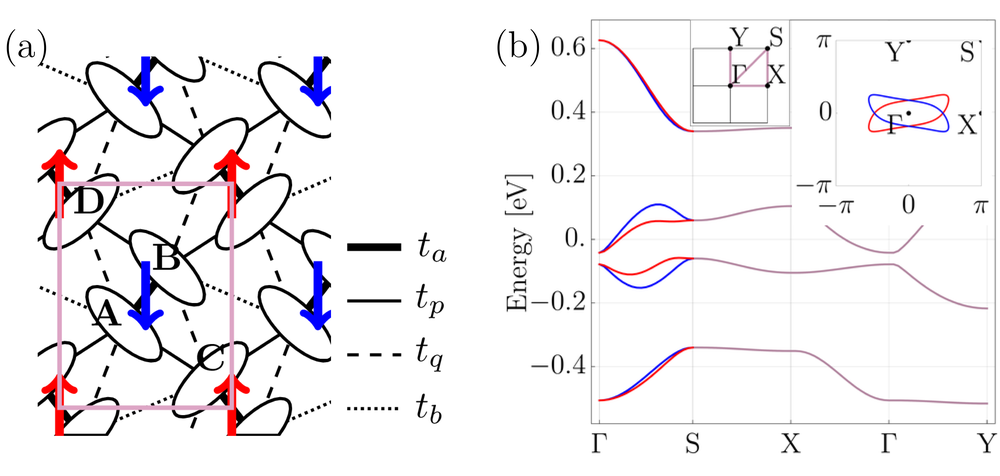}
    \caption{(a) \justifying Quazi-two-dimensional $\kappa$-(BEDT-TTF)$_2$X compound lattice with the magnetic order. (b) The band structure of the compound tight-binding model with $(t_a, t_b, t_p, t_q, h) = (-0.207,\ -0.067,\ -0.102,\ 0.043,\ 0.2)$ eV, where $h$ is the strength of the interaction with the magnetic moments. The BZ and Fermi surface with two spin-polarized pockets are shown in the insets.}
    \label{fig:TB_normal}
\end{figure}

Using the free energy [Eq.~(\ref{2_dim_free_energy})], we obtain stable superconducting states. The first solution (1, 0) [(0, 1)] is the half-and-half
metal-superconductor $|{\upsilon}_1| \neq 0, |{\upsilon}_2| = 0$ [$|{\upsilon}_2| \neq 0, |{\upsilon}_1| = 0$], whereas the second one (1, 1) is the altermagnetic superconductor $|{\upsilon}_1| = |{\upsilon}_2| \neq 0$. We use Bogoliubov-de Gennes formalism and consider the (1, 1) pairing potential  $\text{D}_+ \left(r_x k_x+ r_y k_y\right)  +
 \text{D}_- \left(-r_x k_x+ r_y k_y\right)$ 
 [{see Appendices}
 and Fig.~\ref{fig:TB_SC}(a)]. 
 The excitation spectrum has spin-polarized point nodes protected by a winding number (see Table~\ref{tab:topolo_class_main}). Each of the nodes carries a topological charge. The non-trivial topological invariant leads to localized boundary states between projections of the nodes [see Fig.~\ref{fig:TB_SC}(a)]~\cite{Matsuura_2013,PhysRevB.83.224511}. Depending on the termination direction, there is a region between the nodes, where the localized state is spin-polarized and non-degenerate, i.e., a Majorana boundary mode. The topology of the superconducting state affects physical responses~\cite{Tunnelling_unconSC,PhysRevLett.74.3451,PhysRevB.53.9371}, particularly the Josephson current~\cite{PhysRevB.53.R11957, PhysRevB.56.892, PhysRevLett.86.5369, PhysRevB.71.134520, 10.1063/1.1789931, Shukrinov_2022}.

\textit{Fractional ac spin Josephson effect} ---
 \begin{figure}[t]
    \centering
    \includegraphics[width=\linewidth]{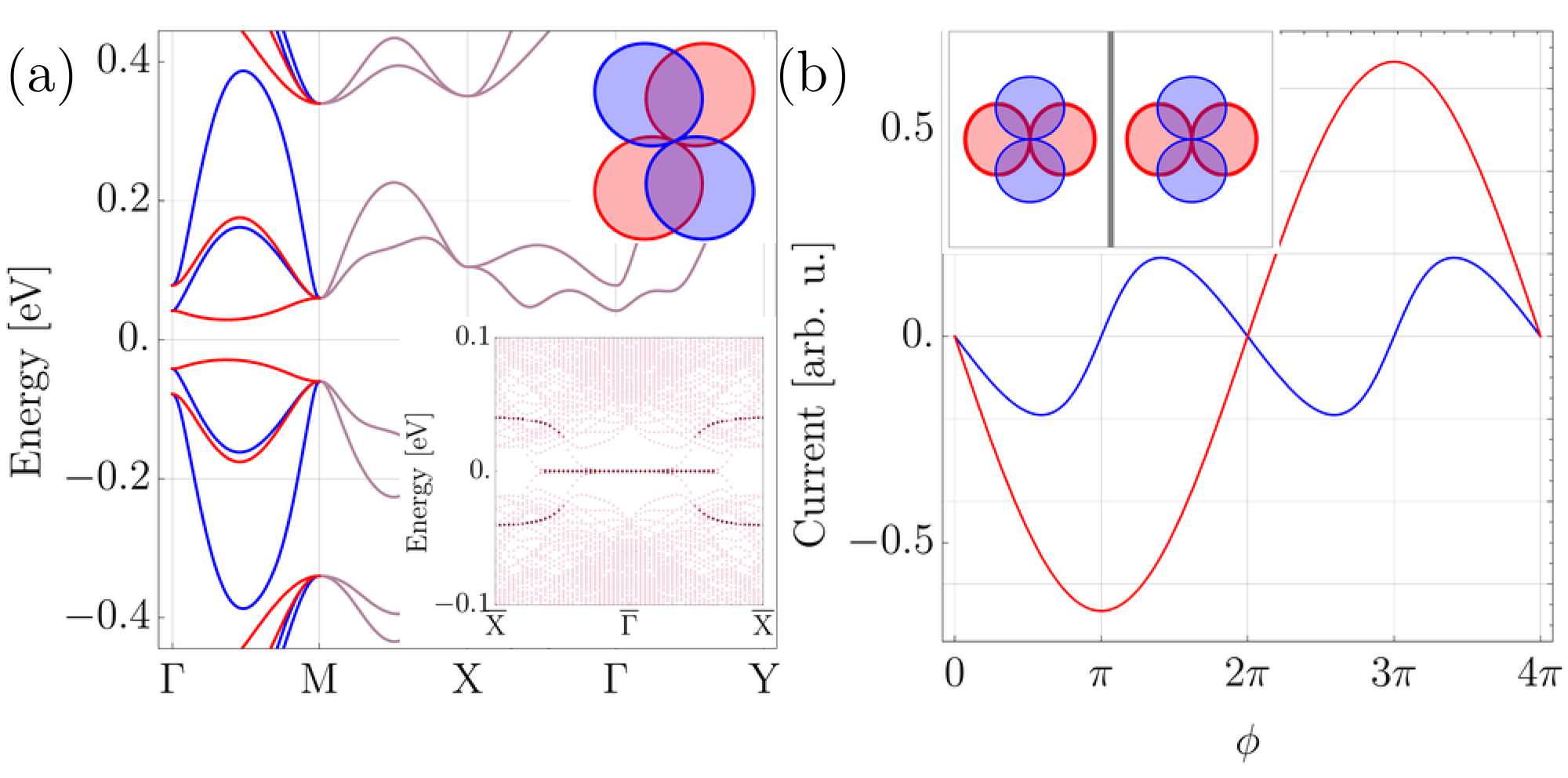}
    \caption{\justifying (a) BdG excitation spectrum for the stable altermagnetic superconducting state (1, 1) in the material example (see Fig.~\ref{fig:TB_normal}). We show the slab model spectrum with 40 layers in the inset. The termination is along $x - y$ line. The pairing potential is shown in the inset. (b)  Josephson current through the junction between two superconducting islands with $\text{D}_+ k_x+\text{D}_- k_y$ and $e^{i{\delta}}(\text{D}_+ k_x+\text{D}_- k_y)$ pairing potentials. The junction geometry is shown in the inset.}
    \label{fig:TB_SC}
\end{figure}
 The altermagnetic superconducting state has an exotic spatial anisotropy. For spin $S = 1$ and $ S =- 1$ pairing potentials, the spatial dependence is related by the {altermagnetic} symmetry. This different anisotropy for spin-up and spin-down states leads to spin-polarized responses related to  the topology of the state. Here we consider the junction between two altermagnetic superconductors with pairing potential $\Delta_0(\text{D}_+ k_x + \text{D}_- k_y)$ and the phase difference ${\delta}$ [see Fig.~\ref{fig:TB_SC}(b)]. 
 We compute the ac Josephson current for small applied voltage $eV \ll |\Delta_0|$ and $T=0$ 
 {(see Appendices)}.
 The phase has the time dependence $\delta(t) = 2eVt/\hbar$, and it changes adiabatically. 
 $S = 1$ pairs contribute similarly as in the 1D p-wave superconductor, leading to $4\pi$ periodicity~\cite{10.1063/1.1789931}. 
 The $4\pi$ periodicity is due to the topological boundary state in the spin-up subspace, whereas the spin-down subspace does not support the boundary modes for the chosen termination, {consequently, the $S = -1$ pairs exhibit a  $2 \pi$ periodicity. }  In the limit of transparent barrier we obtain
  \begin{equation}
    I_{\uparrow}= -\frac{\pi \Delta_0}{4} \frac{2e}{\hbar} \sin{\frac{{\delta}}{2}},\
    I_{\downarrow} = -\frac{ \Delta_0}{2}  \frac{2e}{\hbar}\sin{\frac{{\delta}}{2}} \frac{\cos{\tfrac{{\delta}}{2}}}{|\cos{\tfrac{{\delta}}{2}}|}.
  \end{equation}
In the tunneling regime we have
\begin{equation}
        I_{\uparrow} = -\frac{4\Delta_0}{3Z_0} \frac{2e}{\hbar} \sin{\frac{{\delta}}{2}},\
    I_{\downarrow} = -\frac{ \Delta_0}{2Z_0^2} \frac{2e}{\hbar} \sin{{\delta}},
\end{equation}
with $Z_0= \frac{2m U_0}{\hbar^2 k_F} \gg 1$. In this limit, the main contribution is given by spin-polarized $S=1$ Cooper pairs $I_{\uparrow}$.

\textit{Discussion}---
In this work, we constructed all superconducting basis functions for all spin-point groups describing collinear compensated magnets, including altermagnets.
Based on this complete list, we identified a number of exotic and novel
pairing states that transform under higher-dimensional coreps of 
the spin-point groups.
The states can break or preserve the symmetries of the altermagnet. 
In the later case, the altermagnetic superconducting gap combines anisotropic pairing potentials for $S = \pm 1$. The spin-polarized pairings are related by a spatial (improper) rotation, e.g., a $\pi/2$ rotation. This is in contrast to antiferromagnets, where the potentials overlap fully. 
{The structure of the pairing wave function is different for spin up and spin down}, 
leading to the non-unitary state with a finite vector $id(\mathbf{k})\cross d(\mathbf{k})^*$.
If the order preserves the antiunitary $(\mathcal{T}2_{\perp}||E)$ symmetry, the symmetry protects spin-polarized gapless point excitations in 2D systems. The nodes carry a topological charge, resulting in non-degenerate zero-energy states at the boundary, i.e., Majorana boundary modes. The topological degeneracies and anisotropic spin-space locking lead to exotic spin-polarized topological responses, e.g., fractional ac Josephson currents for only one of the two
spin polarizations. 

The superconducting order can break altermagnetic symmetries. The spin-polarized half-and-half
metal-superconductor breaks $(2_{\perp}||g)$ symmetry and has the gap only in one spin-subspace. Spin chiral state breaks antiunitary symmetry of the collinear order $(\mathcal{T}2_{\perp}||E)$, such that the spin subspaces can carry finite Chern numbers leading to chiral boundary modes.

The discussed pairings are three examples of the large number of novel states.
The complete list of symmetry-allowed pairings 
lays down the foundations for the study of exotic superconductivity in altermagnets. It opens up research avenues, including the following:
(i) Investigation of superconductivity in known altermagnetic candidate materials based on symmetry principles. 
It is a common practice to use point group representations to analyze the possible pairing symmetries~\cite{RevModPhys.63.239}. The approach, however, includes an assumption that SOC is strong so that the spin is coupled to the lattice. For the negligible SOC~\cite{10.1143/PTP.74.221, 10.1143/PTP.75.442}, the framework is too restrictive and misses symmetry-allowed states, particularly the ones discussed in this work. 
(ii)
Exploration of a large variety of novel pairing states, especially those
having order parameters with more than two components. The states described by higher-dimensional coreps lead to intricate physical phenomena corresponding to multicomponent order parameters~\cite{Almoalem2024}. The interplay and competition between several superconducting orders can result in exotic domain physics~\cite{Aoki_2022, PhysRevB.107.085407}. Additionally, the SO(2) spin rotational symmetry comes with large continuous phase degeneracy. The spatial fluctuations of the phases can lead to exotic topological defects~\cite{PismaZhETF, BABAEV2004397, PhysRevLett.99.197002, 10.21468/SciPostPhysCore.5.1.003}, similar to phase vortices in 2D superconductors~\cite{1971JETP...32..493B, Kosterlitz_1973}.
(iii)
Design of novel superconducting devices that exploit the spin-polarization
of the Cooper pair condensates together with the exotic spin-point group symmetries.

\textit{Acknowledgments}--- 
We thank Igor Mazin, Kazushi Kanoda,  Robin Scholle, Rina Tazai, Niclas Heinsdorf, Marcel Franz, Paul McClarty, Hana Rebecca Schiff, Judit Romhanyi for useful discussions.
This work is funded by the Deutsche Forschungsgemeinschaft (DFG, German Research Foundation) – TRR 360 – 492547816.
The irreducible co-representations of all spin-point groups describing the symmetries of compensated collinear magnetic phases are provided in the Zenodo repository~\cite{zenodo} as LaTeX files. For each irreducible co-representation, the corresponding superconducting pairing basis functions are also provided in LaTeX format. The raw data are included in the dataset as CSV files.

\begin{center}
  {\bfseries\large{Appendices}\par}
\end{center}

{\textit{Basis construction---}
Here we present a step-by-step example of how to construct the basis vectors 
for the coreps of the 
$^{\bar{1}}$m$^{\bar{1}}$m$^1$2 spin-point group (SPG).
The halving unitary group is $X = X_{1/2} + (2_{\perp} || M_x) X_{1/2}$, where $X_{1/2} = \text{SO(2)} \times \{(E||E), (E||2_z) \}$ [see Eq.~(\ref{halvinggroup})]. The corresponding $\textbf{B}_{+1}\textbf{B}_{-1}$ irreducible representation (rep) matrices are listed in Table~\ref{reps_mm2}.
\begin{table}[h]
    \fontsize{8}{8}\selectfont
    \centering
{
        \begin{tabular}{c|ccc|c} 
&$R_{\phi}$$||$E&$2_{\perp}||M_x$&$E||2_z$&Type\\ 
 \hline 
$\textbf{B}_{+1}\textbf{B}_{-1}$ & $\left(
\begin{array}{cc}
 e^{i   \phi } & 0 \\
 0 & e^{-i   \phi } \\
\end{array}
\right)$ & $\left(
\begin{array}{cc}
 0 & 1 \\
 1 & 0 \\
\end{array}
\right)$ & $\left(
\begin{array}{cc}
 -1 & 0 \\
  0 &-1 \\
\end{array}
\right)$ &  (a)\\ 
\end{tabular}
}
\caption{
\justifying {$\textbf{B}_{+1}\textbf{B}_{-1}$ rep matrices for the unitary halving group of the $^{\bar{1}}$m$^{\bar{1}}$m$^1$2 SPG. $R_{\phi}$ represents the SO(2) rotation.}}
        \label{reps_mm2}
\end{table}
To construct the basis vectors, we apply the projection operators defined in Eq.~(\ref{projector})
\begin{align}
    \mathbf{P}_{11} &= \tfrac{2}{4\cdot 2\pi}\int_{0}^{2\pi}d \phi e^{-i \phi} [(R_{\phi}||E) -   (R_{\phi}||C_{2z})],\\
    \mathbf{P}_{21} &= \tfrac{2}{4\cdot 2\pi}\int_{0}^{2\pi}d \phi e^{-i \phi} [  (2_{\perp}R_{\phi}||M_{x}) -(2_{\perp}R_{\phi}||M_{y})]. \notag
\end{align}
The action of symmetries on the triplet $d(\mathbf{k})$ vector [see Eq.~(\ref{order_decomp})] is given by
$(s||g)d(\mathbf{k}) = \Lambda(s) d(\Lambda^{-} (g) \mathbf{k})$, 
where $\Lambda(s)$ denotes the $3\times3$ rotation matrix associated with the $s$ operation
and $\Lambda^{-} (g) \mathbf{k}$ is the inverse action of the real-space operation $g$ on the momentum $\mathbf{k}$ vector. 
Applying the projector $\mathbf{P}_{11}$ to a general vector $d(\mathbf{k})$, we obtain  to lowest order,
$d_1^{\textbf{B}_{+1}\textbf{B}_{-1}}(\mathbf{k}) = \left(c_x k_x+ c_y k_y\right) (1, i,0)^T$, with $c_x, c_y \in \mathbb{C}$. 
The second basis vector follows from $ d_2^{\textbf{B}_{+1}\textbf{B}_{-1}}(\mathbf{k}) = \textbf{P}_{21} d_1^{\textbf{B}_{+1}\textbf{B}_{-1}}(\mathbf{k}) =\left(-c_x k_x+ c_y k_y\right)(1,-i,0)^T$.}

{
Using the basis of the unitary halving group rep $\bra{d_1^{\textbf{B}_{+1}\textbf{B}_{-1}}(\mathbf{k}), d_2^{\textbf{B}_{+1}\textbf{B}_{-1}}(\mathbf{k})}$, we now construct the corresponding corep.
The Dimmock indicator~\cite{Dimmock1963} for the rep ${\textbf{B}_{+1}\textbf{B}_{-1}}$ is $
 \frac{1}{|X|} \sum_{R \in X} \chi([R(\mathcal{T}2_{\perp}||E)]^2) =  +1$. This places the rep in class (a) [see Eq.~(\ref{DimmockClasses})], implying that the rep $\mathbf{D}^{\textbf{B}_{+1}\textbf{B}_{-1}}(R)$ and its conjugate $\mathbf{D}^{{\textbf{B}_{+1}\textbf{B}_{-1}}^*}(A^{-1}RA)$ are equivalent.  Moreover, the transformation matrix $N=\big(\begin{smallmatrix}
        1 & 0\\
        0 & 1
    \end{smallmatrix} \big)$
relates these two sets of rep matrices. With this, the corep basis is obtained as $\bra{d_1^{\textbf{B}_{+1}\textbf{B}_{-1}}(\mathbf{k}), d_2^{\textbf{B}_{+1}\textbf{B}_{-1}}(\mathbf{k})} + \bra{(\mathcal{T}2_{\perp}||E)d_1^{\textbf{B}_{+1}\textbf{B}_{-1}}(\mathbf{k}), (\mathcal{T}2_{\perp}||E)d_2^{\textbf{B}_{+1}\textbf{B}_{-1}}(\mathbf{k})}N^{-1}$. The resulting corep $\text{B}_{+1}\text{B}_{-1}$ has the same dimensionality as the original rep, and its basis is given by
    \begin{align}\label{d+-}
    d_+(\mathbf{k}) &= \left(r_x k_x+ r_y k_y\right) (1, i,0)^T, \\
    d_-(\mathbf{k})&=
     \left(-r_x k_x+ r_y k_y\right)(1,-i,0)^T, \ r_x, r_y \in \mathbb{R}.
    \end{align}
\begin{table}[h]
    \fontsize{8}{8}\selectfont
    \centering
{
        \begin{tabular}{c|cccc} 
&$R_{\phi}$$||$E&$2_{\perp}||M_x$&$2_{\perp}||M_y$&$\mathcal{T}2_{\perp}$$||$E\\ 
 \hline 
$\text{A}_{+}$ & $
\begin{array}{c}
 1 \\
\end{array}
$ & $
\begin{array}{c}
 1 \\
\end{array}
$ & $
\begin{array}{c}
 1 \\
\end{array}
$ & $
\begin{array}{c}
 1 \\
\end{array}
$ \\ 
 \hline 
$\text{A}_{-}$ & $
\begin{array}{c}
 1 \\
\end{array}
$ & $
\begin{array}{c}
 -1 \\
\end{array}
$ & $
\begin{array}{c}
 -1 \\
\end{array}
$ & $
\begin{array}{c}
 1 \\
\end{array}
$ \\ 
 \hline 
$\text{B}_{-}$ & $
\begin{array}{c}
 1 \\
\end{array}
$ & $
\begin{array}{c}
 -1 \\
\end{array}
$ & $
\begin{array}{c}
 1 \\
\end{array}
$ & $
\begin{array}{c}
 1 \\
\end{array}
$ \\ 
 \hline 
$\text{B}_{+}$ & $
\begin{array}{c}
 1 \\
\end{array}
$ & $
\begin{array}{c}
 1 \\
\end{array}
$ & $
\begin{array}{c}
 -1 \\
\end{array}
$ & $
\begin{array}{c}
 1 \\
\end{array}
$ \\ 
 \hline 
$\text{A}_{+1}\text{A}_{-1}$ & $\left(
\begin{array}{cc}
 e^{i   \phi } & 0 \\
 0 & e^{-i   \phi } \\
\end{array}
\right)$ & $\left(
\begin{array}{cc}
 0 & 1 \\
 1 & 0 \\
\end{array}
\right)$ & $\left(
\begin{array}{cc}
 0 & 1 \\
 1 & 0 \\
\end{array}
\right)$ & $\left(
\begin{array}{cc}
1 & 0 \\
 0 & 1 \\
\end{array}
\right)$ \\ 
 \hline 
$\text{B}_{+1}\text{B}_{-1}$ & $\left(
\begin{array}{cc}
 e^{i   \phi } & 0 \\
 0 & e^{-i   \phi } \\
\end{array}
\right)$ & $\left(
\begin{array}{cc}
 0 & 1 \\
 1 & 0 \\
\end{array}
\right)$ & $\left(
\begin{array}{cc}
 0 & -1 \\
 -1 & 0 \\
\end{array}
\right)$ & $\left(
\begin{array}{cc}
 1 & 0 \\
 0 & 1 \\
\end{array}
\right)$ \\ 
\end{tabular}
}
        \caption{\justifying {Coreps matrices of SPG $^{\bar{1}}$m$^{\bar{1}}$m$^1$2. }}
        \label{coreps_mm2}
\end{table}
}

{
\textit{Ginzburg-Landau free energy---}A superconducting order parameter transforming according to the $\text{B}_{+1}\text{B}_{-1}$ corep can be written as
$d(\mathbf{k}) = \upsilon_1 d_{+}(\mathbf{k}) +\upsilon_2 d_-(\mathbf{k})$. In the basis $\{d_{+}(\mathbf{k}), d_-(\mathbf{k})\}$, the corep matrices describe how the complex amplitudes $\upsilon_1, \upsilon_2$ transform (see Table~\ref{coreps_mm2}). For example,
we have $(R_{\phi}||E)\{\upsilon_1,\upsilon_2\} \to \{e^{i \phi} \upsilon_1, e^{-i\phi}\upsilon_2 \}$, $(2_{\perp}||M_x)\{\upsilon_1,\upsilon_2\} \to \{\upsilon_2,\upsilon_1 \}$, $(\mathcal{T}2_{\perp}||E)\{\upsilon_1,\upsilon_2\} \to \{\upsilon_1^*,\upsilon_2^* \}$,
and a global 
U(1) phase acts as
$\Phi \{\upsilon_1, \upsilon_2\} \to \{e^{i\omega}\upsilon_1, e^{i\omega}\upsilon_2 \}$.
These symmetry operations constrain the free-energy expansion. The lowest-order invariants are the quadratic term
$\alpha (|\upsilon_1|^2 + |\upsilon_2|^2)$ and the quartic terms $\beta_1 (|\upsilon_1|^2 + |\upsilon_2|^2)^2$, $\beta_2 |\upsilon_1|^2|\upsilon_2|^2$. For positive and large $\beta_1 > 0$, the free energy is bounded from below. If $\beta_2 < 0$, the minimum occurs at $|\upsilon_1| = |\upsilon_2| \neq 0$. This state we call \textit{altermagnetic superconductor}. There is continuous degeneracy associated with the relative phase between the two order parameter components. If instead $\beta_2 > 0$, the minimum corresponds to a single component state $|\upsilon_1| \neq 0, |\upsilon_2| = 0$ or $|\upsilon_1| = 0, |\upsilon_2| \neq 0$, which we term \textit{half-and-half metal-superconductor}.
}

{
We consider spin-orbit coupling (SOC) as a perturbation that lifts the continuous degeneracy of the altermagnetic superconductor.
In the presence of SOC, we fix the magnetic order along the $y$ axis, 
reducing the symmetry group to a subgroup where all operations act on both spin and lattice degrees of freedom.
The resulting magnetic point group (MPG) is $mm'2'$. It contains the unitary mirror $M_x$ and the antiunitary symmetry $\mathcal{T}M_y$. They flip magnetic moments and exchange sublattices. The antiunitary $\mathcal{T}2_z$ leaves the moments unchanged.
The group generators act on the order-parameter components as $(2_x||M_x)\{\upsilon_1,\upsilon_2\} \to \{\upsilon_2,\upsilon_1\}$ and $(\mathcal{T}2_z||2_z)\{\upsilon_1,\upsilon_2\} \to \{-\upsilon_1^*,-\upsilon_2^*\}$.
To lowest order in the free-energy expansion, SOC allows an additional invariant $\alpha_{\text{SOC}}(\upsilon_1 \upsilon^*_2 + \upsilon_2 \upsilon^*_1)$ which lifts the continuous degeneracy of the functional minimum for $\beta_2 < 0$.
For $\alpha_{\text{SOC}} < 0$, the minimum corresponds to the zero-phase state $(1, 1)$, whereas for $\alpha_{\text{SOC}} > 0$, the global minimum corresponds to the $\pi$-phase state $(1, -1)$.
}

{
\textit{Topological
classification---} 
The $(\mathcal{T}2_{\perp}||E)$ symmetry enforces the reality of the gap function (see Eq.~\ref{d+-}). For the $(1,1)$ altermagnetic superconductor we have $\Delta_{(1,1)}(\mathbf{k}) = i([d_+(\mathbf{k}) + d_-(\mathbf{k})]\cdot \tau) \tau_y = -2 (r_x k_x \tau_z + r_y k_y \tau_0)$.
Because the system has SO(2) spin-rotational symmetry, the spin sectors decouple, resulting in real, spin-resolved gap functions
$\Delta_{(1,1)\pm}(\mathbf{k}) = -2 (\pm r_x k_x + r_y k_y)$. 
Each spin sector is then described by a BdG Hamiltonian in the symmetry class {BDI}~\cite{PhysRevB.78.195125}, so the corresponding one-dimensional systems are characterized by an integer $\mathbb{Z}$ winding number. 
In addition, the system possesses the $(2_{\perp}||g)$ symmetries, which relate spin subspaces. The two subspaces have related topological invariants.
}

{
The two spin-resolved gaps $\Delta_{(1,1)\pm}(\mathbf{k})$ have nodal directions related by $(2_{\perp}||M_x)$ symmetry, i.e. $\pm r_x k_x + r_y k_y = 0$. This gap structure results in zero-energy excitations in the BdG spectrum [see Fig.~\ref{fig:TB_SC}(a)] protected by the $\mathbb{Z}$ winding number. 
}

{
\textit{$\kappa$-(BEDT-TTF)$_2$X tight-binding model---}The model has the sublattice degrees of freedom, corresponding to the four sites A, B, C, and D~\cite{PhysRevResearch.5.043171} (see Fig.~\ref{fig:TB_normal})
{\fontsize{8}{8}\selectfont
\begin{align}
H& =\tau_0 \otimes
\begin{pmatrix}
0 & t_b \epsilon_x^*+t_a & t_p \left(\epsilon_x^*+1\right) & t_q \left(\epsilon_y^*+1\right) \\
t_b \epsilon_x+t_a &0 & t_q \left(\epsilon_y+1\right) & t_p \left(\epsilon_x+1\right) \\
t_p \left(\epsilon_x+1\right) & t_q \left(\epsilon_y^*+1\right) & 0 & \epsilon_y^* \left(t_a \epsilon_x+t_b\right) \\
t_q \left(\epsilon_y+1\right) & t_p \left(\epsilon_x^*+1\right) & \epsilon_y \left(t_a \epsilon_x^*+t_b\right) & 0
\end{pmatrix} \notag\\ 
&+ h \tau_z \otimes \text{diag}[1,1,-1,-1], 
\end{align}}where the first term describes the normal state model, and the second term introduces the staggered collinear magnetic order, $\epsilon_x = e^{i k_x}, \epsilon_y=e^{i k_y}$. 
The model parameters are taken $(t_a, t_b, t_p, t_q, h) = (-0.207,\ -0.067,\ -0.102,\ 0.043,\ 0.2)$~\cite{PhysRevResearch.5.043171, PhysRevB.89.045102}. To introduce superconductivity, we use BdG formalism and consider the $(1, 1)$ pairing potential  $\Delta = ([\text{D}_+ \left(r_x k_x+ r_y k_y\right)  +
 \text{D}_- \left(-r_x k_x+ r_y k_y\right)]\cdot \tau) i\tau_y \otimes \text{diag}[1,1,1,1]$ with $(r_x, r_y) = (0.05, 0.1)$.
 }

{
\textit{ac
Josephson effect---}Up to this point, we have considered the $(1,1)$ state of the $\text{B}_{+1}\text{B}_{-1}$ corep given in Eq.~(\ref{d+-}). 
To simplify the forthcoming calculations and make the results more transparent, we now consider the $(1,1)$ state obtained for a system with four-fold rotational symmetry, e.g. the $\text{B}^2_{+1}\text{B}^1_{-1}$ corep of the $ {}^{\bar{1}}4 {}^1m {}^{\bar{1}}m$ spin–point group. We focus on altermagnetic superconductors described by the gap function $\Delta = \Delta_0(\text{D}_{+} k_x + \text{D}_{-} k_y)/k_F$, where $k_F = \sqrt{2 m \mu}/\hbar$, separated by a barrier and characterized by a superconducting phase difference $\delta$ [see Fig.~\ref{fig:TB_SC}(b)]. 
For simplicity, we neglect anisotropies in the electronic band structure of the normal metal and assume a parabolic dispersion
$\epsilon = \hbar^2 (k_x^2 +k_y^2) / 2m - \mu$ for both spin projections. 
The superconductors form an interface normal to the $x$ direction and are periodic along the $y$ direction.
Following Ref.~\cite{10.1063/1.1789931}, we model the contact between them by the potential $U(x) = U_0 \delta(x)$. Because the spin sectors decouple, we treat the Bogoliubov–de Gennes equations for the two spin subspaces separately
\begin{align}
    H_{\uparrow} &= \begin{pmatrix}
\epsilon(\hat{k}_x,k_y) + U(x)& -2 \hat{k}_x \Delta_0/ k_F\\
-2 \hat{k}_x^{\dagger} \Delta_0/ k_F & -\epsilon (\hat{k}_x, k_y) - U(x)
\end{pmatrix},\\
H_{\downarrow} &= \begin{pmatrix}
\epsilon (\hat{k}_x, k_y) + U(x)& 2 k_y \Delta_0/ k_F\\
2 k_y \Delta_0/ k_F & -\epsilon(\hat{k}_x, k_y) - U(x)
\end{pmatrix},
\end{align}
where $\hat{k}_x = -i \partial_x$.
Solving the BdG equations for bound states localized at the interface,
we find the states' energies
\begin{align}
    E_{\uparrow}(k_y, \delta) = \pm \tfrac{2\sqrt{k_F^2-k_y^2}\Delta_0}{k_F} \sqrt{D_{k_y}} \cos \tfrac{\delta }{2},\\
    E_{\downarrow}(k_y, \delta) = \pm \tfrac{2 k_y \Delta_0}{k_F} \sqrt{1- D_{k_y} \sin ^2\tfrac{\delta }{2}},
\end{align}
where $D_{k_y} = 4/(4+Z_{k_y}^2)$, 
with 
$Z_{k_y}=Z_0k_F/ \sqrt{k_F^2 - k_y^2}
$, and $Z_0=2m U_0/(\hbar^2 k_F)$ characterizes the junction transparency. 
We now compute the dynamical ac Josephson effect. Assuming a small applied voltage $eV \ll \Delta_0$, the phase difference across the junction evolves adiabatically as $\delta(t) = 2eVt/\hbar$. At zero temperature $T = 0$ 
the Josephson currents for each spin component are given by
$
    I = \int_{0}^{k_F}\tfrac{d k_y}{k_F} \tfrac{2e}{\hbar}\tfrac{\partial E (k_y, \delta)}{\partial \delta},
$
where contributions from all $k_y$ modes are summed. The integration results are given by
\begin{center}
\begin{align}
I_{\uparrow}
=&-\tfrac{\Delta_0}{8}\tfrac{2e}{\hbar}
\left[
(4-Z_0^2)\arctan\!\left(\tfrac{2}{Z_0}\right)+2Z_0
\right]\sin\tfrac{\delta}{2},
\\
I_{\downarrow}
=&-\tfrac{e\Delta_0}{4\hbar}\tan\tfrac{\delta}{2}
\left[
\sqrt{(Z_0^2+4)(Z_0^2+2\cos\delta+2)}-Z_0^2
\right]
\nonumber\\
&+\tfrac{e\Delta_0}{8\hbar}\tan\tfrac{\delta}{2}\,
Z_0^2\tfrac{\cos\delta+3}{\cos(\delta/2)} \nonumber\\
&\times
\tanh^{-1}\!\left[
\sec\tfrac{\delta}{2}
\sqrt{\tfrac{Z_0^2+2\cos\delta+2}{Z_0^2+4}}
\right]
\nonumber\\
&-\tfrac{e\Delta_0}{8\hbar}\tan\tfrac{\delta}{2}\,
Z_0^2\tfrac{\cos\delta+3}{\cos(\delta/2)}
\tanh^{-1}\!\left(\sec\tfrac{\delta}{2}\right).
\end{align}
\end{center}
}

\textit{Spin chiral superconductor---}We consider the corep $\text{E}^{\text{u}} _{+1}\text{E}^{\text{u}}_{-1}$ of SPG ${}^{\bar{1}}6/{}^{\bar{1}}m{}^1m{}^{\bar{1}}m$. This corep has four basis vectors 
$-\frac{i}{\sqrt2}\text{D}_- (k_x+i k_y)$,
$-\frac{i}{\sqrt2}\text{D}_- (k_x-i k_y)$,
$-\frac{i}{\sqrt2}\text{D}_+(k_x+i k_y)$,
$-\frac{i}{\sqrt2}\text{D}_+(k_x-i k_y)$,
with transformation matrices listed in Table~\ref{tab:EEtransCorep}.
Following the procedure outlined above, we construct the GL free energy [see Eq.~(\ref{f4_6mmm})]. In the regime defined by $\alpha<0, \ \beta_3<0,\
                  4\beta_1+\beta_3>0,\
 \beta_2>\beta_3,\
 \beta_4>\beta_3,\
 \beta_2+\beta_4-\beta_3>|\beta_5|$, the free energy is minimized for $|\upsilon_1| = |\upsilon_3|= r, \ \upsilon_2= \upsilon_4 = 0$ or $|\upsilon_2| = |\upsilon_4|= r, \ \upsilon_1= \upsilon_3 = 0$, with $r^2=-\tfrac{\alpha}{4\beta_1+\beta_3}$.
   The two domains have opposite chiralities and are exchanged by $(\mathcal{T}2_{\perp}||E)$.

\begin{table}[h]
{
\fontsize{8}{8}\selectfont
    \centering
    \begin{tabular}{c|c|c}
$R_{\phi}$$||$E&$2_{\perp}||M_z$&$2_{\perp}||M_y$\\ 
 $\left(
\begin{array}{cccc}
 e^{i   \phi } & 0 & 0 & 0 \\
 0 & e^{i  \phi } & 0 & 0 \\
 0 & 0 & e^{-i  \phi } & 0 \\
 0 & 0 & 0 & e^{-i   \phi } \\
\end{array}
\right)$ & $\left(
\begin{array}{cccc}
 0 & 0 & 1 & 0 \\
 0 & 0 & 0 & 1 \\
 1 & 0 & 0 & 0 \\
 0 & 1 & 0 & 0 \\
\end{array}
\right)$ & $\left(
\begin{array}{cccc}
 0 & 0 & 0& 1  \\
 0 & 0  & 1 &0\\
 0&1  & 0 & 0 \\
  1& 0 & 0 & 0 \\
\end{array}
\right)$ \\
\end{tabular}
\\
\vspace{0.05cm}
 \begin{tabular}{c|c}
$2_{\perp}||C_{6z}$&$\mathcal{T}2_{\perp}$$||$E\\
$\left(
\begin{array}{cccc}
 0 & 0 & e^{i \pi/3} & 0 \\
 0 & 0 & 0 & e^{-i \pi/3} \\
 e^{i \pi/3} & 0 & 0 & 0 \\
 0 & e^{-i \pi/3} & 0 & 0 \\
\end{array}
\right)$ & $\left(
\begin{array}{cccc}
 0&-1 & 0 & 0 \\
 -1 & 0 & 0&0 \\
 0 & 0 &0 & -1  \\
 0 & 0 & -1&0 \\
\end{array}
\right)$ \\ 
    \end{tabular}
}
    \caption{\justifying {$\text{E}^{\text{u}}_{+1}\text{E}^{\text{u}}_{-1}$ corep matrices of SPG ${}^{\bar{1}}6/{}^{\bar{1}}m{}^1m{}^{\bar{1}}m$. 
    }}
    \label{tab:EEtransCorep}
\end{table}

\bibliography{bibliography.bib}

\newpage
\clearpage
\end{document}


\definecolor{myPurple}{RGB}{240, 180, 210}
\definecolor{myGreen}{RGB}{228, 240, 180}
\definecolor{myYellow}{RGB}{240, 240, 160}
\definecolor{myBlue}{RGB}{140, 200, 255}
\title{Supplemental Material for
\\
``Exotic superconducting states in altermagnets"}

\date{\today}

\author{Kirill Parshukov}
\email{k.parshukov@fkf.mpg.de}
\affiliation{Max Planck Institute for Solid State Research, Heisenbergstrasse 1, D-70569 Stuttgart, Germany}

\author{Andreas P. Schnyder}
\email{a.schnyder@fkf.mpg.de}
\affiliation{Max Planck Institute for Solid State Research, Heisenbergstrasse 1, D-70569 Stuttgart, Germany}

\setcounter{figure}{0}
\makeatletter 
\renewcommand{\thefigure}{S\@arabic\c@figure} 

\setcounter{equation}{0}
\makeatletter 
\renewcommand{\theequation}{S\@arabic\c@equation} 

 \setcounter{table}{0}
\makeatletter 
\renewcommand{\thetable}{S\@Roman\c@table} 

 \setcounter{section}{0}
\makeatletter 
\renewcommand{\thesection}{S\@Roman\c@section} 

\makeatother

\maketitle
\onecolumngrid

\widetext

\tableofcontents

 \vspace{0.8cm}

This Supplemental Material (SM) is divided into twelve parts. First, we describe the construction of spin-point group irreducible co-representations (coreps) in Sec.~\ref{coreps_constr}. Using the constructed coreps we discuss two approaches to the construction of their basis functions in Sec.~\ref{basis_constr} and Sec.~\ref{basis_constr_2}. The construction of Ginzburg-Landau free energy is discussed in Sec.~\ref{Free_energy_construction}. The effect of weak spin-orbit coupling on the stable superconducting states is considered in Sec.~\ref{SOC_effect}. In Sec.~\ref{high_dim_corep} we discuss four-dimensional superconducting order parameter leading to spin chiral superconductivity. In Sec.~\ref{topo_class} we describe topological classification of SO(2) symmetric superconducting states. The material example is discussed in Sec.~\ref{material_exampl}. Josephson ac current between two altermagnetic superconductors is computed in Sec.~\ref{junction}. In Sec.~\ref{naming_table} we correspond spin-point groups names from Ref.~\cite{schiff2023spin} to our notations. In the last two sections we tabulate all coreps basis functions for all spin-point groups describing compensated collinear magnets (Sec.~\ref{tables_basis_basis}) and matrices of the coreps (Sec.~\ref{tables_coreps}).

\onecolumngrid
\newpage

\section{Construction of irreducible co-representations}\label{coreps_constr}
In the main text, we discussed two approaches to constructing the basis functions. To use the first one, we need to know the irreducible representations (reps) of the unitary halving group $X$
\begin{equation}\label{GS_decomposition}
    G_S = X + X (\mathcal{T}2_{\perp} || E).
\end{equation}
These reps construction is discussed in Ref.~\cite{schiff2023spin} (see it for more details and examples), but the matrices are not tabulated. Therefore, we derive the reps, and further we construct irresducible co-representations (coreps) for all spin-point groups describing collinear compensated magnets. 

The unitary halving group $X$ can be decomposed into two parts
\begin{align}
    X &= X_{1/2} + (2_{\perp} || g) X_{1/2}, \\
    X_{1/2} &= \text{SO(2)} \times H,
\end{align}
where $X_{1/2}$ is a site-symmetry group.  We use the induction procedure to obtain $X$ reps~\cite{bradley1972mathematical,aroyo2006bilbao}. $X_{1/2} $ is the normal subgroup of $ X $, and its reps matrices $\mathsf{D}(X_{1/2})_{\mu,\nu}$ are labeled by the $\nu$ reps of SO(2) and $\mu$ reps of $H$ groups (the $H$ reps we obtained from the GTPack Mathematica package~\cite{10.3389/fphy.2018.00086,GTPack}). The set of matrices 
\begin{equation}
    (\mathsf{D}(X_{1/2})_{\mu,\nu})_a = \{ \mathsf{D}(a b a^{-1})_{\mu,\nu}, \, b \in X_{1/2}, \, a \in X \}
\end{equation}
form one of the reps of $X_{1/2}$ and is known as conjugate to  $\mathsf{D}(X_{1/2})_{\mu,\nu}$. It is either equivalent, 
\begin{equation}
    (\mathsf{D}(X_{1/2})_{\mu,\nu})_a \sim \mathsf{D}(X_{1/2})_{\mu,\nu}
\end{equation}
or non-equivalent to  $\mathsf{D}(X_{1/2})_{\mu,\nu}$. The set of non-equivalent reps $ (\mathsf{D}(X_{1/2})_{\mu,\nu})_a$  for all $a \in X$ is the orbit of $ \mathsf{D}(X_{1/2})_{\mu,\nu}$. In our case, the orbit can have 1 or 2 sets (the length of the orbit), and the only symmetry we need to consider is $a = (2_{\perp}||g)$. 
The conjugate reps $ (\mathsf{D}(X_{1/2})_{\mu,\nu})_a$ is equivalent to one of the reps $\mathsf{D}(X_{1/2})_{\mu',\nu'}$.

The spin rotation $(2_{\perp}||g)$ relates two reps $\nu$ and $-\nu$ of the SO(2) spin-rotational group. For the group element $b=(R(\varphi)||h) \in X_{1/2}$ we have
\begin{equation}
    (2_{\perp} || g) \, b \, (2_{\perp} || g^{-1}) 
    = (2_{\perp} R(\varphi) 2_{\perp} || g h g^{-1}) 
    = (R(-\varphi) || g h g^{-1}),
\end{equation}
therefore, the conjugated matrices correspond to the $-\nu$ reps. 

If the orbit length is 1, $\nu=-\nu'=0, \ \mu=\mu'$. There are two induced reps (we use the bold font for the unitary group $X$ reps)
\begin{align}
\mathbf{D}_{\mu,\nu,+}(b) &= \mathbf{D}_{\mu,\nu-}(b) = \mathsf{D}_{\mu, \nu}(b), \quad b \in X_{1/2},\\
\mathbf{D}_{\mu,\nu,+}((2_{\perp}||g)) &= -\mathbf{D}_{\mu,\nu,-}((2_{\perp}||g)) = Z,
\end{align}
where $Z$ is determined by the conditions
\begin{align}
\mathsf{D}_{\mu,\nu}((2_{\perp}||g) b (2_{\perp}||g)^{-1}) &= Z \mathsf{D}_{\mu,\nu}(b) Z^{-1}, \quad b  \in X_{1/2},\\
Z^2 &= \mathsf{D}_{\mu,\nu}((2_{\perp}||g)^2).
\end{align}

If the orbit length is 2,  $\nu=-\nu'\neq 0$ or $ \mu \neq\mu'$. The induced reps have the higher dimension, and the matrices are given by
\begin{equation}
    \mathbf{D}_{\mu,\nu}(b) = 
    \begin{pmatrix}
        \mathsf{D}_{\mu,\nu}(b) & 0\\
        0 & (\mathsf{D}_{\mu,\nu})_{(2_{\perp}||g)}(b)
    \end{pmatrix}, \quad b \in X_{1/2}, \quad
    \mathbf{D}_{\mu,\nu}((2_{\perp}||g)) = 
    \begin{pmatrix}
        0 & \mathsf{D}_{\mu,\nu}((2_{\perp}||g)^2)\\
        \mathbf{I} & 0
    \end{pmatrix},
\end{equation}
where $\mathbf{I}$ is the identity matrix. Further we enumerate the new reps with $\mathbf{\Gamma}^{(i)}$ and their matrices $\mathbf{D}^{(i)}$.

$ \mathbf{D}^{(i)}$ will be useful further for the basis functions construction. Here we proceed using the reps to obtain the spin-point group coreps. As we decomposed the group $G_S$ into the unitary halving group $X$ and the anti-unitary elements $XA$ with $A = (\mathcal{T}2_{\perp}||E)$ (see Eq.~\ref{GS_decomposition}), we can use the Dimmock rules~\cite{Dimmock1963} to obtain the coreps~\cite{bradley1972mathematical}. For each of the unitary group $X$ reps we compute the indicator
\begin{equation}\label{DimmockSM}
  \xi^{\mathbf{\Gamma}^{(i)}} = \frac{1}{|X|} \sum_{R \in X} \chi^{\mathbf{\Gamma}^{(i)}}([RA]^2) = 
  \begin{dcases*} 
  +1 & (a) \\ 
  -1 & (b)\\
  0 & (c)
  \end{dcases*},
\end{equation}
where $\chi^{\mathbf{\Gamma}^{(i)}}([RA]^2)$ is the character of the unitary element $[RA]^2$ for the reps $\mathbf{\Gamma}^{(i)}$.

First, we consider the case (a). The matrices $D$ of the spin-point group coreps $G_S$ (we use the standard font for the spin-point group coreps) take the form
\begin{equation}\label{caseA_Dimmock_matrices}
    D(R) = \mathbf{D}(R), \quad D(A) = \pm N, \quad R \in X,
\end{equation}
where coreps with $\pm$ are equivalent. $N$ matrix is determined by
\begin{equation}\label{caseA_Dimmock}
    \mathbf{D}(R) = N \mathbf{D}^*(A^{-1}RA) N^{-1}, \quad NN^* = \mathbf{D}(A^2),
\end{equation}
and can be computed with $N = \tfrac{1}{|G|}\sum_{R \in G} \mathbf{D}^{(i)}(R) X \overline{\mathbf{D}^{(i)}}(R^{-1})$, where $X$ is chosen, so that $N$ is unitary. 
The coreps basis is given by $ \bra{\boldsymbol{\eta}^{(i)} + A\boldsymbol{\eta}^{(i)} N^{-1}}$,
where $\boldsymbol{\eta}^{(i)}$ is the basis for the $\mathbf{\Gamma}^{(i)}$ reps. With the bra notation we highlight that the basis vectors are written in a row.

For the case (b) we have
\begin{align}
    D(R) &= 
    \begin{pmatrix}
        \mathbf{D}(R) & 0 \\
        0 & \mathbf{D}(R) 
    \end{pmatrix},
    \quad 
    D(A) = 
    \begin{pmatrix}
        0 & -N \\
        N & 0 
    \end{pmatrix}, \quad R \in X,\\
    \mathbf{D}(R) &= N \mathbf{D}^*(A^{-1}RA) N^{-1}, \quad NN^* = -\mathbf{D}(A^2),
\end{align}
with the coreps basis row $\bra{\boldsymbol{\eta}^{(i)}, A\boldsymbol{\eta}^{(i)} N^{-1}}$. The number of basis vectors is doubled.

In the case (c) the reps $ \mathbf{D}(R)$  and $ \mathbf{D}^*(A^{-1}RA)$ are not equivalent and the coreps is given by
\begin{equation}
    D(R) = 
    \begin{pmatrix}
        \mathbf{D}(R) & 0 \\
        0 & \mathbf{D}^*(A^{-1}RA) 
    \end{pmatrix},
    \quad 
    D(A) = 
    \begin{pmatrix}
        0 & \mathbf{D}(A^2) \\
        \mathbf{I} & 0 
    \end{pmatrix}, \quad R \in X,\\
\end{equation}
with the doubled coreps basis $\bra{\boldsymbol{\eta}^{(i)}, \boldsymbol{\eta}^{\overline{(i)}}}$.

The summary of the three cases is given by
\begin{align*}
 &(a) \ \bra{\boldsymbol{\eta}^{(i)} + (\mathcal{T}2_{\perp}||E)\boldsymbol{\eta}^{(i)} N^{-1}}, \ \mathbf{\Gamma}^{(i)} \stackrel{N}{\sim} \overline{\mathbf{\Gamma}^{(i)}},\\
 &(b) \ \bra{\boldsymbol{\eta}^{(i)}, (\mathcal{T}2_{\perp}||E)\boldsymbol{\eta}^{(i)} N^{-1}}, \ \mathbf{\Gamma}^{(i)}\stackrel{N}{\sim} \overline{\mathbf{\Gamma}^{(i)}},\\
&(c) \ \bra{\boldsymbol{\eta}^{(i)}, \boldsymbol{\eta}^{\overline{(i)}}}, \ \mathbf{\Gamma}^{(i)}{\not\sim} \overline{\mathbf{\Gamma}^{(i)}}.
\end{align*}

\section{Basis function construction. Approach 1}\label{basis_constr}
To begin with, we consider the first approach discussed in the main text. 
With the known $X$ reps matrices $\mathbf{D}_{\mu,\nu}$, we construct the reps basis functions. Here we consider the spin-point group $^{\bar{1}}$m$^{\bar{1}}$m$^1$2. The X subgroup reps are listed in Teble~\ref{irreps_7}.
\begin{table}[h]
    \centering
        \begin{tabular}{|c|c|c|c|c|c|} 
\hline 
7: C2+IC2x$ \times $C2&$R_{\phi}$$||$Ee&$R_{\phi}$$||$C2z&$2_{\perp}R_{\phi}||$IC2x&$2_{\perp}R_{\phi}||$IC2y& Type\\ 
 \hline 
$\textbf{A}_{+}$ & $\left(
\begin{array}{c}
 1 \\
\end{array}
\right)$ & $\left(
\begin{array}{c}
 1 \\
\end{array}
\right)$ & $\left(
\begin{array}{c}
 1 \\
\end{array}
\right)$ & $\left(
\begin{array}{c}
 1 \\
\end{array}
\right)$
& (a)\\ 
 \hline 
$\textbf{A}_{-}$ & $\left(
\begin{array}{c}
 1 \\
\end{array}
\right)$ & $\left(
\begin{array}{c}
 1 \\
\end{array}
\right)$ & $\left(
\begin{array}{c}
 -1 \\
\end{array}
\right)$ & $\left(
\begin{array}{c}
 -1 \\
\end{array}
\right)$& (a) \\ 
 \hline 
$\textbf{B}_{-}$ & $\left(
\begin{array}{c}
 1 \\
\end{array}
\right)$ & $\left(
\begin{array}{c}
 -1 \\
\end{array}
\right)$ & $\left(
\begin{array}{c}
 -1 \\
\end{array}
\right)$ & $\left(
\begin{array}{c}
 1 \\
\end{array}
\right)$ & (a)\\ 
 \hline 
$\textbf{B}_{+}$ & $\left(
\begin{array}{c}
 1 \\
\end{array}
\right)$ & $\left(
\begin{array}{c}
 -1 \\
\end{array}
\right)$ & $\left(
\begin{array}{c}
 1 \\
\end{array}
\right)$ & $\left(
\begin{array}{c}
 -1 \\
\end{array}
\right)$ & (a)\\ 
 \hline 
$\textbf{A}_{\nu}\textbf{A}_{-\nu}$ & $\left(
\begin{array}{cc}
 e^{i \nu  \phi } & 0 \\
 0 & e^{-i \nu  \phi } \\
\end{array}
\right)$ & $\left(
\begin{array}{cc}
 e^{i \nu  \phi } & 0 \\
 0 & e^{-i \nu  \phi } \\
\end{array}
\right)$ & $\left(
\begin{array}{cc}
 0 & e^{-i \nu  \phi } \\
 e^{i \nu  \phi } & 0 \\
\end{array}
\right)$ & $\left(
\begin{array}{cc}
 0 & e^{-i \nu  \phi } \\
 e^{i \nu  \phi } & 0 \\
\end{array}
\right)$ & (a)\\ 
 \hline 
$\textbf{B}_{\nu}\textbf{B}_{-\nu}$ & $\left(
\begin{array}{cc}
 e^{i \nu  \phi } & 0 \\
 0 & e^{-i \nu  \phi } \\
\end{array}
\right)$ & $\left(
\begin{array}{cc}
 -e^{i \nu  \phi } & 0 \\
 0 & -e^{-i \nu  \phi } \\
\end{array}
\right)$ & $\left(
\begin{array}{cc}
 0 & e^{-i \nu  \phi } \\
 e^{i \nu  \phi } & 0 \\
\end{array}
\right)$ & $\left(
\begin{array}{cc}
 0 & -e^{-i \nu  \phi } \\
 -e^{i \nu  \phi } & 0 \\
\end{array}
\right)$ & (a)\\ 
 \hline 
\end{tabular}
        \caption{\justifying Irreducible representations of the unitary halving group $X$ of the spin-point group $^{\bar{1}}$m$^{\bar{1}}$m$^1$2. In the last two {rows} $\nu$ is the positive integer number $\nu > 0$. Type represents the Dimmock indicator case (see Eq.~\ref{DimmockSM}).}
        \label{irreps_7}
\end{table}

We construct the basis vectors using the projector operators
\begin{equation}\label{projectorSM}
    \mathbf{P}_{kl}^{(i)} = \frac{l_i}{|X|}\sum_{R\in X}\mathbf{D}^{(i)*}(R)_{kl}R,
\end{equation}
where $l_i$ is the dimension of the $i^{\text{th}}$ reps. The explicit form of the first basis vector projector for the $\textbf{B}_{\nu}\textbf{B}_{-\nu}$ reps is given by
\begin{equation}
    \mathbf{P}_{11}^{\textbf{B}_{\nu}\textbf{B}_{-\nu}} = \frac{2}{4\cdot 2\pi}\int \limits_{0}^{2\pi}d \phi [ e^{-i\nu \phi} (R_{\phi}||E) - e^{-i\nu \phi}  (R_{\phi}||C_{2z})]. 
\end{equation}
To obtain the second vector we use the operator
\begin{equation}
    \mathbf{P}_{21}^{\textbf{B}_{\nu}\textbf{B}_{-\nu}} = \frac{2}{4\cdot 2\pi}\int \limits_{0}^{2\pi}d \phi [ e^{-i\nu \phi} (2_{\perp}R_{\phi}||IC_{2x}) -e^{-i\nu \phi}(2_{\perp}R_{\phi}||IC_{2y})]. 
\end{equation}
The action of the symmetries on pairing potentials is presented in Table~\ref{parity_transformationsSM}
\begin{equation}
    \Delta(\mathbf{k}) = i[ \psi(\mathbf{k}) +    (d(\mathbf{k}) \cdot \hat{\tau})] \tau^y,
\end{equation}
where $\tau$ are Pauli matrices. 
\begin{table}[h]
    \centering
    \renewcommand{\arraystretch}{1.5}
    \begin{tabular}{|c|c|c|}
        \hline
        \textbf{Transformation} & \textbf{Even parity } $\psi(\mathbf{k})$ & \textbf{Odd parity} $d(\mathbf{k})$\\
        \hline
        Point group  $R=(E||g)$ & $R \psi(\mathbf{k}) = \psi({\Lambda}^{-} (g) \mathbf{k})$ & $R d(\mathbf{k}) = d({\Lambda}^{-} (g) \mathbf{k})$ \\
        \hline
        Spin rotation  $R=(s||E)$ & $R \psi(\mathbf{k}) = \psi(\mathbf{k})$ & $R d(\mathbf{k}) = {\Lambda} (s) d(\mathbf{k})$ \\
        \hline
        Time reversal $\mathcal{T}=(\mathcal{T}||E)$ & $\mathcal{T} \psi(\mathbf{k}) = \psi^*(-\mathbf{k})$ & $\mathcal{T} d(\mathbf{k}) = -d^*(-\mathbf{k})$ \\
        \hline
        U(1) gauge & $\Phi \psi(\mathbf{k}) = e^{i \omega} \psi(\mathbf{k})$ & $\Phi d(\mathbf{k}) = e^{i \omega} d(\mathbf{k})$ \\
        \hline
    \end{tabular}
    \caption{Transformation properties of even and odd parity pairing states.}
    \label{parity_transformationsSM}
\end{table}
We obtained two basis vectors in the lowest order for $\nu=1$
\begin{equation}
    d_1^{\textbf{B}_{\nu}\textbf{B}_{-\nu}}(\mathbf{k}) = \left(c_x k_x+ c_y k_y\right)
    \begin{pmatrix} 1\\
                            i\\
                            0
    \end{pmatrix},  \quad
    d_2^{\textbf{B}_{\nu}\textbf{B}_{-\nu}}(\mathbf{k}) =
     \left(-c_x k_x+ c_y k_y\right)
     \begin{pmatrix} 1\\
                            -i\\
                            0
    \end{pmatrix},
\end{equation}
with arbitrary complex constants $c_x, c_y$. The singlet pairing potential is always zero. This is a general feature of the $\nu = 1$ irreducible representations, as they describe spin 1 states. 

Now we are interested in the basis functions of the spin-point group coreps. Considering the antiunitary $(2_{\perp}\mathcal{T}||E)$ symmetry, there are Dimmock rules~\cite{Dimmock1963} to construct coreps from the halving unitary subgroup reps (see Eq.~\ref{DimmockSM}). Depending on its value the unitary reps are classified into three categories. 

We obtained the Dimmock indicator $\xi^{\textbf{B}_{\nu}\textbf{B}_{-\nu}} = +1$, and therefore, we follow the case (a)~\ref{caseA_Dimmock_matrices}, and determine the matrices for the the coreps $\text{B}_{\nu}\text{B}_{-\nu}$ (see Table.~\ref{coreps_mm2}). The basis functions take the form 
\begin{equation}
\{d_1^{\text{B}_{\nu}\text{B}_{-\nu}}(\mathbf{k}), d_2^{\text{B}_{\nu}\text{B}_{-\nu}}(\mathbf{k})\} = \{d_1^{\textbf{B}_{\nu}\textbf{B}_{-\nu}}(\mathbf{k}), d_2^{\textbf{B}_{\nu}\textbf{B}_{-\nu}}(\mathbf{k})\} + (\mathcal{T}2_{\perp}||E)\{d_1^{\textbf{B}_{\nu}\textbf{B}_{-\nu}}(\mathbf{k}), d_2^{\textbf{B}_{\nu}\textbf{B}_{-\nu}}(\mathbf{k})\}N^{-1},
\end{equation} 
where  $N$ is the equivalence matrix (see Eq.~\ref{caseA_Dimmock}) between $(2_{\perp}\mathcal{T}||E)$ conjugated reps $\mathbf{D}^{\textbf{B}_{\nu}\textbf{B}_{-\nu}}(R)$ and $\mathbf{D}^{{\textbf{B}_{\nu}\textbf{B}_{-\nu}}*}(A^{-1}RA)$
\begin{equation}
    N = 
    \begin{pmatrix}
        1 & 0\\
        0 & 1
    \end{pmatrix}.
\end{equation}
We obtained the $\text{B}_{\nu}\text{B}_{-\nu}$ coreps basis
\begin{equation}
    d_1^{\text{B}_{\nu}\text{B}_{-\nu}}(\mathbf{k}) = \left(r_x k_x+ r_y k_y\right)
    \begin{pmatrix} 1\\
                            i\\
                            0
    \end{pmatrix},  \quad
    d_2^{\text{B}_{\nu}\text{B}_{-\nu}}(\mathbf{k}) =
     \left(-r_x k_x+ r_y k_y\right)
     \begin{pmatrix} 1\\
                            -i\\
                            0
    \end{pmatrix},
\end{equation}
where $r_x, r_y$ are the real coefficients.
\begin{table}[h]
    \centering
        \begin{tabular}{|c|c|c|c|c|} 
\hline 
7: C2+IC2x$ \times $C2&$R_{\phi}$$||$Ee&$2_{\perp}R_{\phi}||$IC2x&$2_{\perp}R_{\phi}||$IC2y&$R_{\phi}\mathcal{T}2_{\perp}$$||$Ee\\ 
 \hline 
$\text{A}_{+}$ & $\left(
\begin{array}{c}
 1 \\
\end{array}
\right)$ & $\left(
\begin{array}{c}
 1 \\
\end{array}
\right)$ & $\left(
\begin{array}{c}
 1 \\
\end{array}
\right)$ & $\left(
\begin{array}{c}
 1 \\
\end{array}
\right)$ \\ 
 \hline 
$\text{A}_{-}$ & $\left(
\begin{array}{c}
 1 \\
\end{array}
\right)$ & $\left(
\begin{array}{c}
 -1 \\
\end{array}
\right)$ & $\left(
\begin{array}{c}
 -1 \\
\end{array}
\right)$ & $\left(
\begin{array}{c}
 1 \\
\end{array}
\right)$ \\ 
 \hline 
$\text{B}_{-}$ & $\left(
\begin{array}{c}
 1 \\
\end{array}
\right)$ & $\left(
\begin{array}{c}
 -1 \\
\end{array}
\right)$ & $\left(
\begin{array}{c}
 1 \\
\end{array}
\right)$ & $\left(
\begin{array}{c}
 1 \\
\end{array}
\right)$ \\ 
 \hline 
$\text{B}_{+}$ & $\left(
\begin{array}{c}
 1 \\
\end{array}
\right)$ & $\left(
\begin{array}{c}
 1 \\
\end{array}
\right)$ & $\left(
\begin{array}{c}
 -1 \\
\end{array}
\right)$ & $\left(
\begin{array}{c}
 1 \\
\end{array}
\right)$ \\ 
 \hline 
$\text{A}_{\nu}\text{A}_{-\nu}$ & $\left(
\begin{array}{cc}
 e^{i \nu  \phi } & 0 \\
 0 & e^{-i \nu  \phi } \\
\end{array}
\right)$ & $\left(
\begin{array}{cc}
 0 & e^{-i \nu  \phi } \\
 e^{i \nu  \phi } & 0 \\
\end{array}
\right)$ & $\left(
\begin{array}{cc}
 0 & e^{-i \nu  \phi } \\
 e^{i \nu  \phi } & 0 \\
\end{array}
\right)$ & $\left(
\begin{array}{cc}
 e^{i \nu  \phi } & 0 \\
 0 & e^{-i \nu  \phi } \\
\end{array}
\right)$ \\
\hline 
\end{tabular}
        \caption{\justifying Irreducible co-representations of the  spin-point group $^{\bar{1}}$m$^{\bar{1}}$m$^1$2. In the last two {rows} $\nu$ is the positive integer number $\nu > 0$.}
        \label{coreps_mm2}
\end{table}

\section{Basis function construction. Approach 2}\label{basis_constr_2}

There is another approach to construct the coreps basis functions for a group with non-unitary elements. Instead of constructing the basis functions for the unitary halving group reps, the corep projector can be derived from the orthogonality expression for coreps $D^{(i)}$ (see Ref.~\cite{Dimmock1963})
\begin{equation}
    \frac{|G_S|}{l^{(i)}}\delta_{\alpha\beta}\delta_{\mu\nu} = \sum_R [D^{(i)}(R)_{\alpha \mu} D^{(i)}(R)^*_{\beta \nu}\\ 
+ D^{(i)}(RA)_{\alpha \nu} D^{(i)}(RA)^*_{\beta \mu}],
\end{equation}
where $|G_S|$ is the number of group elements, $l_i$ is the dimension of corep.  With this expression we can introduce the projector operator
\begin{equation}
    P_{kl}^{(i)} = \frac{l_i}{|G_S|}\sum_{R\in X}[D^{(i)*}(R)_{kl}R + D^{(i)*}(RA)_{kl}RA],
\end{equation}
where $A=(\mathcal{T}2_{\perp}||E)$ is the antiunitary symmetry, $D^{(i)}$ are the co-representation matrices with the basis vectors $\eta^{(i)}_l$
\begin{align}
    R\eta_l^{(i)} &= \sum_m D^{(i)}(R)_{ml} \eta_m^{(i)},\\
    RA\eta_l^{(i)} &= \sum_m D^{(i)}(RA)_{ml} \eta_m^{(i)}.
\end{align}
We obtain 
\begin{equation}
    P_{kl}^{(i)}\eta_l^{(i)} = \frac{l_i}{|M|}\sum_m\sum_{R}[D^{(i)*}(R)_{kl} D^{(i)}(R)_{ml} \eta_m^{(i)}\\
    + D^{(i)*}(RA)_{kl} D^{(i)}(RA)_{ml} \eta_m^{(i)}] = \eta_k^{(i)}.
\end{equation}

For the $\text{B}_{\nu}\text{B}_{-\nu}$ copers we have
\begin{align}
    P_{11}^{\text{B}_{\nu}\text{B}_{-\nu}} = \frac{2}{8\cdot 2\pi}\int \limits_{0}^{2\pi}d \phi [ &e^{-i\nu \phi} (R_{\phi}||E) - e^{-i\nu \phi}  (R_{\phi}||C_{2z}) + e^{-i\nu \phi} (R_{\phi}||E)(\mathcal{T}2_{\perp}||E) - e^{-i\nu \phi}  (R_{\phi}||C_{2z})(\mathcal{T}2_{\perp}||E) ],\\
    P_{21}^{\text{B}_{\nu}\text{B}_{-\nu}} = \frac{2}{8\cdot 2\pi}\int \limits_{0}^{2\pi}d \phi [ &e^{-i\nu \phi} (2_{\perp}R_{\phi}||IC_{2x}) -e^{-i\nu \phi}(2_{\perp}R_{\phi}||IC_{2y})\\
    &+ e^{-i\nu \phi} (2_{\perp}R_{\phi}||IC_{2x})(\mathcal{T}2_{\perp}||E)  -e^{-i\nu \phi}(2_{\perp}R_{\phi}||IC_{2y})(\mathcal{T}2_{\perp}||E) ]. 
\end{align}

\section{Multisublattice effect}

{
In this work we were interested in the superconducting pairings described by the \textit{novel} irreducible co-representations (coreps) of spin-point groups. These are higher-dimensional coreps $\Gamma^{(i)}_{\nu=1}$ describing states with non-zero spin $S = \pm 1$. Generally, these coreps combine paired representations of SO(2) group. In other words, the basis vectors come in pairs, and their spin parts are $\text{D}_\pm = (1, \pm i, 0)^T$.
These are the fundamental building blocks 
out of which the SC pairing states are composed for any altermagnet
with sublattice / orbital degrees of freedom.
For example, if we explicitly take into account
the two magnetic sublattices of the altermagnet, the SC  pairing function becomes 
\begin{align}
        \Delta^{(i)}_{\nu=\pm1}(\mathbf{k}) &= (\text{D}_\pm \cdot \hat{\tau}) i\tau^y \otimes O_{\pm}^i(\mathbf{k}),\\
O_{\pm}^i(\mathbf{k}) &=        
\begin{pmatrix}
o_{11}^i(\mathbf{k}) & o_{12}^i(\mathbf{k}) \\
o_{21}^i(\mathbf{k}) & o_{22}^i(\mathbf{k}) 
\end{pmatrix},
\end{align}
with a $2 \times 2$ matrix for sublattice degrees of freedom.
}

{
Now, in order to analyze the transformation property of this SC pairing matrix, we
consider the 
unitary halving group $X$ of the altermagnetic spin-point group $G_S$ (as discussed in the main text) 
\begin{equation}
     X = X_{1/2} + (2_{\perp} || g) X_{1/2},
\end{equation}
where $X_{1/2}$ is the sublattice symmetry group, and 
$(2_{\perp} || g) X_{1/2}$ is the set of elements that relate two magnetic sublattices. Elements of $X_{1/2}$ act trivially on the sublattice degrees of freedom, whereas $(2_{\perp} || g) X_{1/2}$ exchange them. We list the transformation properties in the Table~\ref{O_transformations}.
\begin{table}[h]
{
    \centering
    \renewcommand{\arraystretch}{1.5}
    \begin{tabular}{|c|c|}
        \hline
        Sublattice group  $R=(E||g)$ & $R O(\mathbf{k}) = O(\hat{D}^{-} (g) \mathbf{k})$ \\
        \hline
        Sublattice exchange $R=(2_{\perp}||g)$ & $R O(\mathbf{k}) = \sigma_x O(\hat{D}^{-} (g) \mathbf{k}) \sigma_x$\\
        \hline
        Time reversal  $\mathcal{T}=(\mathcal{T}||E)$ & $\mathcal{T} O(\mathbf{k}) = O^*(-\mathbf{k})$ \\
        \hline
    \end{tabular}
    \caption{Transformation properties of the sublattice part $O(\mathbf{k})$ of the pairing potential. Here, $\hat{D}^{-}(g)\mathbf{k}$ represents the inverse action of the symmetry operation $g$ on the momentum vector $\mathbf{k}$.}
    \label{O_transformations}
    }
\end{table}
With the known transformation properties we can apply our algorithm to obtain the coreps basis states for the SC pairing matrix, using projector operators (here, for the unitary halving group rep)
\begin{equation}\label{proj_rep}
    \mathbf{P}_{nn}^{(i)} = \frac{l_i}{|X|}\sum_{R\in X}\mathbf{D}^{(i)*}(R)_{nn}R,
\end{equation}
where $\mathbf{D}^{(i)}_{nn}$ is the $n^{\text{th}}$ diagonal matrix element of the rep $\mathbf{\Gamma}^{(i)}$. However, $\mathbf{D}^{(i)}_{nn}$ is non-zero only for sublattice group elements $X_{1/2}$.
Consider, e.g., the coreps of 
the spin-point group $^{\bar{1}}$m$^{\bar{1}}$m$^12$
given in 
Table~\ref{irreps_7}. 
Matrices of the reps $\textbf{A}_{\nu}\textbf{A}_{-\nu}$ and $\textbf{B}_{\nu}\textbf{B}_{-\nu}$ in the $S = \pm1$ basis have diagonal elements $\mathbf{D}^{(i)}_{nn}$ only for operations $R \in X_{1/2}$. Conversely, all other group elements $R \in (2_{\perp} || g) X_{1/2}$ exchange both the sublattices and the $S = \pm1$ states. Consequently, the diagonal elements of their corresponding matrices are zero $\mathbf{D}^{(i)}_{nn} = 0$.
\begin{table}[h]
{
    \centering

        \caption{Irreducible representations of the unitary halving group $X$ of the spin-point group $^{\bar{1}}$m$^{\bar{1}}$m$^1$2. In the last two rows $\nu$ is the positive integer number $\nu > 0$. Type represents the Dimmock indicator case.}
        \label{irreps_7}
        }
\end{table}
With this, we conclude that the projector~\eqref{proj_rep} does not include terms that transform the sublattice degrees of freedom, and therefore the matrix $O^i(\mathbf{k})$ transforms as a scalar function.  Consequently, all four elements
of the
SC pairing matrix,
$o_{11}^i(\mathbf{k}), o_{12}^i(\mathbf{k}),o_{21}^i(\mathbf{k}), \text{and} \ o_{22}^i(\mathbf{k})$ take the form of the  basis states as constructed in the paper, without any additional restrictions or modifications.
Thus, the constructed basis states of our paper can be directly applied to describe the altermagnetic sublattice degrees of freedom. 
}

{
If, in addition to the magnetic sublattice, there exist further orbital / sublatice degrees of freedom, the situation becomes more subtle, as one needs to know the particular symmetry properties of these additional degrees of freedom. However, a procedure as outlined above using the projector~\eqref{proj_rep} can still be performed. This is an analysis that would
need to be performed case by case, with the knowledge of the particular orbital degrees of freedom of the given altermagnet. 
In multi-orbital altermagnets, where multiple orbitals contribute to the Fermi surfaces, symmetries of the orbital degrees of freedom can create additional constraints, thus limiting the possible superconducting pairings.
}

\section{Ginzburg-Landau free energy construction}\label{Free_energy_construction}
In this section, we describe the construction of Ginzburg-Landau free energy for the spin-point group symmetries.
Taking a corep $\Gamma^{(i)}$ the quadratic term transforms as $\Gamma^{(i)}\otimes\Gamma^{(i)*}$, where $\Gamma^{(i)*}$ is the conjugated corep. The quartic term transforms as $\Gamma^{(i)}\otimes\Gamma^{(i)*}\otimes\Gamma^{(i)}\otimes\Gamma^{(i)*}$. To obtain the invariants we use decomposition of the co-representaions into the irreducible ones (or coreps). An analog of Clebsch-Gordan coefficients is derived by Karavaev~\cite{Karavaev, bradley1972mathematical}
\begin{equation}
c_{i,j,k} = \frac{\frac{1}{|X|} \sum \limits_{R \in X} \chi^i(R) \chi^j(R) \chi^k(R)^*}{\frac{1}{|X|} \sum \limits_{R \in X} \chi^k(R) \chi^k(R)^*},
\end{equation}
where $\chi^k(R)$ is the character of the symmetry operator $R$ for the corep $\Gamma^{(k)}$. In our case the summation is understood as the integration for the elements of the SO(2) spin rotational group.
The coefficients $c_{i,j,k}$ determine how many times the corep $\Gamma^{(k)}$ appears in the decomposition of the co-representation $\Gamma^{(i)}\otimes\Gamma^{(j)}$. 

We consider $^{\bar{1}}$m$^{\bar{1}}$m$^1$2 spin-point group, and its corep $\text{B}_{\nu}\text{B}_{-\nu}$ with $\nu = 1$ (see Table 7 in Sec.~\ref{tables_coreps} and Sec.~\ref{tables_basis_basis}) . The quadratic term takes the form 
\begin{equation}
\text{B}_{+1}\text{B}_{-1}\otimes(\text{B}_{+1}\text{B}_{-1})^*  = \text{A}_{+} \oplus \text{A}_{-} \oplus \text{A}_{+2}\text{A}_{-2}.
\end{equation}
The trivial corep $\text{A}_{+}$ appears once, therefore, there is only one invariant term $\alpha (|{\upsilon}_1|^2 + |{\upsilon}_2|^2)$. 

To obtain the quartic terms, we note that only the products of identical coreps can give the trivial one. Therefore, in the expression
\begin{equation}
(\text{A}_{+} \oplus \text{A}_{-} \oplus \text{A}_{+2}\text{A}_{-2}) \otimes (\text{A}_{+} \oplus \text{A}_{-} \oplus \text{A}_{+2}\text{A}_{-2}),
\end{equation}
we consider only $\text{A}_{+} \otimes \text{A}_{+}$, $\text{A}_{-} \otimes \text{A}_{-}$ and $\text{A}_{+2}\text{A}_{-2} \otimes \text{A}_{+2}\text{A}_{-2}$ co-representations. Computing their decomposition, we obtained that each of the co-representations has one trivial corep $\text{A}_{+}$. In total, there are three trivial coreps, however, there are only two independent invariant terms $\beta_1 (|{\upsilon}_1|^2 + |{\upsilon}_2|^2)^2 + \beta_2 |{\upsilon}_1|^2|{\upsilon}_2|^2$. The free energy takes the form
\begin{equation}
    \label{free_energy_2D_Corep}
    f[|{\upsilon}_1|, |{\upsilon}_2|] = \alpha (|{\upsilon}_1|^2 +|{\upsilon}_2|^2) + \beta_1 (|{\upsilon}_1|^2 + |{\upsilon}_2|^2)^2 + \beta_2 |{\upsilon}_1|^2|{\upsilon}_2|^2.
\end{equation}
The form of the free energy is general for the two-dimensional altermagnetic coreps describing spin-1 states.
For positive $\beta_1 > 0$, there is a minimum. If $\beta_2 < 0$, the solution is $|{\upsilon}_1| = |{\upsilon}_2| \neq 0$. For positive  $\beta_2 > 0$, the minimum is at $|{\upsilon}_1| \neq 0, |{\upsilon}_2| = 0$ or $|{\upsilon}_1| = 0, |{\upsilon}_2| \neq 0$.

In the case $\beta_2 < 0$, the superconducting state does not break altermagnetic symmetries. At the same time, there is continuous degeneracy for states with different relative phases between two order parameters. To lift this degeneracy, we consider further the effect of spin-orbit coupling.

\section{Effect of spin-orbit coupling}\label{SOC_effect}
Spin-point group symmetries are not exact symmetries. In real materials spin-orbit coupling (SOC) is always present, and its effect has to be discussed. Here we consider it as a perturbation that does not create new minima of the free energy, but it lifts the degeneracy. Particularly for the functional Eq. (\ref{free_energy_2D_Corep}), we are allowed to add additional terms. 

In the presence of SOC, the symmetries act both on spin and lattice degrees of freedom. In the main text we discussed the spin-point group $^{\bar{1}}$m$^{\bar{1}}$m$^1$2 for $\kappa$-(BEDT-TTF)$_2$X~\cite{PhysRevLett.89.017003,doi:10.1143/JPSJ.74.2351,PhysRevB.76.054509}. In the presence of SOC, the magnetic order points along the $y$ axis. The magnetic point group is given by $mm'2'$, where the unitary mirror $M_x$ and antiunitary mirror $\mathcal{T}M_y$ flip the magnetic moments and relate the sublattices. $\mathcal{T}2_z$ leaves the magnetic moments invariant. 

Without SOC, the spin part of the basis vector does not correspond to the real space coordinates, it has it's own space x, y, z. To construct the basis vectors, we have chosen the quantization axis along the z axis. This direction must point now along the real space $y$ axis. Therefore, in the real space coordinates, vectors $\text{D}_{+}$ and $\text{D}_{-}$ take the form $(i,0,1)$, $(-i,0,1)$.
With this we can determine the action of the magnetic symmetries on the $\text{B}_{+1}\text{B}_{-1}$ basis vectors $\{\text{D}_- \left(-r_x k_x+ r_y k_y\right), \text{D}_+ \left(r_x k_x+ r_y k_y\right)\}$. 
\begin{table}[h]
    \centering
    \renewcommand{\arraystretch}{1.3}
    \begin{tabular}{c|c |c |c}
    $E$ & $M_x$ & $\mathcal{T}M_y$ & $\mathcal{T}2_z$\\

        $\begin{bmatrix} 1 & 0 \\ 0 & 1 \end{bmatrix}$ & 
        $\begin{bmatrix} 0 & 1 \\ 1 & 0 \end{bmatrix}$ & 
        $\begin{bmatrix} 0 & -1 \\ -1 & 0 \end{bmatrix} \mathcal{K}$ & 
        $\begin{bmatrix} -1 & 0 \\ 0 & -1 \end{bmatrix}\mathcal{K}$ \\
    \end{tabular}
    \caption{$mm'2'$ magnetic point group symmetry action on vectors $\{\text{D}_- \left(-r_x k_x+ r_y k_y\right), \text{D}_+ \left(r_x k_x+ r_y k_y\right)\}$. }\label{mm2_coreps}
\end{table}

With the use of Table \ref{mm2_coreps}, we find that the magnetic point group symmetry allow for the quadratic invariant ${\upsilon}_1 {\upsilon}^*_2 + {\upsilon}_2 {\upsilon}^*_1$ in the free energy. It's the lowest order term due to SOC effect. The functional takes the form
\begin{equation}
    f[|{\upsilon}_1|, |{\upsilon}_2|] = \alpha (|{\upsilon}_1|^2 +|{\upsilon}_2|^2) + \alpha_{\text{SOC}}({\upsilon}_1 {\upsilon}^*_2 + {\upsilon}_2 {\upsilon}^*_1) + \beta_1 (|{\upsilon}_1|^2 + |{\upsilon}_2|^2)^2 + \beta_2 |{\upsilon}_1|^2|{\upsilon}_2|^2.
\end{equation}

The new term lifts the degeneracy of the functional minimum. In the case $\beta_2 < 0$, the solution without SOC is $|{\upsilon}_1|=|{\upsilon}_2|$ with any possible relative phase. The SOC lifts the degeneracy leaving two possible minima. If $\alpha_{\text{SOC}} < 0$ the state with zero relative phase $(1, 1)$ has the minimal free energy. When $\alpha_{\text{SOC}} > 0$, the global minimum is for the state with the $\pi$ relative phase $(1, -1)$. 
\section{Higher-dimensional coreps and spin chiral superconductors}\label{high_dim_corep}
The phenomenological consideration of superconducting basis functions for four- and higher-dimensional coreps is more complicated, as the number of parameters in the Ginzburg-Landau free energy grows fast. We will discuss one example and leave the comprehensive study for the future work. Here our goal is to study the interplay of several basis functions in each of the spin subspaces. We want to give a constructive proof of the existence of chiral superconducting states for each of the spin-polarized subspaces.

The spin point group of the confirmed altermagnets MnTe~\cite{krempasky_jungwirth_nature_24} and CrSb~\cite{Reimers2024,Yang2025} is given by ${}^{\bar{1}}6/{}^{\bar{1}}m{}^1m{}^{\bar{1}}m$ (see also Table \ref{tab:group_classifications}). Among the coreps (see Table 48 in Sec.~\ref{tables_coreps} and Sec.~\ref{tables_basis_basis}) there is the four-dimensional $\text{E}^{\text{u}}_{\nu}\text{E}^{\text{u}}_{-\nu}$ (see Table~\ref{tab:EECoerp}) with four basis vectors $ \text{D}_- k_x, \text{D}_- k_y, \text{D}_+ k_x \ \text{and} \ \text{D}_+ k_y$. 
\begin{table}[h]
    \centering
    \begin{tabular}{cccc}
$R_{\phi}$$||$Ee&$2_{\perp}R_{\phi}||$IC2z&$2_{\perp}R_{\phi}||$IC2y\\ 
 \hline 
 $\left(
\begin{array}{cccc}
 e^{i \nu  \phi } & 0 & 0 & 0 \\
 0 & e^{i \nu  \phi } & 0 & 0 \\
 0 & 0 & e^{-i \nu  \phi } & 0 \\
 0 & 0 & 0 & e^{-i \nu  \phi } \\
\end{array}
\right)$ & $\left(
\begin{array}{cccc}
 0 & 0 & e^{-i \nu  \phi } & 0 \\
 0 & 0 & 0 & e^{-i \nu  \phi } \\
 e^{i \nu  \phi } & 0 & 0 & 0 \\
 0 & e^{i \nu  \phi } & 0 & 0 \\
\end{array}
\right)$ & $\left(
\begin{array}{cccc}
 0 & 0 & e^{-i \nu  \phi } & 0 \\
 0 & 0 & 0 & -e^{-i \nu  \phi } \\
 e^{i \nu  \phi } & 0 & 0 & 0 \\
 0 & -e^{i \nu  \phi } & 0 & 0 \\
\end{array}
\right)$ \\
\end{tabular}
\\
\vspace{0.5cm}
 \begin{tabular}{cc}
$2_{\perp}R_{\phi}||$C6z&$R_{\phi}\mathcal{T}2_{\perp}$$||$Ee\\
\hline
  $\left(
\begin{array}{cccc}
 0 & 0 & \frac{1}{2} e^{-i \nu  \phi } & \frac{1}{2} \sqrt{3} e^{-i \nu  \phi } \\
 0 & 0 & -\frac{1}{2} \sqrt{3} e^{-i \nu  \phi } & \frac{1}{2} e^{-i \nu  \phi } \\
 \frac{1}{2} e^{i \nu  \phi } & \frac{1}{2} \sqrt{3} e^{i \nu  \phi } & 0 & 0 \\
 -\frac{1}{2} \sqrt{3} e^{i \nu  \phi } & \frac{1}{2} e^{i \nu  \phi } & 0 & 0 \\
\end{array}
\right)$ & $\left(
\begin{array}{cccc}
 e^{i \nu  \phi } & 0 & 0 & 0 \\
 0 & e^{i \nu  \phi } & 0 & 0 \\
 0 & 0 & e^{-i \nu  \phi } & 0 \\
 0 & 0 & 0 & e^{-i \nu  \phi } \\
\end{array}
\right)$ \\ 
    \end{tabular}
    \caption{Matrices of $\text{E}^{\text{u}}_{\nu}\text{E}^{\text{u}}_{-\nu}$ coreps for the group ${}^{\bar{1}}6/{}^{\bar{1}}m{}^1m{}^{\bar{1}}m$ (see Table 48 in Sec.~\ref{tables_coreps}). Here, $R_{\phi}$ denotes an SO(2) spin rotation by angle $\phi$ around the quantization axis, Ee is the identity operator, IC2z represents reflection through the $z$-mirror plane, and C6z is a sixfold rotation about the $z$-axis. In this notation, the operation $2_{\perp}R_{\phi}\parallel\text{C6z}$ corresponds to a 60 degrees rotation of the lattice (via C6z), a spin rotation by angle $\phi$ about the quantization axis (via $R_{\phi}$), and a spin flip encoded by the unitary operator $2_{\perp}$.}
    \label{tab:EECoerp}
\end{table}
To simplify expressions, we perform a unitary transformation $U$ and change the coreps matrices
\begin{align}
U &=\frac{1}{\sqrt{2}}
    \begin{pmatrix}
  i & 1 & 0 & 0 \\
  i & -1 & 0 & 0\\
  0 & 0 & i & 1\\
  0 & 0 &i & -1\\
\end{pmatrix},\label{unitary_tr_coreps}\\
\tilde{D}(R) &= U D (R)U^{\dagger},\label{tr_coreps_R}\\
\tilde{D}(A) &= U D (A)U^{T},\label{tr_coreps_A}\\
\bra{\tilde{\eta}} &= \bra{\eta} U^{{\dagger}},
\end{align}
where $R$ ($A$) are (anti-) unitary symmetry operations, $D$ are the coreps matrices, $\bra{\eta}$ is the row of the coreps basis states. We list $\tilde{D}$ matrices in Table~\ref{tab:EEtransCorep}.
\begin{table}[h]
    \centering
    \begin{tabular}{cccc}
$R_{\phi}$$||$Ee&$2_{\perp}R_{\phi}||$IC2z&$2_{\perp}R_{\phi}||$IC2y\\ 
 \hline 
 $\left(
\begin{array}{cccc}
 e^{i \nu  \phi } & 0 & 0 & 0 \\
 0 & e^{i \nu  \phi } & 0 & 0 \\
 0 & 0 & e^{-i \nu  \phi } & 0 \\
 0 & 0 & 0 & e^{-i \nu  \phi } \\
\end{array}
\right)$ & $\left(
\begin{array}{cccc}
 0 & 0 & e^{-i \nu  \phi } & 0 \\
 0 & 0 & 0 & e^{-i \nu  \phi } \\
 e^{i \nu  \phi } & 0 & 0 & 0 \\
 0 & e^{i \nu  \phi } & 0 & 0 \\
\end{array}
\right)$ & $\left(
\begin{array}{cccc}
 0 & 0 & 0& e^{-i \nu  \phi }  \\
 0 & 0  & e^{-i \nu  \phi } &0\\
 0&e^{i \nu  \phi }  & 0 & 0 \\
  e^{i \nu  \phi }& 0 & 0 & 0 \\
\end{array}
\right)$ \\
\end{tabular}
\\
\vspace{0.5cm}
 \begin{tabular}{cc}
$2_{\perp}R_{\phi}||$C6z&$R_{\phi}\mathcal{T}2_{\perp}$$||$Ee\\
\hline
$\left(
\begin{array}{cccc}
 0 & 0 & e^{i \pi/3}e^{-i \nu  \phi } & 0 \\
 0 & 0 & 0 & e^{-i \pi/3}e^{-i \nu  \phi } \\
 e^{i \pi/3}e^{i \nu  \phi } & 0 & 0 & 0 \\
 0 & e^{-i \pi/3}e^{i \nu  \phi } & 0 & 0 \\
\end{array}
\right)$ & $\left(
\begin{array}{cccc}
 0&-e^{i \nu  \phi } & 0 & 0 \\
 -e^{i \nu  \phi } & 0 & 0&0 \\
 0 & 0 &0 & -e^{-i \nu  \phi }  \\
 0 & 0 & -e^{-i \nu  \phi }&0 \\
\end{array}
\right)$ \\ 
    \end{tabular}
    \caption{$\tilde{D}$ matrices [see Eqs.~\eqref{unitary_tr_coreps}, \eqref{tr_coreps_R}, and \eqref{tr_coreps_A}] of $\text{E}^{\text{u}}_{\nu}\text{E}^{\text{u}}_{-\nu}$ coreps for the group ${}^{\bar{1}}6/{}^{\bar{1}}m{}^1m{}^{\bar{1}}m$ (see Table~\ref{tab:EECoerp}).}
    \label{tab:EEtransCorep}
\end{table}

For the new basis states 
$-\frac{i}{\sqrt2}\text{D}_- (k_x+i k_y)$,
$-\frac{i}{\sqrt2}\text{D}_- (k_x-i k_y)$,
$-\frac{i}{\sqrt2}\text{D}_+(k_x+i k_y)$,
$-\frac{i}{\sqrt2}\text{D}_+(k_x-i k_y)$, we introduce the order parameters ${\upsilon}_1, {\upsilon}_2, {\upsilon}_3$ and ${\upsilon}_4$, and construct all invariants up to the fourth order
\begin{align}
   f[{\upsilon}_1, {\upsilon}_2, {\upsilon}_3, {\upsilon}_4] &=  \alpha (|{\upsilon}_1|^2 + |{\upsilon}_2|^2 + |{\upsilon}_3|^2 + |{\upsilon}_4|^2) + \beta_1 (|{\upsilon}_1|^2 + |{\upsilon}_2|^2 + |{\upsilon}_3|^2 + |{\upsilon}_4|^2)^2\\ &+  \beta_2 (|{\upsilon}_1|^2 |{\upsilon}_2|^2 + |{\upsilon}_3|^2 |{\upsilon}_4|^2) + \beta_3 (|{\upsilon}_1|^2 |{\upsilon}_3|^2 + |{\upsilon}_2|^2 |{\upsilon}_4|^2)\\ &+  \beta_4 (|{\upsilon}_2|^2 |{\upsilon}_3|^2 + |{\upsilon}_1|^2 |{\upsilon}_4|^2) + 2 \beta_5 |{\upsilon}_1||{\upsilon}_2||{\upsilon}_3||{\upsilon}_4| \cos{\theta},
\end{align}
where $\theta = \theta_1 -\theta_2 -\theta_3+\theta_4$ is given by the phases of ${\upsilon}_n = |{\upsilon}_n|e^{i\theta_n}$. The minimum of the free energy has a large degeneracy in the relative phases. In the regime, when parameters are in the range 
\begin{equation}
\alpha<0,\qquad \beta_3<0,\qquad
                  4\beta_1+\beta_3>0,\qquad
 \beta_2>\beta_3,\qquad
 \beta_4>\beta_3,\qquad
 \beta_2+\beta_4-\beta_3>|\beta_5|,
\end{equation}
there is a global minimum of the free energy with
\begin{equation}
 \begin{aligned}
 v_+&=r(1,0,e^{i\xi},0)^T,\qquad
 v_-=r(0,1,0,e^{i\xi})^T,\\
 r^2&=-\frac{\alpha}{4\beta_1+\beta_3},\qquad
 f_{\min}=-\frac{\alpha^2}{4\beta_1+\beta_3}.
 \end{aligned}
\end{equation}
{For these states, the relative spin phase $\xi$
is arbitrary. The corresponding pairing potentials are}
\begin{equation}
d_\pm(\mathbf{k}) =-\frac{i r}{\sqrt2}
            \bigl(\text{D}_-+e^{i\xi}\text{D}_+\bigr)(k_x\pm i k_y).\\
\end{equation}
{Each state} breaks the antiunitary
$(\mathcal{T}2_{\perp}||E)$ symmetry leading to the chiral state in each
spin subspace $\text{D}_-(k_x{\pm}i k_y)$ and
$\text{D}_+(k_x{\pm}i k_y)$.
{The two domains have opposite chiralities and are
exchanged by this antiunitary operation.}
The topological classification of the superconducting states is
considered in Sec.~\ref{topo_class}.

\section{Topological classification}\label{topo_class}

In this section, we construct the topological classification of the altermagnetic superconducting states. To implement the symmetry classification, we adopt the consideration of Ref.~\cite{PhysRevB.78.195125}. 
Generally we can separate the spin subspaces, the BdG Hamiltonian takes the form
\begin{equation}
    H = (\mathbf{c}^{\dagger}_{\uparrow}, \mathbf{c}^{\dagger}_{\downarrow},\mathbf{c}_{\uparrow}, \mathbf{c}_{\downarrow} )
    \begin{pmatrix}
        a_{\uparrow} & 0 & \Delta_{\uparrow} & 0\\
        0 & a_{\downarrow} & 0 & \Delta_{\downarrow}\\
        \Delta^{\dagger}_{\uparrow} & 0 & -a^T_{\uparrow} & 0\\
        0 & \Delta^{\dagger}_{\downarrow} & 0 & -a^T_{\downarrow}
    \end{pmatrix} 
    \begin{pmatrix}
        \mathbf{c}_{\uparrow}\\
        \mathbf{c}_{\downarrow}\\
        \mathbf{c}^{\dagger}_{\uparrow}\\
        \mathbf{c}^{\dagger}_{\downarrow}
    \end{pmatrix},
\end{equation}
with $N\cross N$ matrices $a_{\sigma} = a_{\sigma}^{\dagger}$ and $\Delta_{\sigma}=-\Delta_{\sigma}^{T}$ (Fermi statistics) for a system with $N$ orbital/lattice degrees of freedom. 
Considering independently the spin subspaces we need to analyse the Hamiltonians
\begin{equation}
    H_{\sigma} = (\mathbf{c}^{\dagger}_{\sigma},\mathbf{c}_{\sigma})
    \begin{pmatrix}
        a_{\sigma} & \Delta_{\sigma} \\
        \Delta^{\dagger}_{\sigma} & -a^T_{\sigma}
    \end{pmatrix} 
    \begin{pmatrix}
        \mathbf{c}_{\sigma}\\
        \mathbf{c}^{\dagger}_{\sigma}
    \end{pmatrix} = (\mathbf{c}^{\dagger}_{\sigma},\mathbf{c}_{\sigma})
    \mathcal{H}_{\sigma}
    \begin{pmatrix}
        \mathbf{c}_{\sigma}\\
        \mathbf{c}^{\dagger}_{\sigma}
    \end{pmatrix},
\end{equation}
where we defined $\mathcal{H}_{\sigma}$ matrix.
{The Hamiltonian $\mathcal{H}_{\sigma}$ has particle-hole symmetry with the transformation $\mathcal{C} = \tau_x\mathcal{K}$, where $\mathcal{C}^2 = +1$}.
Without further constraints, the Hamiltonian is a member of the symmetry class {D}. In 2D this can be classified by the Chern number. A possible realization of the state is the  superconducting state described by the four-dimensional coreps (see Sec.~\ref{high_dim_corep}). In one of the spin-subspaces the relative phase between two basis functions can lead to the chiral state breaking the altermagnetic $(2_{\perp}\mathcal{T}||E)$ symmetry. 

The superconducting state can also preserve the altermagnetic antiunitary symmetry of the magnetic order, i.e.,  $(2_{\perp}\mathcal{T}||E) {= \mathcal{K}}$ { and  $(2_{\perp}\mathcal{T}||E)^2 = +1$}. It restricts the Hamiltonian to be the real matrix $a_{\sigma}=a_{\sigma}^*,\  \Delta_{\sigma}=\Delta_{\sigma}^*$. Using the hermiticity of the $a_{\sigma}$, and the Fermi statistics $\Delta_{\sigma}=-\Delta_{\sigma}^{T}$ we obtain $a_{\sigma}=a_{\sigma}^{T}, \ \Delta_{\sigma}^{\dagger} = - \Delta_{\sigma}$. Therefore, the Hamiltonian has the sublattice symmetry
\begin{equation}
\tau_x 
\mathcal{H}_{\sigma}
\tau_x = - 
\mathcal{H}_{\sigma},
\end{equation}
and belongs to the {BDI class}. 1D systems with this symmetry are classified by the $\mathbb{Z}$ winding number. 
We summarize the topological classification in Table~\ref{tab:topolo_class}.

Up to now, we discussed spin subspaces separately. However, there is the symmetry that relates them. If the superconducting state does not break the altermagntic symmetry, the two subspaces have related topological invariants.

\begin{table}[h]
    \centering
        \renewcommand{\arraystretch}{1.5}
    \begin{tabular}{c|c|c|c|c}
         Class  & $(\mathcal{T}2_{\perp}||E)$ & 1 & 2 & 3  \\
         \hline
         {D}  & 0 & {$\mathbb{Z}_2$} & $\mathbb{Z}$&0\\
         \hline
         {BDI} & \checkmark  & $\mathbb{Z}$ & 0&{0}\\
    \end{tabular}
    \caption{\justifying
    Classification of topological phases of the altermagnetic superconducting states. The invariants are determined for each of the spin-subspaces. If the anti-unitary symmetry $(\mathcal{T}2_{\perp}||E)$ is broken by the complex order parameter, the two-dimensional invariant is a Chern number computed for each of the spin subspaces. In $(\mathcal{T}2_{\perp}||E)$ symmetric superconductor, the invariant is one-dimensional winding number.}
    \label{tab:topolo_class}
\end{table}
\section{ Material example}\label{material_exampl}
\begin{figure}[h]
\begin{minipage}{0.32\textwidth}(a)
    \begin{tikzpicture}[scale=2]

    \clip (0,0) rectangle (1.7, 2.2);

    \def\a{1}  
    \def\arr{0.2}  
    \def\ay{1.3}
    \def\ellipseW{0.285} 
    \def\ellipseH{0.11}

    \foreach \x in {-1, 0, 1, 2} {
        \foreach \y in {-1, 0, 1, 2} {
            \pgfmathsetmacro\yshift{\y * \ay}

            \draw[thick] (\x*\a+0.5, \yshift+0.5*1.3) -- (\x*\a+0, \yshift+0.25*1.3);

            \draw[thick] (\x*\a+0.5, \yshift+0.5*1.3) -- (\x*\a+\a+0, \yshift+0.25*1.3);

            \draw[line width=0.7mm] (\x*\a+0.5, \yshift+0.5*1.3) -- (\x*\a+0.75, \yshift+0.75*1.3);
            \draw[thick, densely dotted] (\x*\a+0.5 , \yshift+0.5*1.3) -- (\x*\a+0.75 - 1, \yshift+0.75*1.3);
            
            \draw[line width=0.7mm] (\x*\a+0, \yshift+0.25*1.3) -- (\x*\a+0.25, \yshift+0);
            \draw[thick, densely dotted] (\x*\a+0+1, \yshift+0.25*1.3) -- (\x*\a+0.25, \yshift+0);

            \draw[dashed, thick] (\x*\a+0.5, \yshift+0.5*1.3) -- (\x*\a+0.25, \yshift+0);

                \draw[dashed, thick] (\x*\a+0.5, \yshift+0.5*1.3) -- (\x*\a+0.25, \yshift+\ay);

            \draw[dashed, thick] (\x*\a+0.75, \yshift+0.75*1.3) -- (\x*\a + 1, \yshift+0.25*1.3);

            \draw[dashed, thick] (\x*\a+0.75, \yshift+0.75*1.3) -- (\x*\a + 1, \yshift+1.25*1.3);

            \draw[thick] (\x*\a+0.75, \yshift+0.75*1.3) -- (\x*\a + 1.25, \yshift+1.*1.3);

            \draw[thick] (\x*\a+0.75, \yshift+0.75*1.3) -- (\x*\a +0.25, \yshift+1.*1.3);
        }
    }
        \foreach \x in {-1, 0, 1, 2} {
        \foreach \y in {-1, 0, 1, 2} {
            \pgfmathsetmacro\yshift{\y * \ay}

            \foreach \dx/\dy/\angle in {
                0.25/0/45, 
                0/0.25/45, 
                0.5/0.5/-45, 
                0.75/0.75/-45} 
            {
                \draw[thick,fill=white] (\x*\a+\dx, \yshift+\dy*1.3) 
                    ellipse [x radius=\ellipseW, y radius=\ellipseH, rotate=\angle];
            }

            \draw[red, line width=0.8mm, ->] (\x*\a+0.125, \yshift+1.3*0.125-\arr) -- (\x*\a+0.125, \yshift+1.3*0.125+\arr);

            \draw[blue, line width=0.8mm, ->] (\x*\a+0.625, \yshift+1.3*0.625+\arr) -- (\x*\a+0.625, \yshift+1.3*0.625-\arr);}}
    \draw[myPurple, line width=0.4mm] (0.125,0.125*1.3) rectangle (1.125,1.125*1.3);
    \node at (0.4,0.55*1.3) {\textbf{A}};
    \node at (0.75,0.8*1.3) {\textbf{B}};
    \node at (1.,0.35*1.3) {\textbf{C}};
    \node at (.3,1.05*1.3) {\textbf{D}};
\end{tikzpicture}
\begin{tikzpicture}
    \draw[line width=0.7mm] (0,0) -- (0.5,0) node[right] {$t_a$};
    \draw[thick] (0,-0.5) -- (0.5,-0.5) node[right] {$t_p$};
    \draw[thick, dashed] (0,-1) -- (0.5,-1) node[right] {$t_q$};
    \draw[thick,  densely dotted] (0,-1.5) -- (0.5,-1.5) node[right] {$t_b$};
\end{tikzpicture}
\end{minipage}
\hfill
    \begin{minipage}{0.32\textwidth}
    \begin{tikzpicture}

        \node[inner sep=0pt] (img) at (0,0) {\includegraphics[width=\linewidth,valign=t]{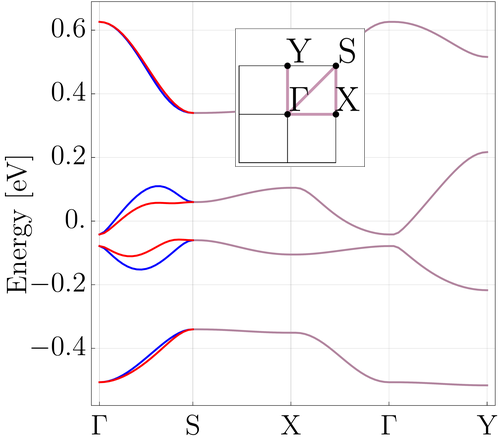}};
        \node[anchor=south west, yshift=5pt] at (img.south west) {(b)};
    \end{tikzpicture}
        \end{minipage}
    \hfill
    \begin{minipage}{0.32\textwidth}
    \begin{tikzpicture}
        \node[inner sep=0pt] (img) at (0,0) {\includegraphics[width=\linewidth,valign=t]{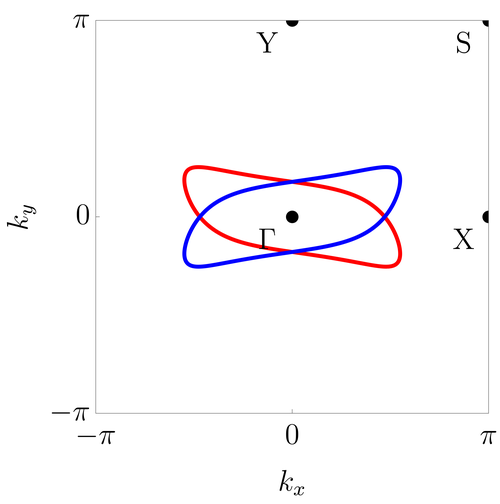}};
        \node[anchor=south west, yshift=15pt] at (img.south west) {(c)};
    \end{tikzpicture}
    \end{minipage}
    \caption{(a) \justifying Quazi-two -dimensional $\kappa$-(BEDT-TTF)$_2$X compound lattice with it's magnetic order. (b) The band structure of the compound tight-binding model and it's Brillouin zone. (c) The Fermi surface with two spin-polarized pockets. }
    \label{fig:TB_normal}
\end{figure}

In this section, we discuss the tight-binding model of the $\kappa$-(BEDT-TTF)$_2$X compound family~\cite{PhysRevResearch.5.043171}. The model has the sublattice degrees of freedom, corresponding to the four sites A, B, C, and D
\begin{equation}
\sigma_0 \otimes
\begin{pmatrix}
0 & t_b e^{-i k_x}+t_a & t_p \left(e^{-i k_x}+1\right) & t_q \left(e^{-i k_y}+1\right) \\
t_b e^{i k_x}+t_a &0 & t_q \left(e^{i k_y}+1\right) & t_p \left(e^{i k_x}+1\right) \\
t_p \left(e^{i k_x}+1\right) & t_q \left(e^{-i k_y}+1\right) & 0 & e^{-i k_y} \left(t_a e^{i k_x}+t_b\right) \\
t_q \left(e^{i k_y}+1\right) & t_p \left(e^{-i k_x}+1\right) & e^{i k_y} \left(t_a e^{-i k_x}+t_b\right) & 0
\end{pmatrix} + h \sigma_z \otimes \text{diag}[1,1,-1,-1], 
\end{equation}
where the first term describes hopping integrals between the sites, and the second term introduces the staggered collinear magnetic order. The lattice is shown in Fig.~\ref{fig:TB_normal}(a) in the main text. Here we consider the model values as in the Refs.~\cite{PhysRevResearch.5.043171, PhysRevB.89.045102} $(t_a, t_b, t_p, t_q, h) = (-0.207,\ -0.067,\ -0.102,\ 0.043,\ 0.2)$.

\begin{figure}[h]
\begin{minipage}{0.32\textwidth}
    \begin{tikzpicture}
        \node[inner sep=0pt] (img) at (0,0) {\includegraphics[width=\linewidth,valign=t]{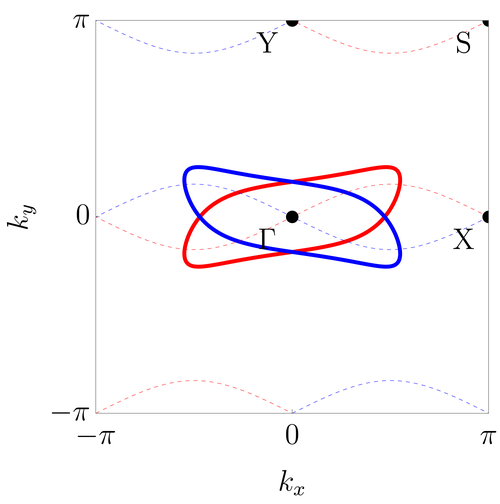}};
        \node[anchor=south west, yshift=8pt] at (img.south west) {(a)};
    \end{tikzpicture}
        \end{minipage}
    \hfill
    \begin{minipage}{0.32\textwidth}
    \begin{tikzpicture}
        \node[inner sep=0pt] (img) at (0,0) {\includegraphics[width=\linewidth,valign=t]{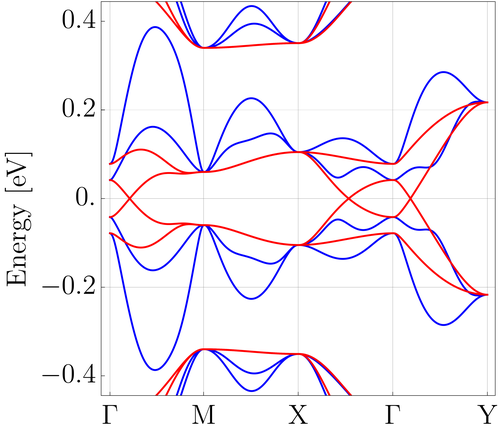}};
        \node[anchor=south west, yshift=0pt] at (img.south west) {(b)};
    \end{tikzpicture}
    \end{minipage}
    \begin{minipage}{0.32\textwidth}
    \begin{tikzpicture}
        \node[inner sep=0pt] (img) at (0,0) {\includegraphics[width=\linewidth,valign=t]{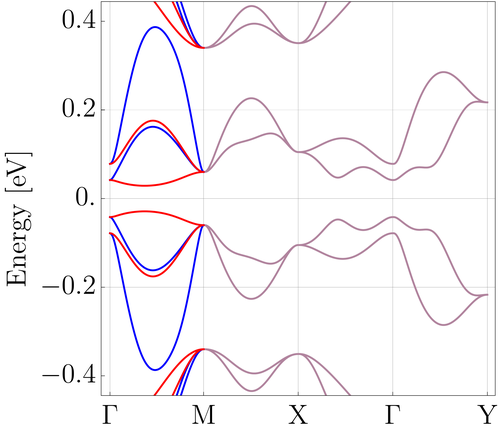}};
        \node[anchor=south west, yshift=0pt] at (img.south west) {(c)};
    \end{tikzpicture}
    \end{minipage}
    \caption{(a) Fermi surface with two spin-
polarized pockets. The gap nodes are shown with the dashed line. (b)-(c) BdG excitation spectrum for the
altermagnetic superconducting state (1, 0) and (1, 1)}
    \label{fig:TB_SC}
\end{figure}

\begin{figure}
\begin{minipage}{0.32\textwidth}
    \begin{tikzpicture}
        \node[inner sep=0pt] (img) at (0,0) {\includegraphics[width=\linewidth,valign=t]{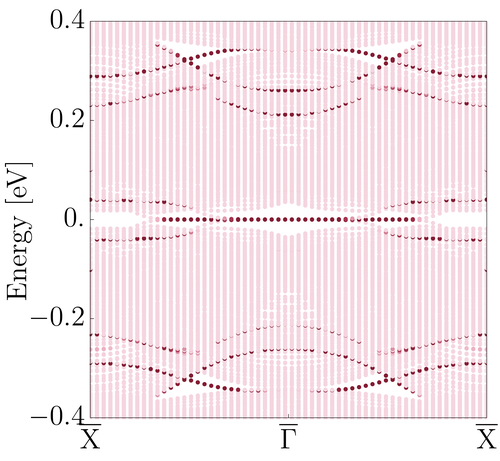}};
        \node[anchor=south west, yshift=0pt] at (img.south west) {(a)};
    \end{tikzpicture}
        \end{minipage}
    \hfill
    \begin{minipage}{0.32\textwidth}
    \begin{tikzpicture}
        \node[inner sep=0pt] (img) at (0,0) {\includegraphics[width=\linewidth,valign=t]{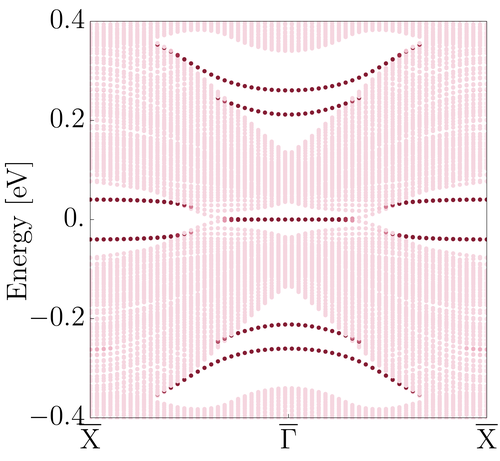}};
        \node[anchor=south west, yshift=0pt] at (img.south west) {(b)};
    \end{tikzpicture}
    \end{minipage}
    \begin{minipage}{0.32\textwidth}
    \begin{tikzpicture}
        \node[inner sep=0pt] (img) at (0,0) {\includegraphics[width=\linewidth,valign=t]{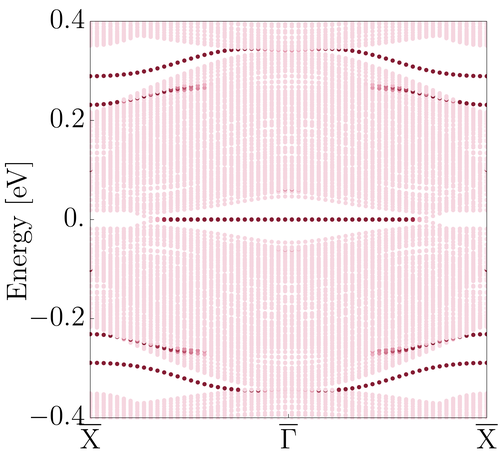}};
        \node[anchor=south west, yshift=0pt] at (img.south west) {(c)};
    \end{tikzpicture}
    \end{minipage}
    \caption{\justifying Slab model with 40 layers.The termination is along $x-y$ line. (a) Both spins up and down subspaces, (b) only spin up, (c) only spin down.}
    \label{fig:TB_SC_slab}
\end{figure}

The superconducting pairing potential also gets the sublattice degrees of freedom. The procedure discussed in this work can be extended to describe the basis functions in this case. Here, we only discuss the p-wave paring potential that transforms as $\text{B}_{+1}\text{B}_{-1}$ coreps, but acts trivially on the lattice degrees of freedom $\text{diag} [1, 1, 1, 1]$. First, we consider the $(1, 1)$ state discussed previously in Sec.~\ref{SOC_effect}
\begin{align}
    \Delta_{p(1,1)}(k) &= i[\left(r_x k_x+ r_y k_y\right) (\tau^x + i \tau^y) +  \left(r_x k_x - r_y k_y\right) (\tau^x - i \tau^y)] \tau^y\otimes \text{diag} [1, 1, 1, 1]\\
    &= 2
    \begin{pmatrix}
        -r_x k_x - r_y k_y&0\\
        0& r_x k_x - r_y k_y
    \end{pmatrix}
    \otimes \text{diag} [1, 1, 1, 1].
\end{align}
The potential preserves the altermagnetic symmetries. The superconductor has gapless spin-polarized excitations, i.e., Dirac points at zero energy of the Bogoliubov-de Gennes spectrum (see Fig.~\ref{fig:TB_SC} and~\ref{fig:TB_SC_slab}). The degenearcy is protected by chiral symmetry discussed in Sec.~\ref{topo_class}. The symmetry quantizes  the winding number for each of the spin-subspaces. 

In a model with the edge, there are spin-polarized boundary states connecting the Dirac points (see Fig.~\ref{fig:TB_SC_slab}). Choosing the termination direction along the diagonal $x - y$, the edge states with opposite spin do not overlap with each other for some of the $k$ points. This effect will be used in the next section~\ref{junction}, where we discuss the ac Josephson current through the junction between altermagnetic superconductors. 

In Sec.~\ref{free_energy_2D_Corep}, we obtain another minimum of the free energy, i.e., $(1, 0)$
\begin{equation}
    \Delta_{p(1,0)}(k) = i\left(r_x k_x+ r_y k_y\right) (\tau^x + i \tau^y)  \tau^y\otimes \text{diag} [1, 1, 1, 1].
\end{equation}
In this case, the superconducting gap is opened only for one of the spin subspaces. The other subspace stays metallic. This state is half-and-half metal-superconductor, and it breaks the altermagnetic symmetry relating two spin sublattices [see Fig.~\ref{fig:TB_SC}(b)].

\section{Junction between altermagnetic superconductors}\label{junction}
\begin{figure}
\begin{tikzpicture}
\centering
  \node[inner sep=0pt] (img) at (0,0) {\includegraphics[width=0.5\textwidth]{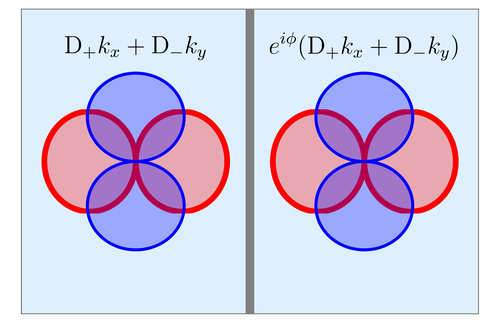}};

  \draw[->, thick, black] (0.1,-2) -- (1,-2) node[anchor=west] {\Large $x$};
  \draw[->, thick, black] (0.1,-2) -- (0.1,-1) node[anchor=south, xshift=10pt, yshift=-10pt] {\Large $y$};
\end{tikzpicture}
    \caption{\justifying Junction geometry.}
    \label{fig:junction}
\end{figure}
Here, we discuss the contact effect between two altermagnetic superconductors separated by a barrier (see Fig.~\ref{fig:junction}). We neglect the anisotropy of the electronic band structure in the normal metal and assume the parabolic dispersion $\epsilon = \hbar^2 (k_x^2 +k_y^2) / 2m - \mu$ for both spin projections. We are interested in the effect of the altermagnetic pairing potential $\text{D}_{+}\Delta_0 k_x/k_F + \text{D}_{-}\Delta_0k_y/k_F$, with $k_F = \sqrt{2 m \mu} /\hbar$. The superconductors are stack in the $x$ direction, and we assume the periodicity in the $y$ direction. 
We follow Ref.~\cite{10.1063/1.1789931} and describe the contact between superconductors by the potential $U(x) = U_0 \delta(x)$. We solve the Bogoliubov -- de Gennes equations for two spin subspaces separately
\begin{equation}\label{BdG}
\begin{pmatrix}
[\epsilon_{k_y} (\hat{k}_x) + U(x)]\sigma_0 & \Delta_{k_y} (x, \hat{k}_x) \\
\Delta_{ k_y}^{\dagger} (x, \hat{k}_x) & [-\epsilon_{k_y} (\hat{k}_x) - U(x)]\sigma_0
\end{pmatrix}
\psi_n = E_n \psi_n,
\end{equation}
where $\hat{k}_x = -i \partial_x$. The gap function takes the form
\begin{equation}
     \Delta_{k_y} (x, \hat{k}_x) = i (\sigma_x + i \sigma_y) \sigma_y \Delta_0 \hat{k}_x / k_F + i (\sigma_x - i \sigma_y) \sigma_y \Delta_0 k_y / k_F = 
     \begin{pmatrix}
        2 k_y \Delta_0/ k_F & 0 \\
        0 & - 2 \hat{k}_x \Delta_0/ k_F
     \end{pmatrix},
\end{equation} 
and we can consider separately spin up and down subspaces
\begin{equation}
    H_{\uparrow} = \begin{pmatrix}
\epsilon_{k_y} (\hat{k}_x) + U(x)& -2 \hat{k}_x \Delta_0/ k_F\\
-2 \hat{k}_x^{\dagger} \Delta_0/ k_F & -\epsilon_{k_y} (\hat{k}_x) - U(x)
\end{pmatrix}, \  H_{\downarrow} = \begin{pmatrix}
\epsilon_{k_y} (\hat{k}_x) + U(x)& 2 k_y \Delta_0/ k_F\\
2 k_y \Delta_0/ k_F & -\epsilon_{k_y} (\hat{k}_x) - U(x)
\end{pmatrix}.
\end{equation}
To solve the Bogoliubov-de Gennes equations, we use the bound state ansatz for the left and right sides of the junction
\begin{align}\label{ansatz}
    \psi_{L \sigma}(k_y, x) &= e^{\kappa x} \left[ A_L 
        \begin{pmatrix}
        u_{L  \sigma +} \\
        v_{L  \sigma +}
        \end{pmatrix}
        e^{i k_{F_{k_y}}  x} + B_L 
        \begin{pmatrix}
        u_{L  \sigma -} \\
        v_{L  \sigma -}
        \end{pmatrix}
        e^{-i k_{F_{k_y}}  x} \right],\\
    \psi_{R \sigma }(k_y, x) &= e^{-\kappa x} \left[ A_R 
        \begin{pmatrix}
        u_{R  \sigma +} \\
        v_{R  \sigma  +}
        \end{pmatrix}
        e^{i k_{F_{k_y}}  x} + B_R
        \begin{pmatrix}
        u_{R  \sigma -} \\
        v_{R  \sigma -}
        \end{pmatrix}
        e^{-i k_{F_{k_y}}  x} \right],
\end{align}
where we separate the left- and right-moving waves, and encode spin with $ \sigma$. The $k_y$-dependent $k_{F_{k_y}} = \sqrt{k_F^2 - k_y^2}$ is the Fermi velocity along the $x$ direction, $\kappa$ correspond to the exponential localization of the bound state. 
Sewing the wave-functions at the two sides (R and L) of the junction for each of the spin components, we obtain
\begin{equation}\label{sew}
\psi_{L \sigma}(k_y,0) = \psi_{R \sigma}(k_y,0), \quad \partial_x \psi_{R \sigma}(k_y,0) - \partial_x \psi_{L \sigma}(k_y,0) = k_{F_{k_y}} Z \psi_{R \sigma}(k_y,0), \quad
Z= \frac{2m U_0}{\hbar^2 k_{F_{k_y}}}.
\end{equation}
Pugging the ansatz wave functions \ref{ansatz} separately for left- and right- moving waves into the Eq.(\ref{BdG}), we obtain the expressions 
\begin{equation}\label{rations_u}
    \frac{u_{L \sigma  \alpha}}{v_{L \sigma \alpha}} = \frac{\Delta_{\sigma \alpha}}{E + i \alpha \kappa \hbar k_{F_{k_y}}^2/m}, \quad
     \frac{u_{R \sigma  \alpha}}{v_{R \sigma \alpha}} = \frac{\Delta_{\sigma \alpha}}{E - i \alpha \kappa \hbar k_{F_{k_y}}^2/m}, \quad
\kappa = \frac{\sqrt{\Delta_{ \sigma \alpha} ^2 - E^2}}{\hbar^2 k_{F_{k_y}}/m},
\end{equation}
where for spin $\uparrow$ we have $ \Delta_{\uparrow \alpha} = -2\alpha \sqrt{k_F^2-k_y^2}\Delta_0/k_F $ and $ \Delta_{\downarrow \alpha} = 2 k_y \Delta_0/k_F $. The expressions are obtained for $ k_{F_{k_y}} \gg \kappa$.

Using the expressions \ref{rations_u} and substituting the wave function \ref{ansatz} into Eq.~\ref{sew}, we obtain two systems of linear equations on $A_L, A_R, B_L, B_R$ for each of the spin subspaces. The system has a solution if its determinant is zero. From this we get the expressions for bound state energies
\begin{align}
    E_{\downarrow}(k_y, {\delta}) = \pm \frac{2 k_y \Delta_0}{k_F} \sqrt{1- D_{k_y} \sin ^2\left(\frac{{\delta} }{2}\right)},\\
    E_{\uparrow}(k_y, {\delta}) = \pm \frac{2\sqrt{k_F^2-k_y^2}\Delta_0}{k_F} \sqrt{D_{k_y}} \cos \left(\frac{{\delta} }{2}\right),
\end{align}
where $D_{k_y} = \tfrac{4}{4+Z^2}$, with $k_y$-dependent $Z= \frac{2m U_0}{\hbar^2 k_{F_{k_y}}} = \tfrac{k_F}{k_{F_{k_y}}}Z_0$. $Z_0$ is independent of $k_y$ and characterizes the transparency of the junction.  

We proceed with computing the dynamical fractional ac Josephson effect. We assume an applied small voltage $eV \ll \Delta_0$ at the junction. The phase has the time dependence ${\delta}(t) = 2eVt/\hbar$. We also assume that the state does not relax into the thermodynamic limit, and the phase changes adiabatically. With this we compute the currents for both spin components as 
\begin{equation}
    I = \int \limits_{0}^{k_F}\frac{d k_y}{k_F} \frac{2e}{\hbar}\frac{\partial E (k_y, {\delta})}{\partial {\delta}},
\end{equation}
where we sum contributions from all $k_y$ points and assume the zero temperature $T=0$.
\begin{align}
    I_{\uparrow} =&-\frac{\Delta_0}{8}  \frac{2e}{\hbar} (-Z_0^2 \arctan{\frac{2}{Z_0}} + 2Z_0+4 \arctan{\frac{2}{Z_0}} ) \sin{\frac{{\delta} }{2}},\\
    I_{\downarrow}
    =&-\frac{e\Delta_0}{4\hbar}\tan\frac{\delta}{2}
    \left[
    \sqrt{(Z_0^2+4)(Z_0^2+2\cos\delta+2)}-Z_0^2
    \right]
    \nonumber\\
    &+\frac{e\Delta_0}{8\hbar}\tan\frac{\delta}{2}\,
    Z_0^2\frac{\cos\delta+3}{\cos(\delta/2)}
    \tanh^{-1}\!\left[
    \sec\frac{\delta}{2}
    \sqrt{\frac{Z_0^2+2\cos\delta+2}{Z_0^2+4}}
    \right]
    \nonumber\\
    &-\frac{e\Delta_0}{8\hbar}\tan\frac{\delta}{2}\,
    Z_0^2\frac{\cos\delta+3}{\cos(\delta/2)}
    \tanh^{-1}\!\left(\sec\frac{\delta}{2}\right).
\end{align}
In the limit of the transparent barrier $Z_0 = 0$, the expressions take the form
  \begin{equation}
    I_{\uparrow}= -\frac{\pi \Delta_0}{4} \frac{2e}{\hbar} \sin{\frac{{\delta}}{2}},\
    I_{\downarrow} = -\frac{ \Delta_0}{2}  \frac{2e}{\hbar}\sin{\frac{{\delta}}{2}} \frac{\cos{\tfrac{{\delta}}{2}}}{|\cos{\tfrac{{\delta}}{2}}|}.
  \end{equation}

\begin{figure}[t]
    \centering
    \includegraphics[width=0.32\linewidth]{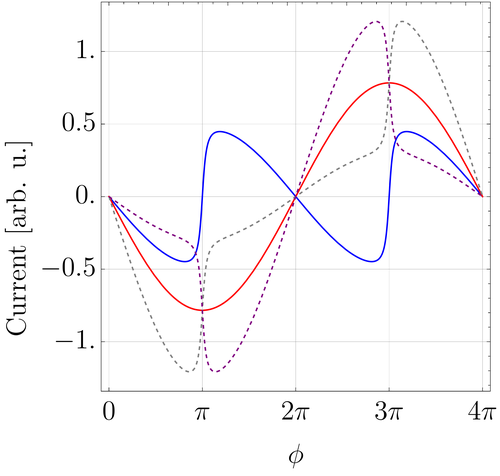}
    \includegraphics[width=0.32\linewidth]{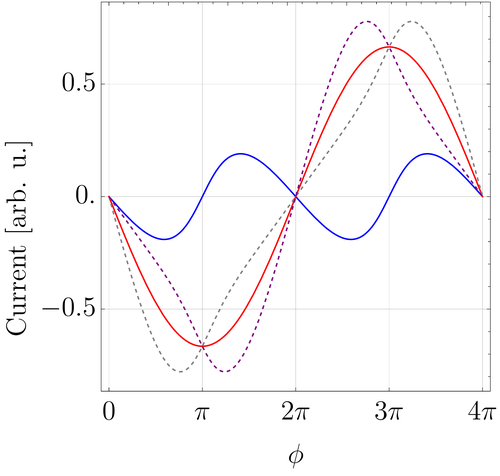}
    \includegraphics[width=0.32\linewidth]{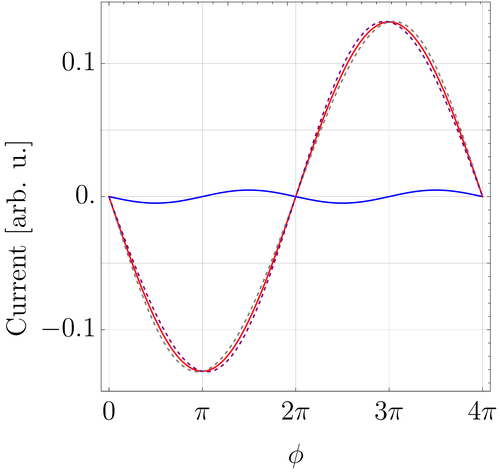}
    \caption{\justifying Josephson current for the case of (a) $Z_0 = 0.1$, (b) $Z_0 = 1$ and (c) $Z_0 =10$. The red line is the spin-up current, and the blue corresponds to the spin-down current. The total current is shown with a gray dashed line, and the difference with the purple dashed line.}
    \label{fig:Josephson_current}
\end{figure}

In the tunneling limit $Z_0 \gg 1$, we obtain
\begin{align}
    I_{\uparrow} &= -\frac{4\Delta}{3Z_0}  \frac{2e}{\hbar}\sin{\frac{{\delta}}{2}},\\
    I_{\downarrow} &= -\frac{ \Delta}{2Z_0^2}  \frac{2e}{\hbar}\sin{{\delta}}.
\end{align}
For the spin up we extract the $4\pi$ periodicity and the $1/Z_0$ scaling, however, spin down current is $2\pi$ periodic and scales with $1/Z_0^2$.
This different behaviour allows to use the junction for the generation of ac spin-polarized superconducting currents (see Fig.~\ref{fig:Josephson_current}).

\clearpage

\bibliographystyle{apsrev4-1}
\bibliography{bibliography.bib}
\maxdeadcycles=200

\newpage
\section{Spin-point group naming}\label{naming_table}
Here, we introduce the naming convention used in the tables. We explicitly write the structure of the non-trivial spin-point subgroup as $H + (2_{\perp}||g)H$. 
\begin{table}[h]
    \centering
    \begin{tabular}{|c|c|c|c|}
        \hline
         & Group~\cite{schiff2023spin} & Sublattice symmetry $H$ & Compensation symmetry $(2_{\perp}||g)$ \\
        \hline
        1 & $\mathbf{b}^{\infty} \times {}^{\bar{1}} \bar{1}$ & C1 & IEe \\
        2 & $\mathbf{b}^{\infty} \times {}^{\bar{1}} 2$ & C1 & C2z \\
        3 & $\mathbf{b}^{\infty} \times {}^{\bar{1}}m$ & C1 & IC2z \\
        4 & $\mathbf{b}^{\infty} \times {}^12/{}^{\bar{1}}m$ & C2 & IC2z \\
        5 & $\mathbf{b}^{\infty} \times {}^{\bar{1}}2/^1m$ & Cs & C2z \\
        6 & $\mathbf{b}^{\infty} \times {}^{\bar{1}}2/{}^{\bar{1}}m$ & Ci & C2z \\
        7 & $\mathbf{b}^{\infty} \times {}^{\bar{1}}m {}^{\bar{1}}m ^12$ & C2 & IC2x \\
        8 & $\mathbf{b}^{\infty} \times {}^1m {}^{\bar{1}}m {}^{\bar{1}}2$ & Cs & C2x \\
        9 & $\mathbf{b}^{\infty} \times  {}^1 2 {}^{\bar{1}}2 {}^{\bar{1}}2$ & C2 & C2x \\
        10 & $\mathbf{b}^{\infty} \times {}^1m {}^{\bar{1}}m {}^{\bar{1}}m$ & C2h & IC2x \\
        11 & $\mathbf{b}^{\infty} \times {}^1m {}^1m {}^{\bar{1}}m$ & C2v & IC2z \\
        12 & $\mathbf{b}^{\infty} \times {}^{\bar{1}}m {}^{\bar{1}}m {}^{\bar{1}}m$ & D2 & IEe \\
        13 & $\mathbf{b}^{\infty} \times {}^{\bar{1}}4$ & C2 & C4z \\
        14 & $\mathbf{b}^{\infty} \times {}^{\bar{1}}\bar{4}$ & C2 & IC4z \\
        15 & $\mathbf{b}^{\infty} \times {}^{\bar{1}}4/{}^1m$ & C2h & C4z \\
        16 & $\mathbf{b}^{\infty} \times {}^{\bar{1}}4/{}^{\bar{1}}m$ & S4 & C4z \\
        17 & $\mathbf{b}^{\infty} \times {}^{1}4/{}^{\bar{1}}m$ & C4 & IC2z \\
        18 & $\mathbf{b}^{\infty} \times {}^{1}4 {}^{\bar{1}}2 {}^{\bar{1}}2$ & C4 & C2x \\
        19 & $\mathbf{b}^{\infty} \times {}^{\bar{1}}4 {}^{1}2 {}^{\bar{1}}2$ & D2 & C4z \\
        20 & $\mathbf{b}^{\infty} \times {}^{\bar{1}}4 {}^1m {}^{\bar{1}}m$ & C2v & C4z \\
        21 & $\mathbf{b}^{\infty} \times {}^{\bar{1}}\bar{4} {}^{\bar{1}}2 {}^{1}m$ & C2v & IC4z \\
        22 & $\mathbf{b}^{\infty} \times{}^{\bar{1}}\bar{4} {}^{1}2 {}^{\bar{1}}m$ & D2 & IC4z \\
        23 & $\mathbf{b}^{\infty} \times {}^{1}4 {}^{\bar{1}}m {}^{\bar{1}}m$ & C4 & IC2x \\
        24 & $\mathbf{b}^{\infty} \times {}^{1}{\bar{4}}{}^{\bar{1}}m {}^{\bar{1}}m$ & S4 & C2x \\
        25 & $\mathbf{b}^{\infty} \times {}^{\bar{1}}4/ {}^1m {}^{\bar{1}}m{}^{1}m$ & D2d & IEe \\
        26 & $\mathbf{b}^{\infty} \times {}^{1}4/ {}^{\bar{1}}m {}^{1}m{}^{1}m$ & C4v & IEe \\
        27 & $\mathbf{b}^{\infty} \times {}^{\bar{1}}4/ {}^1m {}^{1}m{}^{\bar{1}}m$ & D2h & C4z \\
        28 & $\mathbf{b}^{\infty} \times  {}^{1}4/ {}^{1}m {}^{\bar{1}}m{}^{\bar{1}}m$ & C4h & C2x \\
        29 & $\mathbf{b}^{\infty} \times {}^{1}4/ {}^{\bar{1}}m {}^{\bar{1}}m{}^{\bar{1}}m$ & D4 & IEe \\
        30 & $\mathbf{b}^{\infty} \times {}^{\bar{1}}\bar{3}$ & C3 & IEe \\
        31 & $\mathbf{b}^{\infty} \times {}^13{}^{\bar{1}}2$ & C3 & C2x \\
        32 & $\mathbf{b}^{\infty} \times  {}^13{}^{\bar{1}}m$ & C3 & IC2x \\
        33 & $\mathbf{b}^{\infty} \times {}^1\bar{3}{}^{\bar{1}}m$ & C3i & IC2x \\
        \hline
    \end{tabular}   
\end{table}
\begin{table}[h]
    \centering
    \begin{tabular}{|c|c|c|c|}
        \hline
         & Group~\cite{schiff2023spin} & Sublattice symmetry H & Compensation symmetry $(2_{\perp}||g)$ \\
        \hline
        34 &  $\mathbf{b}^{\infty} \times {}^{\bar{1}}\bar{3}{}^1m$& C3v & IEe \\ 
        35 &  $\mathbf{b}^{\infty} \times {}^{\bar{1}}\bar{3}{}^{\bar{1}}m$& D3 & IEe \\ 
        36 &  $\mathbf{b}^{\infty} \times {}^{\bar{1}}\bar{6}$& C3 & IC2z \\ 
        37 &  $\mathbf{b}^{\infty} \times {}^{\bar{1}}6$& C3 & C2z \\ 
        38 &  $\mathbf{b}^{\infty} \times {}^1 6{}^{\bar{1}}2{}^{\bar{1}}2$& C6 & C2x \\ 
        39 &  $\mathbf{b}^{\infty} \times {}^{\bar{1}}6{}^{1}2{}^{\bar{1}}2$& D3 & C2z \\ 
        40 &  $\mathbf{b}^{\infty} \times {}^{\bar{1}}6/{}^{\bar{1}}m$& C3i & IC2z \\ 
        41 &  $\mathbf{b}^{\infty} \times {}^{\bar{1}}6/{}^{1}m$& C3h & IEe \\ 
        42 &  $\mathbf{b}^{\infty} \times {}^{1}6/{}^{\bar{1}}m$& C6 & IEe \\ 
        43 &  $\mathbf{b}^{\infty} \times {}^{1}6{}^{\bar{1}}m{}^{\bar{1}}m$& C6 & IC2y \\ 
        44 &  $\mathbf{b}^{\infty} \times {}^{\bar{1}}6{}^{1}m{}^{\bar{1}}m$& C3v & C2z \\ 
        45 &  $\mathbf{b}^{\infty} \times {}^{1}6{}^{\bar{1}}m{}^{\bar{1}}2$& C3h & IC2x \\ 
        46 &  $\mathbf{b}^{\infty} \times {}^{\bar{1}}6{}^{1}m{}^{\bar{1}}2$& C3v & IC2z \\ 
        47 &  $\mathbf{b}^{\infty} \times {}^{\bar{1}}6{}^{\bar{1}}m{}^{1}2$& D3 & IC2z \\ 
        48 &  $\mathbf{b}^{\infty} \times {}^{\bar{1}}6/{}^{\bar{1}}m{}^1m{}^{\bar{1}}m$& D3d & IC2z \\ 
        49 &  $\mathbf{b}^{\infty} \times {}^{\bar{1}}6/{}^1m{}^{\bar{1}}m{}^{\bar{1}}m$& D3h & IEe \\ 
        50 &  $\mathbf{b}^{\infty} \times {}^{1}6/{}^1m{}^{\bar{1}}m{}^{\bar{1}}m$& C3h & C2x \\ 
        51 &  $\mathbf{b}^{\infty} \times {}^{1}6/{}^{\bar{1}}m{}^{1}m{}^{1}m$& C6v & IEe \\ 
        52 &  $\mathbf{b}^{\infty} \times {}^{1}6/{}^{\bar{1}}m{}^{\bar{1}}m{}^{\bar{1}}m$& D6 & IEe \\ 
        53 &  $\mathbf{b}^{\infty} \times {}^{1}2/{}^{\bar{1}}m {}^{\bar{1}}\bar{3}$& T & IEe \\ 
        54 &  $\mathbf{b}^{\infty} \times {}^{\bar{1}}\bar{4}{}^1 3{}^{\bar{1}}m$& Th & IC4z \\ 
        55 &  $\mathbf{b}^{\infty} \times {}^{\bar{1}}\bar{4}{}^1 3{}^{\bar{1}}2$& T & C4z \\ 
        
        56 &  $\mathbf{b}^{\infty} \times {}^{\bar{1}} 4 /{}^1 m {}^1\bar{3} {}^{\bar{1}}2/{}^{\bar{1}}m$& Th & C4z \\ 
        57 &  $\mathbf{b}^{\infty} \times {}^{\bar{1}} 4 /{}^{\bar{1}} m {}^{\bar{1}}\bar{3} {}^{\bar{1}}2/{}^{1}m$& Td & IEe \\ 
        58 &  $\mathbf{b}^{\infty} \times {}^{1} 4 /{}^{\bar{1}} m {}^{\bar{1}}\bar{3} {}^{1}2/{}^{\bar{1}}m$& O & IEe \\ 
        \hline
    \end{tabular}
    \caption{Groups specifications.}
    \label{tab:group_classifications}
\end{table}

\clearpage
\newpage

\section{Tables of corep basis functions}\label{tables_basis_basis}
\noindent
Here, we list pairing basis functions for coreps of the spin-point groups describing compensated collinear order. All coreps are listed in Sec.~\ref{tables_coreps}. We use the following notations  $\text{D}_{\pm} = (1, \pm i, 0)^T$, $\text{D}_{\text{z}} = (0, 0, 1)^T$. With $r, r_x, r_y, r_z$ we represent real numbers, $C$ and $C_1$ correspond to complex numbers. The data is available in the digital form at DOI: 10.5281/zenodo.16279047.
\newline
\FloatBarrier
\begin{table}[h]
    \fontsize{9}{11}\selectfont
    \centering
    \hspace*{\fill}
            \begin{tabular}{|c|c|c|} 
\hline 
1: C1+IEe$ \times $C1&$\psi$ & $d$  \\ 
 \hline 
$\text{A}_{+}$ & $\begin{array}{c}
 1 \\
\end{array}$ & $\begin{array}{c}
 i \text{D}_{\text{z}} \left(r_z k_z+r_y k_y+r_x k_x\right) \\
\end{array}$\\ 
 \hline 
$\text{A}_{-}$ & $\begin{array}{c}
 0 \\
\end{array}$ & $\begin{array}{c}
 0 \\
\end{array}$\\ 
 \hline 
$\text{A}_{\nu}\text{A}_{-\nu}$ & $\begin{array}{c}
 0 \\
 0 \\
\end{array}$ & $\begin{array}{c}
 \text{D}_- \left(r_z k_z+r_y k_y+r_x k_x\right) \\
 -\text{D}_+ \left(r_z k_z+r_y k_y+r_x k_x\right) \\
\end{array}$\\ 
 \hline 
\end{tabular} 
            \hspace*{\fill}
             \newline
    \vspace*{0.5 cm}
    \newline
    \hspace*{\fill}
    \begin{tabular}{|c|c|c|} 
\hline 
2: C1+C2z$ \times $C1&$\psi$ & $d$  \\ 
 \hline 
$\text{A}_{+}$ & $\begin{array}{c}
 1 \\
\end{array}$ & $\begin{array}{c}
i \text{D}_{\text{z}} (r_y k_y+r_x k_x) \\
\end{array}$\\ 
 \hline 
$\text{A}_{-}$ & $\begin{array}{c}
 k_z \left(r k_x+k_y\right) \\
\end{array}$ & $\begin{array}{c}
 i \text{D}_{\text{z}} k_z \\
\end{array}$\\ 
 \hline 
$\text{A}_{\nu}\text{A}_{-\nu}$ & $\begin{array}{c}
 0 \\
 0 \\
\end{array}$ & $\begin{array}{c}
 \text{D}_- \left( r_y k_y+r_x k_x + r_z k_z\right) \\
 \text{D}_+  \left( -r_y k_y-r_x k_x + r_z k_z\right) \\
\end{array}$\\ 
 \hline 
\end{tabular} 
    \hspace*{\fill}
    \newline
    \vspace*{0.5 cm}
    \newline
    \hspace*{\fill}
        \begin{tabular}{|c|c|c|} 
\hline 
3: C1+IC2z$ \times $C1&$\psi$ & $d$  \\ 
 \hline 
$\text{A}_{+}$ & $\begin{array}{c}
 1 \\
\end{array}$ & $\begin{array}{c}
 i \text{D}_{\text{z}} k_z \\
\end{array}$\\ 
 \hline 
$\text{A}_{-}$ & $\begin{array}{c}
 k_z \left(r k_x+k_y\right) \\
\end{array}$ & $\begin{array}{c}
i \text{D}_{\text{z}} (r_y k_y+r_x k_x) \\
\end{array}$\\ 
 \hline 
$\text{A}_{\nu}\text{A}_{-\nu}$ & $\begin{array}{c}
 0 \\
 0 \\
\end{array}$ & $\begin{array}{c}
 \text{D}_-  \left( r_y k_y+r_x k_x + r_z k_z\right) \\
\text{D}_+ \left( r_y k_y+r_x k_x - r_z k_z\right) \\
\end{array}$\\ 
 \hline 
\end{tabular} 
        \hspace*{\fill}
    \newline
    \vspace*{0.5 cm}
    \newline
    \hspace*{\fill}
        \begin{tabular}{|c|c|c|} 
\hline 
4: C2+IC2z$ \times $C2&$\psi$ & $d$  \\ 
 \hline 
$\text{A}_{+}$ & $\begin{array}{c}
 1 \\
\end{array}$ & $\begin{array}{c}
 i \text{D}_{\text{z}} k_z \\
\end{array}$\\ 
 \hline 
$\text{A}_{-}$ & $\begin{array}{c}
 0 \\
\end{array}$ & $\begin{array}{c}
 0 \\
\end{array}$\\ 
 \hline 
$\text{B}_{-}$ & $\begin{array}{c}
 k_z \left(r k_x+k_y\right) \\
\end{array}$ & $\begin{array}{c}
i \text{D}_{\text{z}} (r_y k_y+r_x k_x) \\
\end{array}$\\ 
 \hline 
$\text{B}_{+}$ & $\begin{array}{c}
 0 \\
\end{array}$ & $\begin{array}{c}
 0 \\
\end{array}$\\ 
 \hline 
$\text{A}_{\nu}\text{A}_{-\nu}$ & $\begin{array}{c}
 0 \\
 0 \\
\end{array}$ & $\begin{array}{c}
 \text{D}_- k_z \\
 -\text{D}_+ k_z \\
\end{array}$\\ 
 \hline 
$\text{B}_{\nu}\text{B}_{-\nu}$ & $\begin{array}{c}
 0 \\
 0 \\
\end{array}$ & $\begin{array}{c}
 \text{D}_- \left(r k_x+k_y\right) \\
 \text{D}_+ \left(r k_x+k_y\right) \\
\end{array}$\\ 
 \hline 
\end{tabular}
        \hspace*{\fill}
    \newline
    \vspace*{0.5 cm}
    \newline
    \hspace*{\fill}
        \begin{tabular}{|c|c|c|} 
\hline 
5: Cs+C2z$ \times $Cs&$\psi$ & $d$  \\ 
 \hline 
$\text{A}^{\text{'}}_{+}$ & $\begin{array}{c}
 1 \\
\end{array}$ & $\begin{array}{c}
i \text{D}_{\text{z}} (r_y k_y+r_x k_x) \\
\end{array}$\\ 
 \hline 
$\text{A}^{\text{'}}_{-}$ & $\begin{array}{c}
 0 \\
\end{array}$ & $\begin{array}{c}
 0 \\
\end{array}$\\ 
 \hline 
$\text{A}^{\text{{\text{'}}{\text{'}}}}_{-}$ & $\begin{array}{c}
 k_z \left(r k_x+k_y\right) \\
\end{array}$ & $\begin{array}{c}
 i \text{D}_{\text{z}} k_z \\
\end{array}$\\ 
 \hline 
$\text{A}^{\text{{\text{'}}{\text{'}}}}_{+}$ & $\begin{array}{c}
 0 \\
\end{array}$ & $\begin{array}{c}
 0 \\
\end{array}$\\ 
 \hline 
$\text{A}^{\text{'}}_{\nu}\text{A}^{\text{'}}_{-\nu}$ & $\begin{array}{c}
 0 \\
 0 \\
\end{array}$ & $\begin{array}{c}
 \text{D}_- \left(r k_x+k_y\right) \\
 -\text{D}_+ \left(r k_x+k_y\right) \\
\end{array}$\\ 
 \hline 
$\text{A}^{\text{{\text{'}}{\text{'}}}}_{\nu}\text{A}^{\text{{\text{'}}{\text{'}}}}_{-\nu}$ & $\begin{array}{c}
 0 \\
 0 \\
\end{array}$ & $\begin{array}{c}
 \text{D}_- k_z \\
 \text{D}_+ k_z \\
\end{array}$\\ 
 \hline 
\end{tabular}
        \hspace*{\fill}
\end{table}

\begin{table}[h]
    \fontsize{9}{10}\selectfont
    \centering
    \hspace*{\fill}
            \begin{tabular}{|c|c|c|} 
\hline 
6: Ci+C2z$ \times $Ci&$\psi$ & $d$  \\ 
 \hline 
$\text{A}^{\text{g}}_{+}$ & $\begin{array}{c}
 1 \\
\end{array}$ & $\begin{array}{c}
 0 \\
\end{array}$\\ 
 \hline 
$\text{A}^{\text{g}}_{-}$ & $\begin{array}{c}
 k_z \left(r k_x+k_y\right) \\
\end{array}$ & $\begin{array}{c}
 0 \\
\end{array}$\\ 
 \hline 
$\text{A}^{\text{u}}_{-}$ & $\begin{array}{c}
 0 \\
\end{array}$ & $\begin{array}{c}
 i \text{D}_{\text{z}} k_z \\
\end{array}$\\ 
 \hline 
$\text{A}^{\text{u}}_{+}$ & $\begin{array}{c}
 0 \\
\end{array}$ & $\begin{array}{c}
i \text{D}_{\text{z}} (r_y k_y+r_x k_x) \\
\end{array}$\\ 
 \hline 
$\text{A}^{\text{g}}_{\nu}\text{A}^{\text{g}}_{-\nu}$ & $\begin{array}{c}
 0 \\
 0 \\
\end{array}$ & $\begin{array}{c}
 0 \\
 0 \\
\end{array}$\\ 
 \hline 
$\text{A}^{\text{u}}_{\nu}\text{A}^{\text{u}}_{-\nu}$ & $\begin{array}{c}
 0 \\
 0 \\
\end{array}$ & $\begin{array}{c}
 \text{D}_- \left( r_y k_y+r_x k_x + r_z k_z\right) \\
 \text{D}_+  \left( -r_y k_y-r_x k_x + r_z k_z\right) \\
\end{array}$\\ 
 \hline 
\end{tabular} 
            \hspace*{\fill}
             \newline
    \vspace*{0.5 cm}
    \newline
    \hspace*{\fill}
    \begin{tabular}{|c|c|c|} 
\hline 
7: C2+IC2x$ \times $C2&$\psi$ & $d$  \\ 
 \hline 
$\text{A}_{+}$ & $\begin{array}{c}
 1 \\
\end{array}$ & $\begin{array}{c}
 i \text{D}_{\text{z}} k_x k_y k_z \\
\end{array}$\\ 
 \hline 
$\text{A}_{-}$ & $\begin{array}{c}
 k_x k_y \\
\end{array}$ & $\begin{array}{c}
 i \text{D}_{\text{z}} k_z \\
\end{array}$\\ 
 \hline 
$\text{B}_{-}$ & $\begin{array}{c}
 k_x k_z \\
\end{array}$ & $\begin{array}{c}
 i \text{D}_{\text{z}} k_y \\
\end{array}$\\ 
 \hline 
$\text{B}_{+}$ & $\begin{array}{c}
 k_y k_z \\
\end{array}$ & $\begin{array}{c}
 i \text{D}_{\text{z}} k_x \\
\end{array}$\\ 
 \hline 
$\text{A}_{\nu}\text{A}_{-\nu}$ & $\begin{array}{c}
 0 \\
 0 \\
\end{array}$ & $\begin{array}{c}
 \text{D}_- k_z \\
 \text{D}_+ k_z \\
\end{array}$\\ 
 \hline 
$\text{B}_{\nu}\text{B}_{-\nu}$ & $\begin{array}{c}
 0 \\
 0 \\
\end{array}$ & $\begin{array}{c}
 \text{D}_- \left(r k_x+k_y\right) \\
 -\text{D}_+ \left(r k_x-k_y\right) \\
\end{array}$\\ 
 \hline 
\end{tabular}
    \hspace*{\fill}
    \newline
    \vspace*{0.5 cm}
    \newline
    \hspace*{\fill}
        \begin{tabular}{|c|c|c|} 
\hline 
8: Cs+C2x$ \times $Cs&$\psi$ & $d$  \\ 
 \hline 
$\text{A}^{\text{'}}_{+}$ & $\begin{array}{c}
 1 \\
\end{array}$ & $\begin{array}{c}
 i \text{D}_{\text{z}} k_y \\
\end{array}$\\ 
 \hline 
$\text{A}^{\text{'}}_{-}$ & $\begin{array}{c}
 k_x k_y \\
\end{array}$ & $\begin{array}{c}
 i \text{D}_{\text{z}} k_x \\
\end{array}$\\ 
 \hline 
$\text{A}^{\text{{\text{'}}{\text{'}}}}_{-}$ & $\begin{array}{c}
 k_x k_z \\
\end{array}$ & $\begin{array}{c}
 i \text{D}_{\text{z}} k_x k_y k_z \\
\end{array}$\\ 
 \hline 
$\text{A}^{\text{{\text{'}}{\text{'}}}}_{+}$ & $\begin{array}{c}
 k_y k_z \\
\end{array}$ & $\begin{array}{c}
 i \text{D}_{\text{z}} k_z \\
\end{array}$\\ 
 \hline 
$\text{A}^{\text{'}}_{\nu}\text{A}^{\text{'}}_{-\nu}$ & $\begin{array}{c}
 0 \\
 0 \\
\end{array}$ & $\begin{array}{c}
 \text{D}_- \left(r k_x+k_y\right) \\
 \text{D}_+ \left(r k_x-k_y\right) \\
\end{array}$\\ 
 \hline 
$\text{A}^{\text{{\text{'}}{\text{'}}}}_{\nu}\text{A}^{\text{{\text{'}}{\text{'}}}}_{-\nu}$ & $\begin{array}{c}
 0 \\
 0 \\
\end{array}$ & $\begin{array}{c}
 \text{D}_- k_z \\
 -\text{D}_+ k_z \\
\end{array}$\\ 
 \hline 
\end{tabular}
        \hspace*{\fill}
    \newline
    \vspace*{0.5 cm}
    \newline
    \hspace*{\fill}
        \begin{tabular}{|c|c|c|} 
\hline 
9: C2+C2x$ \times $C2&$\psi$ & $d$  \\ 
 \hline 
$\text{A}_{+}$ & $\begin{array}{c}
 1 \\
\end{array}$ & $\begin{array}{c}
 i \text{D}_{\text{z}} k_z \\
\end{array}$\\ 
 \hline 
$\text{A}_{-}$ & $\begin{array}{c}
 k_x k_y \\
\end{array}$ & $\begin{array}{c}
 i \text{D}_{\text{z}} k_x k_y k_z \\
\end{array}$\\ 
 \hline 
$\text{B}_{-}$ & $\begin{array}{c}
 k_x k_z \\
\end{array}$ & $\begin{array}{c}
 i \text{D}_{\text{z}} k_x \\
\end{array}$\\ 
 \hline 
$\text{B}_{+}$ & $\begin{array}{c}
 k_y k_z \\
\end{array}$ & $\begin{array}{c}
 i \text{D}_{\text{z}} k_y \\
\end{array}$\\ 
 \hline 
$\text{A}_{\nu}\text{A}_{-\nu}$ & $\begin{array}{c}
 0 \\
 0 \\
\end{array}$ & $\begin{array}{c}
 \text{D}_- k_z \\
 -\text{D}_+ k_z \\
\end{array}$\\ 
 \hline 
$\text{B}_{\nu}\text{B}_{-\nu}$ & $\begin{array}{c}
 0 \\
 0 \\
\end{array}$ & $\begin{array}{c}
 \text{D}_- \left(r k_x+k_y\right) \\
 \text{D}_+ \left(r k_x-k_y\right) \\
\end{array}$\\ 
 \hline 
\end{tabular}
        \hspace*{\fill}
    \newline
    \vspace*{0.5 cm}
    \newline
    \hspace*{\fill}
        \begin{tabular}{|c|c|c|} 
\hline 
10: C2h+IC2x$ \times $C2h&$\psi$ & $d$  \\ 
 \hline 
$\text{A}^{\text{g}}_{+}$ & $\begin{array}{c}
 1 \\
\end{array}$ & $\begin{array}{c}
 0 \\
\end{array}$\\ 
 \hline 
$\text{A}^{\text{g}}_{-}$ & $\begin{array}{c}
 k_x k_y \\
\end{array}$ & $\begin{array}{c}
 0 \\
\end{array}$\\ 
 \hline 
$\text{A}^{\text{u}}_{-}$ & $\begin{array}{c}
 0 \\
\end{array}$ & $\begin{array}{c}
 i \text{D}_{\text{z}} k_z \\
\end{array}$\\ 
 \hline 
$\text{A}^{\text{u}}_{+}$ & $\begin{array}{c}
 0 \\
\end{array}$ & $\begin{array}{c}
 i \text{D}_{\text{z}} k_x k_y k_z \\
\end{array}$\\ 
 \hline 
$\text{B}^{\text{g}}_{-}$ & $\begin{array}{c}
 k_x k_z \\
\end{array}$ & $\begin{array}{c}
 0 \\
\end{array}$\\ 
 \hline 
$\text{B}^{\text{g}}_{+}$ & $\begin{array}{c}
 k_y k_z \\
\end{array}$ & $\begin{array}{c}
 0 \\
\end{array}$\\ 
 \hline 
$\text{B}^{\text{u}}_{-}$ & $\begin{array}{c}
 0 \\
\end{array}$ & $\begin{array}{c}
 i \text{D}_{\text{z}} k_y \\
\end{array}$\\ 
 \hline 
$\text{B}^{\text{u}}_{+}$ & $\begin{array}{c}
 0 \\
\end{array}$ & $\begin{array}{c}
 i \text{D}_{\text{z}} k_x \\
\end{array}$\\ 
 \hline 
$\text{A}^{\text{g}}_{\nu}\text{A}^{\text{g}}_{-\nu}$ & $\begin{array}{c}
 0 \\
 0 \\
\end{array}$ & $\begin{array}{c}
 0 \\
 0 \\
\end{array}$\\ 
 \hline 
$\text{A}^{\text{u}}_{\nu}\text{A}^{\text{u}}_{-\nu}$ & $\begin{array}{c}
 0 \\
 0 \\
\end{array}$ & $\begin{array}{c}
 \text{D}_- k_z \\
 \text{D}_+ k_z \\
\end{array}$\\ 
 \hline 
$\text{B}^{\text{g}}_{\nu}\text{B}^{\text{g}}_{-\nu}$ & $\begin{array}{c}
 0 \\
 0 \\
\end{array}$ & $\begin{array}{c}
 0 \\
 0 \\
\end{array}$\\ 
 \hline 
$\text{B}^{\text{u}}_{\nu}\text{B}^{\text{u}}_{-\nu}$ & $\begin{array}{c}
 0 \\
 0 \\
\end{array}$ & $\begin{array}{c}
 \text{D}_- \left(r k_x+k_y\right) \\
 -\text{D}_+ \left(r k_x-k_y\right) \\
\end{array}$\\ 
 \hline 
\end{tabular}
        \hspace*{\fill}
\end{table}

\begin{table}[h]
    \fontsize{7}{8}\selectfont
    \centering
    \hspace*{\fill}
            \begin{tabular}{|c|c|c|} 
\hline 
11: C2v+IC2z$ \times $C2v&$\psi$ & $d$  \\ 
 \hline 
$\text{A}^1_{+}$ & $\begin{array}{c}
 1 \\
\end{array}$ & $\begin{array}{c}
 i \text{D}_{\text{z}} k_z \\
\end{array}$\\ 
 \hline 
$\text{A}^1_{-}$ & $\begin{array}{c}
 0 \\
\end{array}$ & $\begin{array}{c}
 0 \\
\end{array}$\\ 
 \hline 
$\text{A}^2_{-}$ & $\begin{array}{c}
 0 \\
\end{array}$ & $\begin{array}{c}
 0 \\
\end{array}$\\ 
 \hline 
$\text{A}^2_{+}$ & $\begin{array}{c}
 k_x k_y \\
\end{array}$ & $\begin{array}{c}
 i \text{D}_{\text{z}} k_x k_y k_z \\
\end{array}$\\ 
 \hline 
$\text{B}^1_{-}$ & $\begin{array}{c}
 k_y k_z \\
\end{array}$ & $\begin{array}{c}
 i \text{D}_{\text{z}} k_y \\
\end{array}$\\ 
 \hline 
$\text{B}^1_{+}$ & $\begin{array}{c}
 0 \\
\end{array}$ & $\begin{array}{c}
 0 \\
\end{array}$\\ 
 \hline 
$\text{B}^2_{-}$ & $\begin{array}{c}
 k_x k_z \\
\end{array}$ & $\begin{array}{c}
 i \text{D}_{\text{z}} k_x \\
\end{array}$\\ 
 \hline 
$\text{B}^2_{+}$ & $\begin{array}{c}
 0 \\
\end{array}$ & $\begin{array}{c}
 0 \\
\end{array}$\\ 
 \hline 
$\text{A}^1_{\nu}\text{A}^1_{-\nu}$ & $\begin{array}{c}
 0 \\
 0 \\
\end{array}$ & $\begin{array}{c}
 \text{D}_- k_z \\
 -\text{D}_+ k_z \\
\end{array}$\\ 
 \hline 
$\text{A}^2_{\nu}\text{A}^2_{-\nu}$ & $\begin{array}{c}
 0 \\
 0 \\
\end{array}$ & $\begin{array}{c}
 \text{D}_- k_x k_y k_z \\
 -\text{D}_+ k_x k_y k_z \\
\end{array}$\\ 
 \hline 
$\text{B}^1_{\nu}\text{B}^1_{-\nu}$ & $\begin{array}{c}
 0 \\
 0 \\
\end{array}$ & $\begin{array}{c}
 \text{D}_- k_y \\
 \text{D}_+ k_y \\
\end{array}$\\ 
 \hline 
$\text{B}^2_{\nu}\text{B}^2_{-\nu}$ & $\begin{array}{c}
 0 \\
 0 \\
\end{array}$ & $\begin{array}{c}
 \text{D}_- k_x \\
 \text{D}_+ k_x \\
\end{array}$\\ 
 \hline 
\end{tabular} 
            \hspace*{\fill}
             \newline
    \vspace*{0.5 cm}
    \newline
    \hspace*{\fill}
    \begin{tabular}{|c|c|c|} 
\hline 
12: D2+IEe$ \times $D2&$\psi$ & $d$  \\ 
 \hline 
$\text{A}^1_{+}$ & $\begin{array}{c}
 1 \\
\end{array}$ & $\begin{array}{c}
 i \text{D}_{\text{z}} k_x k_y k_z \\
\end{array}$\\ 
 \hline 
$\text{A}^1_{-}$ & $\begin{array}{c}
 0 \\
\end{array}$ & $\begin{array}{c}
 0 \\
\end{array}$\\ 
 \hline 
$\text{B}^1_{-}$ & $\begin{array}{c}
 0 \\
\end{array}$ & $\begin{array}{c}
 0 \\
\end{array}$\\ 
 \hline 
$\text{B}^1_{+}$ & $\begin{array}{c}
 k_y k_z \\
\end{array}$ & $\begin{array}{c}
 i \text{D}_{\text{z}} k_x \\
\end{array}$\\ 
 \hline 
$\text{B}^2_{-}$ & $\begin{array}{c}
 0 \\
\end{array}$ & $\begin{array}{c}
 0 \\
\end{array}$\\ 
 \hline 
$\text{B}^2_{+}$ & $\begin{array}{c}
 k_x k_z \\
\end{array}$ & $\begin{array}{c}
 i \text{D}_{\text{z}} k_y \\
\end{array}$\\ 
 \hline 
$\text{B}^3_{-}$ & $\begin{array}{c}
 0 \\
\end{array}$ & $\begin{array}{c}
 0 \\
\end{array}$\\ 
 \hline 
$\text{B}^3_{+}$ & $\begin{array}{c}
 k_x k_y \\
\end{array}$ & $\begin{array}{c}
 i \text{D}_{\text{z}} k_z \\
\end{array}$\\ 
 \hline 
$\text{A}^1_{\nu}\text{A}^1_{-\nu}$ & $\begin{array}{c}
 0 \\
 0 \\
\end{array}$ & $\begin{array}{c}
 \text{D}_- k_x k_y k_z \\
 -\text{D}_+ k_x k_y k_z \\
\end{array}$\\ 
 \hline 
$\text{B}^1_{\nu}\text{B}^1_{-\nu}$ & $\begin{array}{c}
 0 \\
 0 \\
\end{array}$ & $\begin{array}{c}
 \text{D}_- k_x \\
 -\text{D}_+ k_x \\
\end{array}$\\ 
 \hline 
$\text{B}^2_{\nu}\text{B}^2_{-\nu}$ & $\begin{array}{c}
 0 \\
 0 \\
\end{array}$ & $\begin{array}{c}
 \text{D}_- k_y \\
 -\text{D}_+ k_y \\
\end{array}$\\ 
 \hline 
$\text{B}^3_{\nu}\text{B}^3_{-\nu}$ & $\begin{array}{c}
 0 \\
 0 \\
\end{array}$ & $\begin{array}{c}
 \text{D}_- k_z \\
 -\text{D}_+ k_z \\
\end{array}$\\ 
 \hline 
\end{tabular}
    \hspace*{\fill}
    \newline
    \vspace*{0.5 cm}
    \newline
    \hspace*{\fill}
        \begin{tabular}{|c|c|c|} 
\hline 
13: C2+C4z$ \times $C2&$\psi$ & $d$  \\ 
 \hline 
$\text{A}_{+}$ & $\begin{array}{c}
 1 \\
\end{array}$ & $\begin{array}{c}
i \text{D}_{\text{z}} k_z (r k_x k_y-k_x^2+k_y^2)\\
\end{array}$\\ 
 \hline 
$\text{A}_{-}$ & $\begin{array}{c}
 r k_x k_y-k_x^2+k_y^2 \\
\end{array}$ & $\begin{array}{c}
 i \text{D}_{\text{z}} k_z \\
\end{array}$\\ 
 \hline 
$\text{B}_{+}$ & $\begin{array}{c}
 C k_z \left(k_y-i k_x\right) \\
 C^* k_z \left(k_y+i k_x\right) \\
\end{array}$ & $\begin{array}{c}
 C \text{D}_{\text{z}} \left(k_y+i k_x\right) \\
 i C^* \text{D}_{\text{z}} \left(k_x+i k_y\right) \\
\end{array}$\\ 
 \hline 
$\text{A}_{\nu}\text{A}_{-\nu}$ & $\begin{array}{c}
 0 \\
 0 \\
\end{array}$ & $\begin{array}{c}
 \text{D}_- k_z \\
 \text{D}_+ k_z \\
\end{array}$\\ 
 \hline 
$\text{B}_{\nu}\text{B}_{-\nu}$ & $\begin{array}{c}
 0 \\
 0 \\
\end{array}$ & $\begin{array}{c}
 \text{D}_- \left(r k_x+k_y\right) \\
 \text{D}_+ \left(k_x-r k_y\right) \\
\end{array}$\\ 
 \hline 
\end{tabular}
        \hspace*{\fill}
    \newline
    \vspace*{0.5 cm}
    \newline
    \hspace*{\fill}
        \begin{tabular}{|c|c|c|} 
\hline 
14: C2+IC4z$ \times $C2&$\psi$ & $d$  \\ 
 \hline 
$\text{A}_{+}$ & $\begin{array}{c}
 1 \\
\end{array}$ & $\begin{array}{c}
 i \text{D}_{\text{z}} k_z \\
\end{array}$\\ 
 \hline 
$\text{A}_{-}$ & $\begin{array}{c}
 r k_x k_y-k_x^2+k_y^2 \\
\end{array}$ & $\begin{array}{c}
 i \text{D}_{\text{z}} k_z (r k_x k_y-k_x^2+k_y^2)\\
\end{array}$\\ 
 \hline 
$\text{B}_{+}$ & $\begin{array}{c}
 C k_z \left(k_y-i k_x\right) \\
 C^* k_z \left(k_y+i k_x\right) \\
\end{array}$ & $\begin{array}{c}
 C \text{D}_{\text{z}} \left(k_y-i k_x\right) \\
 C^* \text{D}_{\text{z}} \left(-k_y-i k_x\right) \\
\end{array}$\\ 
 \hline 
$\text{A}_{\nu}\text{A}_{-\nu}$ & $\begin{array}{c}
 0 \\
 0 \\
\end{array}$ & $\begin{array}{c}
 \text{D}_- k_z \\
 -\text{D}_+ k_z \\
\end{array}$\\ 
 \hline 
$\text{B}_{\nu}\text{B}_{-\nu}$ & $\begin{array}{c}
 0 \\
 0 \\
\end{array}$ & $\begin{array}{c}
 \text{D}_- \left(r k_x+k_y\right) \\
 -\text{D}_+ \left(k_x-r k_y\right) \\
\end{array}$\\ 
 \hline 
\end{tabular}
        \hspace*{\fill}
    \newline
    \vspace*{0.5 cm}
    \newline
    \hspace*{\fill}
        \begin{tabular}{|c|c|c|} 
\hline 
15: C2h+C4z$ \times $C2h&$\psi$ & $d$  \\ 
 \hline 
$\text{A}^{\text{g}}_{+}$ & $\begin{array}{c}
 1 \\
\end{array}$ & $\begin{array}{c}
 0 \\
\end{array}$\\ 
 \hline 
$\text{A}^{\text{g}}_{-}$ & $\begin{array}{c}
 r k_x k_y-k_x^2+k_y^2 \\
\end{array}$ & $\begin{array}{c}
 0 \\
\end{array}$\\ 
 \hline 
$\text{A}^{\text{u}}_{-}$ & $\begin{array}{c}
 0 \\
\end{array}$ & $\begin{array}{c}
 i \text{D}_{\text{z}} k_z \\
\end{array}$\\ 
 \hline 
$\text{A}^{\text{u}}_{+}$ & $\begin{array}{c}
 0 \\
\end{array}$ & $\begin{array}{c}
i \text{D}_{\text{z}} k_z (r k_x k_y-k_x^2+k_y^2)\\
\end{array}$\\ 
 \hline 
$\text{B}^{\text{g}}_{+}$ & $\begin{array}{c}
 C k_z \left(k_y-i k_x\right) \\
 C^* k_z \left(k_y+i k_x\right) \\
\end{array}$ & $\begin{array}{c}
 0 \\
 0 \\
\end{array}$\\ 
 \hline 
$\text{B}^{\text{u}}_{+}$ & $\begin{array}{c}
 0 \\
 0 \\
\end{array}$ & $\begin{array}{c}
 C \text{D}_{\text{z}} \left(k_y+i k_x\right) \\
 i C^* \text{D}_{\text{z}} \left(k_x+i k_y\right) \\
\end{array}$\\ 
 \hline 
$\text{A}^{\text{g}}_{\nu}\text{A}^{\text{g}}_{-\nu}$ & $\begin{array}{c}
 0 \\
 0 \\
\end{array}$ & $\begin{array}{c}
 0 \\
 0 \\
\end{array}$\\ 
 \hline 
$\text{A}^{\text{u}}_{\nu}\text{A}^{\text{u}}_{-\nu}$ & $\begin{array}{c}
 0 \\
 0 \\
\end{array}$ & $\begin{array}{c}
 \text{D}_- k_z \\
 \text{D}_+ k_z \\
\end{array}$\\ 
 \hline 
$\text{B}^{\text{g}}_{\nu}\text{B}^{\text{g}}_{-\nu}$ & $\begin{array}{c}
 0 \\
 0 \\
\end{array}$ & $\begin{array}{c}
 0 \\
 0 \\
\end{array}$\\ 
 \hline 
$\text{B}^{\text{u}}_{\nu}\text{B}^{\text{u}}_{-\nu}$ & $\begin{array}{c}
 0 \\
 0 \\
\end{array}$ & $\begin{array}{c}
 \text{D}_- \left(r k_x+k_y\right) \\
 \text{D}_+ \left(k_x-r k_y\right) \\
\end{array}$\\ 
 \hline 
\end{tabular}
        \hspace*{\fill}
\end{table}

\begin{table}[h]
    \fontsize{7}{8}\selectfont
    \centering
    \hspace*{\fill}
            \begin{tabular}{|c|c|c|} 
\hline 
16: S4+C4z$ \times $S4&$\psi$ & $d$  \\ 
 \hline 
$\text{A}_{+}$ & $\begin{array}{c}
 1 \\
\end{array}$ & $\begin{array}{c}
i \text{D}_{\text{z}} k_z (r k_x k_y-k_x^2+k_y^2)\\
\end{array}$\\ 
 \hline 
$\text{A}_{-}$ & $\begin{array}{c}
 0 \\
\end{array}$ & $\begin{array}{c}
 0 \\
\end{array}$\\ 
 \hline 
$\text{B}_{-}$ & $\begin{array}{c}
 r k_x k_y-k_x^2+k_y^2 \\
\end{array}$ & $\begin{array}{c}
 i \text{D}_{\text{z}} k_z \\
\end{array}$\\ 
 \hline 
$\text{B}_{+}$ & $\begin{array}{c}
 0 \\
\end{array}$ & $\begin{array}{c}
 0 \\
\end{array}$\\ 
 \hline 
$\text{E}^2_{-}$ & $\begin{array}{c}
 0 \\
 0 \\
\end{array}$ & $\begin{array}{c}
 0 \\
 0 \\
\end{array}$\\ 
 \hline 
$\text{E}^2_{+}$ & $\begin{array}{c}
 C k_z \left(k_y-i k_x\right) \\
 C^* k_z \left(k_y+i k_x\right) \\
\end{array}$ & $\begin{array}{c}
 C \text{D}_{\text{z}} \left(k_y+i k_x\right) \\
 i C^* \text{D}_{\text{z}} \left(k_x+i k_y\right) \\
\end{array}$\\ 
 \hline 
$\text{A}_{\nu}\text{A}_{-\nu}$ & $\begin{array}{c}
 0 \\
 0 \\
\end{array}$ & $\begin{array}{c}
 -\text{D}_- k_z \left(-r k_x k_y+k_x^2-k_y^2\right) \\
 \text{D}_+ k_z \left(-r k_x k_y+k_x^2-k_y^2\right) \\
\end{array}$\\ 
 \hline 
$\text{B}_{\nu}\text{B}_{-\nu}$ & $\begin{array}{c}
 0 \\
 0 \\
\end{array}$ & $\begin{array}{c}
 \text{D}_- k_z \\
 \text{D}_+ k_z \\
\end{array}$\\ 
 \hline 
$\text{E}^2_{\nu}\text{E}^2_{-\nu}$ & $\begin{array}{c}
 0 \\
 0 \\
 0 \\
 0 \\
\end{array}$ & $\begin{array}{c}
 C \text{D}_- \left(k_y+i k_x\right) \\
 C \text{D}_+ \left(k_x-i k_y\right) \\
 \text{D}_- C^* \left(k_y-i k_x\right) \\
 \text{D}_+ C^* \left(k_x+i k_y\right) \\
\end{array}$\\ 
 \hline 
\end{tabular}
            \hspace*{\fill}
             \newline
    \vspace*{0.5 cm}
    \newline
    \hspace*{\fill}
    \begin{tabular}{|c|c|c|} 
\hline 
17: C4+IC2z$ \times $C4&$\psi$ & $d$  \\ 
 \hline 
$\text{A}_{+}$ & $\begin{array}{c}
 1 \\
\end{array}$ & $\begin{array}{c}
 i \text{D}_{\text{z}} k_z \\
\end{array}$\\ 
 \hline 
$\text{A}_{-}$ & $\begin{array}{c}
 0 \\
\end{array}$ & $\begin{array}{c}
 0 \\
\end{array}$\\ 
 \hline 
$\text{B}_{-}$ & $\begin{array}{c}
 0 \\
\end{array}$ & $\begin{array}{c}
 0 \\
\end{array}$\\ 
 \hline 
$\text{B}_{+}$ & $\begin{array}{c}
 r k_x k_y-k_x^2+k_y^2 \\
\end{array}$ & $\begin{array}{c}
i \text{D}_{\text{z}} k_z (r k_x k_y-k_x^2+k_y^2)\\
\end{array}$\\ 
 \hline 
$\text{E}^2_{-}$ & $\begin{array}{c}
 C k_z \left(k_y+i k_x\right) \\
 C^* k_z \left(k_y-i k_x\right) \\
\end{array}$ & $\begin{array}{c}
 C \text{D}_{\text{z}} \left(k_y+i k_x\right) \\
 i C^* \text{D}_{\text{z}} \left(k_x+i k_y\right) \\
\end{array}$\\ 
 \hline 
$\text{E}^2_{+}$ & $\begin{array}{c}
 0 \\
 0 \\
\end{array}$ & $\begin{array}{c}
 0 \\
 0 \\
\end{array}$\\ 
 \hline 
$\text{A}_{\nu}\text{A}_{-\nu}$ & $\begin{array}{c}
 0 \\
 0 \\
\end{array}$ & $\begin{array}{c}
 \text{D}_- k_z \\
 -\text{D}_+ k_z \\
\end{array}$\\ 
 \hline 
$\text{B}_{\nu}\text{B}_{-\nu}$ & $\begin{array}{c}
 0 \\
 0 \\
\end{array}$ & $\begin{array}{c}
 -\text{D}_- k_z \left(-r k_x k_y+k_x^2-k_y^2\right) \\
 \text{D}_+ k_z \left(-r k_x k_y+k_x^2-k_y^2\right) \\
\end{array}$\\ 
 \hline 
$\text{E}^2_{\nu}\text{E}^2_{-\nu}$ & $\begin{array}{c}
 0 \\
 0 \\
 0 \\
 0 \\
\end{array}$ & $\begin{array}{c}
 C \text{D}_- \left(k_y+i k_x\right) \\
 C \text{D}_+ \left(k_y+i k_x\right) \\
 \text{D}_- C^* \left(k_y-i k_x\right) \\
 \text{D}_+ C^* \left(k_y-i k_x\right) \\
\end{array}$\\ 
 \hline 
\end{tabular}
    \hspace*{\fill}
    \newline
    \vspace*{0.5 cm}
    \newline
    \hspace*{\fill}
        \begin{tabular}{|c|c|c|} 
\hline 
18: C4+C2x$ \times $C4&$\psi$ & $d$  \\ 
 \hline 
$\text{A}_{+}$ & $\begin{array}{c}
 1 \\
\end{array}$ & $\begin{array}{c}
 i \text{D}_{\text{z}} k_z \\
\end{array}$\\ 
 \hline 
$\text{A}_{-}$ & $\begin{array}{c}
 k_x k_y^3-k_x^3 k_y \\
\end{array}$ & $\begin{array}{c}
 -i \text{D}_{\text{z}} k_x k_y k_z \left(k_x^2-k_y^2\right) \\
\end{array}$\\ 
 \hline 
$\text{B}_{-}$ & $\begin{array}{c}
 k_x k_y \\
\end{array}$ & $\begin{array}{c}
 i \text{D}_{\text{z}} k_x k_y k_z \\
\end{array}$\\ 
 \hline 
$\text{B}_{+}$ & $\begin{array}{c}
 k_y^2-k_x^2 \\
\end{array}$ & $\begin{array}{c}
 -i \text{D}_{\text{z}} k_z \left(k_x^2-k_y^2\right) \\
\end{array}$\\ 
 \hline 
$\text{E}^1_{0}\text{E}^2_{0}$ & $\begin{array}{c}
 k_z \left(k_x+i k_y\right) \\
 -\left(k_z \left(k_x-i k_y\right)\right) \\
\end{array}$ & $\begin{array}{c}
 \text{D}_{\text{z}} \left(k_y-i k_x\right) \\
 \text{D}_{\text{z}} \left(k_y+i k_x\right) \\
\end{array}$\\ 
 \hline 
$\text{A}_{\nu}\text{A}_{-\nu}$ & $\begin{array}{c}
 0 \\
 0 \\
\end{array}$ & $\begin{array}{c}
 \text{D}_- k_z \\
 -\text{D}_+ k_z \\
\end{array}$\\ 
 \hline 
$\text{B}_{\nu}\text{B}_{-\nu}$ & $\begin{array}{c}
 0 \\
 0 \\
\end{array}$ & $\begin{array}{c}
 -\text{D}_- k_z \left(-r k_x k_y+k_x^2-k_y^2\right) \\
 \text{D}_+ k_z \left(r k_x k_y+k_x^2-k_y^2\right) \\
\end{array}$\\ 
 \hline 
$\text{E}^1_{\mu}\text{E}^2_{-\mu}$ & $\begin{array}{c}
 0 \\
 0 \\
 0 \\
 0 \\
\end{array}$ & $\begin{array}{c}
 C \text{D}_- \left(k_y-i k_x\right) \\
 -i C \text{D}_+ \left(k_x-i k_y\right) \\
 \text{D}_- C^* \left(k_y+i k_x\right) \\
 i \text{D}_+ C^* \left(k_x+i k_y\right) \\
\end{array}$\\ 
 \hline 
\end{tabular}
        \hspace*{\fill}
    \newline
    \vspace*{0.5 cm}
    \newline
    \hspace*{\fill}
        \begin{tabular}{|c|c|c|} 
\hline 
19: D2+C4z$ \times $D2&$\psi$ & $d$  \\ 
 \hline 
$\text{A}^1_{+}$ & $\begin{array}{c}
 1 \\
\end{array}$ & $\begin{array}{c}
 i \text{D}_{\text{z}} k_x k_y k_z \\
\end{array}$\\ 
 \hline 
$\text{A}^1_{-}$ & $\begin{array}{c}
 k_y^2-k_x^2 \\
\end{array}$ & $\begin{array}{c}
 -i \text{D}_{\text{z}} k_x k_y k_z \left(k_x^2-k_y^2\right) \\
\end{array}$\\ 
 \hline 
$\text{B}^3_{-}$ & $\begin{array}{c}
 k_x k_y \\
\end{array}$ & $\begin{array}{c}
 i \text{D}_{\text{z}} k_z \\
\end{array}$\\ 
 \hline 
$\text{B}^3_{+}$ & $\begin{array}{c}
 k_x k_y^3-k_x^3 k_y \\
\end{array}$ & $\begin{array}{c}
 -i \text{D}_{\text{z}} k_z \left(k_x^2-k_y^2\right) \\
\end{array}$\\ 
 \hline 
$\text{B}^1_{0}\text{B}^2_{0}$ & $\begin{array}{c}
 k_y k_z \\
 k_x k_z \\
\end{array}$ & $\begin{array}{c}
 i \text{D}_{\text{z}} k_x \\
 i \text{D}_{\text{z}} k_y \\
\end{array}$\\ 
 \hline 
$\text{A}^1_{\nu}\text{A}^1_{-\nu}$ & $\begin{array}{c}
 0 \\
 0 \\
\end{array}$ & $\begin{array}{c}
 \text{D}_- k_x k_y k_z \\
 -\text{D}_+ k_x k_y k_z \\
\end{array}$\\ 
 \hline 
$\text{B}^1_{\mu}\text{B}^2_{-\mu}$ & $\begin{array}{c}
 0 \\
 0 \\
\end{array}$ & $\begin{array}{c}
 \text{D}_- k_x \\
 -\text{D}_+ k_y \\
\end{array}$\\ 
 \hline 
$\text{B}^3_{\nu}\text{B}^3_{-\nu}$ & $\begin{array}{c}
 0 \\
 0 \\
\end{array}$ & $\begin{array}{c}
 \text{D}_- k_z \\
 \text{D}_+ k_z \\
\end{array}$\\ 
 \hline 
\end{tabular}
        \hspace*{\fill}
\end{table}

\begin{table}[h]
    \fontsize{7}{8}\selectfont
    \centering
    \hspace*{\fill}
    \begin{tabular}{|c|c|c|} 
\hline 
20: C2v+C4z$ \times $C2v&$\psi$ & $d$  \\ 
 \hline 
$\text{A}^1_{+}$ & $\begin{array}{c}
 1 \\
\end{array}$ & $\begin{array}{c}
 -i \text{D}_{\text{z}} k_z \left(k_x^2-k_y^2\right) \\
\end{array}$\\ 
 \hline 
$\text{A}^1_{-}$ & $\begin{array}{c}
 k_y^2-k_x^2 \\
\end{array}$ & $\begin{array}{c}
 i \text{D}_{\text{z}} k_z \\
\end{array}$\\ 
 \hline 
$\text{A}^2_{-}$ & $\begin{array}{c}
 k_x k_y \\
\end{array}$ & $\begin{array}{c}
 -i \text{D}_{\text{z}} k_x k_y k_z \left(k_x^2-k_y^2\right) \\
\end{array}$\\ 
 \hline 
$\text{A}^2_{+}$ & $\begin{array}{c}
 k_x k_y^3-k_x^3 k_y \\
\end{array}$ & $\begin{array}{c}
 i \text{D}_{\text{z}} k_x k_y k_z \\
\end{array}$\\ 
 \hline 
$\text{B}^2_{0}\text{B}^1_{0}$ & $\begin{array}{c}
 k_x k_z \\
 -k_y k_z \\
\end{array}$ & $\begin{array}{c}
 i \text{D}_{\text{z}} k_x \\
 i \text{D}_{\text{z}} k_y \\
\end{array}$\\ 
 \hline 
$\text{A}^1_{\nu}\text{A}^1_{-\nu}$ & $\begin{array}{c}
 0 \\
 0 \\
\end{array}$ & $\begin{array}{c}
 \text{D}_- k_z \\
 \text{D}_+ k_z \\
\end{array}$\\ 
 \hline 
$\text{A}^2_{\nu}\text{A}^2_{-\nu}$ & $\begin{array}{c}
 0 \\
 0 \\
\end{array}$ & $\begin{array}{c}
 \text{D}_- k_x k_y k_z \\
 -\text{D}_+ k_x k_y k_z \\
\end{array}$\\ 
 \hline 
$\text{B}^2_{\mu}\text{B}^1_{-\mu}$ & $\begin{array}{c}
 0 \\
 0 \\
\end{array}$ & $\begin{array}{c}
 \text{D}_- k_x \\
 -\text{D}_+ k_y \\
\end{array}$\\ 
 \hline 
\end{tabular}
                \hspace*{\fill}
             \newline
    \vspace*{0.5 cm}
    \newline
    \hspace*{\fill}\begin{tabular}{|c|c|c|} 
\hline 
21: C2v+IC4z$ \times $C2v&$\psi$ & $d$  \\ 
 \hline 
$\text{A}^1_{+}$ & $\begin{array}{c}
 1 \\
\end{array}$ & $\begin{array}{c}
 i \text{D}_{\text{z}} k_z \\
\end{array}$\\ 
 \hline 
$\text{A}^1_{-}$ & $\begin{array}{c}
 k_y^2-k_x^2 \\
\end{array}$ & $\begin{array}{c}
 -i \text{D}_{\text{z}} k_z \left(k_x^2-k_y^2\right) \\
\end{array}$\\ 
 \hline 
$\text{A}^2_{-}$ & $\begin{array}{c}
 k_x k_y \\
\end{array}$ & $\begin{array}{c}
 i \text{D}_{\text{z}} k_x k_y k_z \\
\end{array}$\\ 
 \hline 
$\text{A}^2_{+}$ & $\begin{array}{c}
 k_x k_y^3-k_x^3 k_y \\
\end{array}$ & $\begin{array}{c}
 -i \text{D}_{\text{z}} k_x k_y k_z \left(k_x^2-k_y^2\right) \\
\end{array}$\\ 
 \hline 
$\text{B}^2_{0}\text{B}^1_{0}$ & $\begin{array}{c}
 k_x k_z \\
 -k_y k_z \\
\end{array}$ & $\begin{array}{c}
 i \text{D}_{\text{z}} k_x \\
 -i \text{D}_{\text{z}} k_y \\
\end{array}$\\ 
 \hline 
$\text{A}^1_{\nu}\text{A}^1_{-\nu}$ & $\begin{array}{c}
 0 \\
 0 \\
\end{array}$ & $\begin{array}{c}
 \text{D}_- k_z \\
 -\text{D}_+ k_z \\
\end{array}$\\ 
 \hline 
$\text{A}^2_{\nu}\text{A}^2_{-\nu}$ & $\begin{array}{c}
 0 \\
 0 \\
\end{array}$ & $\begin{array}{c}
 \text{D}_- k_x k_y k_z \\
 \text{D}_+ k_x k_y k_z \\
\end{array}$\\ 
 \hline 
$\text{B}^2_{\mu}\text{B}^1_{-\mu}$ & $\begin{array}{c}
 0 \\
 0 \\
\end{array}$ & $\begin{array}{c}
 \text{D}_- k_x \\
 \text{D}_+ k_y \\
\end{array}$\\ 
 \hline 
\end{tabular} 
            \hspace*{\fill}
             \newline
    \vspace*{0.5 cm}
    \newline
    \hspace*{\fill}
    \begin{tabular}{|c|c|c|} 
\hline 
22: D2+IC4z$ \times $D2&$\psi$ & $d$  \\ 
 \hline 
$\text{A}^1_{+}$ & $\begin{array}{c}
 1 \\
\end{array}$ & $\begin{array}{c}
 -i \text{D}_{\text{z}} k_x k_y k_z \left(k_x^2-k_y^2\right) \\
\end{array}$\\ 
 \hline 
$\text{A}^1_{-}$ & $\begin{array}{c}
 k_y^2-k_x^2 \\
\end{array}$ & $\begin{array}{c}
 i \text{D}_{\text{z}} k_x k_y k_z \\
\end{array}$\\ 
 \hline 
$\text{B}^3_{-}$ & $\begin{array}{c}
 k_x k_y \\
\end{array}$ & $\begin{array}{c}
 -i \text{D}_{\text{z}} k_z \left(k_x^2-k_y^2\right) \\
\end{array}$\\ 
 \hline 
$\text{B}^3_{+}$ & $\begin{array}{c}
 k_x k_y^3-k_x^3 k_y \\
\end{array}$ & $\begin{array}{c}
 i \text{D}_{\text{z}} k_z \\
\end{array}$\\ 
 \hline 
$\text{B}^1_{0}\text{B}^2_{0}$ & $\begin{array}{c}
 k_y k_z \\
 k_x k_z \\
\end{array}$ & $\begin{array}{c}
 i \text{D}_{\text{z}} k_x \\
 -i \text{D}_{\text{z}} k_y \\
\end{array}$\\ 
 \hline 
$\text{A}^1_{\nu}\text{A}^1_{-\nu}$ & $\begin{array}{c}
 0 \\
 0 \\
\end{array}$ & $\begin{array}{c}
 \text{D}_- k_x k_y k_z \\
 \text{D}_+ k_x k_y k_z \\
\end{array}$\\ 
 \hline 
$\text{B}^1_{\mu}\text{B}^2_{-\mu}$ & $\begin{array}{c}
 0 \\
 0 \\
\end{array}$ & $\begin{array}{c}
 \text{D}_- k_x \\
 \text{D}_+ k_y \\
\end{array}$\\ 
 \hline 
$\text{B}^3_{\nu}\text{B}^3_{-\nu}$ & $\begin{array}{c}
 0 \\
 0 \\
\end{array}$ & $\begin{array}{c}
 \text{D}_- k_z \\
 -\text{D}_+ k_z \\
\end{array}$\\ 
 \hline 
\end{tabular}
    \hspace*{\fill}
    \newline
    \vspace*{0.5 cm}
    \newline
    \hspace*{\fill}
        \begin{tabular}{|c|c|c|} 
\hline 
23: C4+IC2x$ \times $C4&$\psi$ & $d$  \\ 
 \hline 
$\text{A}_{+}$ & $\begin{array}{c}
 1 \\
\end{array}$ & $\begin{array}{c}
 -i \text{D}_{\text{z}} k_x k_y k_z \left(k_x^2-k_y^2\right) \\
\end{array}$\\ 
 \hline 
$\text{A}_{-}$ & $\begin{array}{c}
 k_x k_y^3-k_x^3 k_y \\
\end{array}$ & $\begin{array}{c}
 i \text{D}_{\text{z}} k_z \\
\end{array}$\\ 
 \hline 
$\text{B}_{-}$ & $\begin{array}{c}
 k_x k_y \\
\end{array}$ & $\begin{array}{c}
 -i \text{D}_{\text{z}} k_z \left(k_x^2-k_y^2\right) \\
\end{array}$\\ 
 \hline 
$\text{B}_{+}$ & $\begin{array}{c}
 k_y^2-k_x^2 \\
\end{array}$ & $\begin{array}{c}
 i \text{D}_{\text{z}} k_x k_y k_z \\
\end{array}$\\ 
 \hline 
$\text{E}^1_{0}\text{E}^2_{0}$ & $\begin{array}{c}
 k_z \left(k_x+i k_y\right) \\
 -\left(k_z \left(k_x-i k_y\right)\right) \\
\end{array}$ & $\begin{array}{c}
 \text{D}_{\text{z}} \left(k_x+i k_y\right) \\
 \text{D}_{\text{z}} \left(k_x-i k_y\right) \\
\end{array}$\\ 
 \hline 
$\text{A}_{\nu}\text{A}_{-\nu}$ & $\begin{array}{c}
 0 \\
 0 \\
\end{array}$ & $\begin{array}{c}
 \text{D}_- k_z \\
 \text{D}_+ k_z \\
\end{array}$\\ 
 \hline 
$\text{B}_{\nu}\text{B}_{-\nu}$ & $\begin{array}{c}
 0 \\
 0 \\
\end{array}$ & $\begin{array}{c}
 -\text{D}_- k_z \left(-r k_x k_y+k_x^2-k_y^2\right) \\
 -\text{D}_+ k_z \left(r k_x k_y+k_x^2-k_y^2\right) \\
\end{array}$\\ 
 \hline 
$\text{E}^1_{\mu}\text{E}^2_{-\mu}$ & $\begin{array}{c}
 0 \\
 0 \\
 0 \\
 0 \\
\end{array}$ & $\begin{array}{c}
 C \text{D}_- \left(k_y-i k_x\right) \\
 C \text{D}_+ \left(k_y+i k_x\right) \\
 \text{D}_- C^* \left(k_y+i k_x\right) \\
 \text{D}_+ C^* \left(k_y-i k_x\right) \\
\end{array}$\\ 
 \hline 
\end{tabular}
        \hspace*{\fill}
    \newline
    \vspace*{0.5 cm}
    \newline
    \hspace*{\fill}
        \begin{tabular}{|c|c|c|} 
\hline 
24: S4+C2x$ \times $S4&$\psi$ & $d$  \\ 
 \hline 
$\text{A}_{+}$ & $\begin{array}{c}
 1 \\
\end{array}$ & $\begin{array}{c}
 -i \text{D}_{\text{z}} k_z \left(k_x^2-k_y^2\right) \\
\end{array}$\\ 
 \hline 
$\text{A}_{-}$ & $\begin{array}{c}
 k_x k_y^3-k_x^3 k_y \\
\end{array}$ & $\begin{array}{c}
 i \text{D}_{\text{z}} k_x k_y k_z \\
\end{array}$\\ 
 \hline 
$\text{B}_{-}$ & $\begin{array}{c}
 k_x k_y \\
\end{array}$ & $\begin{array}{c}
 -i \text{D}_{\text{z}} k_x k_y k_z \left(k_x^2-k_y^2\right) \\
\end{array}$\\ 
 \hline 
$\text{B}_{+}$ & $\begin{array}{c}
 k_y^2-k_x^2 \\
\end{array}$ & $\begin{array}{c}
 i \text{D}_{\text{z}} k_z \\
\end{array}$\\ 
 \hline 
$\text{E}^1_{0}\text{E}^2_{0}$ & $\begin{array}{c}
 -\left(k_z \left(k_x-i k_y\right)\right) \\
 k_z \left(k_x+i k_y\right) \\
\end{array}$ & $\begin{array}{c}
 \text{D}_{\text{z}} \left(k_y-i k_x\right) \\
 \text{D}_{\text{z}} \left(k_y+i k_x\right) \\
\end{array}$\\ 
 \hline 
$\text{A}_{\nu}\text{A}_{-\nu}$ & $\begin{array}{c}
 0 \\
 0 \\
\end{array}$ & $\begin{array}{c}
 -\text{D}_- k_z \left(-r k_x k_y+k_x^2-k_y^2\right) \\
 \text{D}_+ k_z \left(r k_x k_y+k_x^2-k_y^2\right) \\
\end{array}$\\ 
 \hline 
$\text{B}_{\nu}\text{B}_{-\nu}$ & $\begin{array}{c}
 0 \\
 0 \\
\end{array}$ & $\begin{array}{c}
 \text{D}_- k_z \\
 -\text{D}_+ k_z \\
\end{array}$\\ 
 \hline 
$\text{E}^1_{\mu}\text{E}^2_{-\mu}$ & $\begin{array}{c}
 0 \\
 0 \\
 0 \\
 0 \\
\end{array}$ & $\begin{array}{c}
 C \text{D}_- \left(k_y-i k_x\right) \\
 -i C \text{D}_+ \left(k_x-i k_y\right) \\
 \text{D}_- C^* \left(k_y+i k_x\right) \\
 i \text{D}_+ C^* \left(k_x+i k_y\right) \\
\end{array}$\\ 
 \hline 
\end{tabular}
        \hspace*{\fill}
\end{table}

\begin{table}[h]
    \fontsize{7}{7.5}\selectfont
    \centering
    \hspace*{\fill}
      \begin{tabular}{|c|c|c|} 
\hline 
25: D2d+IEe$ \times $D2d&$\psi$ & $d$  \\ 
 \hline 
$\text{A}^1_{+}$ & $\begin{array}{c}
 1 \\
\end{array}$ & $\begin{array}{c}
 i \text{D}_{\text{z}} k_x k_y k_z \\
\end{array}$\\ 
 \hline 
$\text{A}^1_{-}$ & $\begin{array}{c}
 0 \\
\end{array}$ & $\begin{array}{c}
 0 \\
\end{array}$\\ 
 \hline 
$\text{A}^2_{-}$ & $\begin{array}{c}
 0 \\
\end{array}$ & $\begin{array}{c}
 0 \\
\end{array}$\\ 
 \hline 
$\text{A}^2_{+}$ & $\begin{array}{c}
 k_x k_z^3-k_x^3 k_z \\
\end{array}$ & $\begin{array}{c}
 -i \text{D}_{\text{z}} k_y \left(k_x^2-k_z^2\right) \\
\end{array}$\\ 
 \hline 
$\text{B}^1_{-}$ & $\begin{array}{c}
 0 \\
\end{array}$ & $\begin{array}{c}
 0 \\
\end{array}$\\ 
 \hline 
$\text{B}^1_{+}$ & $\begin{array}{c}
 k_x k_z \\
\end{array}$ & $\begin{array}{c}
 i \text{D}_{\text{z}} k_y \\
\end{array}$\\ 
 \hline 
$\text{B}^2_{-}$ & $\begin{array}{c}
 0 \\
\end{array}$ & $\begin{array}{c}
 0 \\
\end{array}$\\
\hline 
\end{tabular}
    \hspace*{\fill}
             \newline
    \vspace*{0.5 cm}
    \newline
    \hspace*{\fill}\begin{tabular}{|c|c|c|} 
\hline 
26: C4v+IEe$ \times $C4v&$\psi$ & $d$  \\ 
 \hline 
$\text{A}^1_{+}$ & $\begin{array}{c}
 1 \\
\end{array}$ & $\begin{array}{c}
 i \text{D}_{\text{z}} k_z \\
\end{array}$\\ 
 \hline 
$\text{A}^1_{-}$ & $\begin{array}{c}
 0 \\
\end{array}$ & $\begin{array}{c}
 0 \\
\end{array}$\\ 
 \hline 
$\text{A}^2_{-}$ & $\begin{array}{c}
 0 \\
\end{array}$ & $\begin{array}{c}
 0 \\
\end{array}$\\ 
 \hline 
$\text{A}^2_{+}$ & $\begin{array}{c}
 k_x k_y^3-k_x^3 k_y \\
\end{array}$ & $\begin{array}{c}
 -i \text{D}_{\text{z}} k_x k_y k_z \left(k_x^2-k_y^2\right) \\
\end{array}$\\ 
 \hline 
$\text{B}^1_{-}$ & $\begin{array}{c}
 0 \\
\end{array}$ & $\begin{array}{c}
 0 \\
\end{array}$\\ 
 \hline 
$\text{B}^1_{+}$ & $\begin{array}{c}
 k_y^2-k_x^2 \\
\end{array}$ & $\begin{array}{c}
 -i \text{D}_{\text{z}} k_z \left(k_x^2-k_y^2\right) \\
\end{array}$\\ 
 \hline 
$\text{B}^2_{-}$ & $\begin{array}{c}
 0 \\
\end{array}$ & $\begin{array}{c}
 0 \\
\end{array}$\\
\hline 
\end{tabular}
            \hspace*{\fill}
             \newline
    \vspace*{0.5 cm}
    \newline
    \hspace*{\fill}
    \begin{tabular}{|c|c|c|} 
\hline 
27:D2h+C4z$ \times $D2h&$\psi$ & $d$  \\ 
 \hline 
$\text{A}^{\text{1g}}_{+}$ & $\begin{array}{c}
 1 \\
\end{array}$ & $\begin{array}{c}
 0 \\
\end{array}$\\ 
 \hline 
$\text{A}^{\text{1g}}_{-}$ & $\begin{array}{c}
 k_y^2-k_x^2 \\
\end{array}$ & $\begin{array}{c}
 0 \\
\end{array}$\\ 
 \hline 
$\text{A}^{\text{1u}}_{-}$ & $\begin{array}{c}
 0 \\
\end{array}$ & $\begin{array}{c}
 -i \text{D}_{\text{z}} k_x k_y k_z \left(k_x^2-k_y^2\right) \\
\end{array}$\\ 
 \hline 
$\text{A}^{\text{1u}}_{+}$ & $\begin{array}{c}
 0 \\
\end{array}$ & $\begin{array}{c}
 i \text{D}_{\text{z}} k_x k_y k_z \\
\end{array}$\\ 
 \hline 
$\text{B}^{\text{2g}}_{-}$ & $\begin{array}{c}
 k_x k_y \\
\end{array}$ & $\begin{array}{c}
 0 \\
\end{array}$\\ 
 \hline 
$\text{B}^{\text{2g}}_{+}$ & $\begin{array}{c}
 k_x k_y^3-k_x^3 k_y \\
\end{array}$ & $\begin{array}{c}
 0 \\
\end{array}$\\ 
 \hline 
$\text{B}^{\text{2u}}_{-}$ & $\begin{array}{c}
 0 \\
\end{array}$ & $\begin{array}{c}
 i \text{D}_{\text{z}} k_z \\
\end{array}$\\
\hline 
\end{tabular}
        \hspace*{\fill}
\end{table}

\begin{table}[h]
    \fontsize{8}{9}\selectfont
    \centering
    \hspace*{\fill}
    \begin{tabular}{|c|c|c|} 
\hline 
28: C4h+C2x$ \times $C4h&$\psi$ & $d$  \\ 
 \hline 
$\Gamma ^1_{+}$ & $\begin{array}{c}
 1 \\
\end{array}$ & $\begin{array}{c}
 0 \\
\end{array}$\\ 
 \hline 
$\Gamma ^1_{-}$ & $\begin{array}{c}
 k_x k_y^3-k_x^3 k_y \\
\end{array}$ & $\begin{array}{c}
 0 \\
\end{array}$\\ 
 \hline 
$\Gamma ^4_{-}$ & $\begin{array}{c}
 0 \\
\end{array}$ & $\begin{array}{c}
 -i \text{D}_{\text{z}} k_x k_y k_z \left(k_x^2-k_y^2\right) \\
\end{array}$\\ 
 \hline 
$\Gamma ^4_{+}$ & $\begin{array}{c}
 0 \\
\end{array}$ & $\begin{array}{c}
 i \text{D}_{\text{z}} k_z \\
\end{array}$\\ 
 \hline 
$\Gamma ^5_{-}$ & $\begin{array}{c}
 0 \\
\end{array}$ & $\begin{array}{c}
 i \text{D}_{\text{z}} k_x k_y k_z \\
\end{array}$\\ 
 \hline 
$\Gamma ^5_{+}$ & $\begin{array}{c}
 0 \\
\end{array}$ & $\begin{array}{c}
 -i \text{D}_{\text{z}} k_z \left(k_x^2-k_y^2\right) \\
\end{array}$\\ 
 \hline 
$\Gamma ^8_{-}$ & $\begin{array}{c}
 k_x k_y \\
\end{array}$ & $\begin{array}{c}
 0 \\
\end{array}$\\
\hline 
\end{tabular}
    \hspace*{\fill}
             \newline
    \vspace*{0.5 cm}
    \newline
    \hspace*{\fill}\begin{tabular}{|c|c|c|} 
\hline 
29: D4+IEe$ \times $D4&$\psi$ & $d$  \\ 
 \hline 
$\Gamma ^1_{+}$ & $\begin{array}{c}
 1 \\
\end{array}$ & $\begin{array}{c}
 -i \text{D}_{\text{z}} k_x k_y k_z \left(k_x^2-k_y^2\right) \\
\end{array}$\\ 
 \hline 
$\Gamma ^1_{-}$ & $\begin{array}{c}
 0 \\
\end{array}$ & $\begin{array}{c}
 0 \\
\end{array}$\\ 
 \hline 
$\Gamma ^2_{-}$ & $\begin{array}{c}
 0 \\
\end{array}$ & $\begin{array}{c}
 0 \\
\end{array}$\\ 
 \hline 
$\Gamma ^2_{+}$ & $\begin{array}{c}
 k_x k_y^3-k_x^3 k_y \\
\end{array}$ & $\begin{array}{c}
 i \text{D}_{\text{z}} k_z \\
\end{array}$\\ 
 \hline 
$\Gamma ^3_{-}$ & $\begin{array}{c}
 0 \\
\end{array}$ & $\begin{array}{c}
 0 \\
\end{array}$\\ 
 \hline 
$\Gamma ^3_{+}$ & $\begin{array}{c}
 k_x k_y \\
\end{array}$ & $\begin{array}{c}
 -i \text{D}_{\text{z}} k_z \left(k_x^2-k_y^2\right) \\
\end{array}$\\ 
 \hline 
$\Gamma ^4_{-}$ & $\begin{array}{c}
 0 \\
\end{array}$ & $\begin{array}{c}
 0 \\
\end{array}$\\
\hline 
\end{tabular}
            \hspace*{\fill}
             \newline
    \vspace*{0.5 cm}
    \newline
    \hspace*{\fill}
    \begin{tabular}{|c|c|c|} 
\hline 
30: C3+IEe$ \times $C3&$\psi$ & $d$  \\ 
 \hline 
$\text{A}_{+}$ & $\begin{array}{c}
 1 \\
\end{array}$ & $\begin{array}{c}
 i \text{D}_{\text{z}} k_z \\
\end{array}$\\ 
 \hline 
$\text{A}_{-}$ & $\begin{array}{c}
 0 \\
\end{array}$ & $\begin{array}{c}
 0 \\
\end{array}$\\ 
 \hline 
$\text{E}^2_{-}$ & $\begin{array}{c}
 0 \\
 0 \\
\end{array}$ & $\begin{array}{c}
 0 \\
 0 \\
\end{array}$\\ 
 \hline 
$\text{E}^2_{+}$ & $\begin{array}{c}
 2 C k_z \left(k_y+i k_x\right)-C_1 \left(k_x+i k_y\right){}^2 \\
 2 C^* k_z \left(k_y-i k_x\right)-C_1 \left(k_x-i k_y\right){}^2 \\
\end{array}$ & $\begin{array}{c}
 C \text{D}_{\text{z}} \left(k_y+i k_x\right) \\
 i C^* \text{D}_{\text{z}} \left(k_x+i k_y\right) \\
\end{array}$\\ 
 \hline 
$\text{A}_{\nu}\text{A}_{-\nu}$ & $\begin{array}{c}
 0 \\
 0 \\
\end{array}$ & $\begin{array}{c}
 \text{D}_- k_z \\
 -\text{D}_+ k_z \\
\end{array}$\\ 
 \hline 
$\text{E}^2_{\nu}\text{E}^2_{-\nu}$ & $\begin{array}{c}
 0 \\
 0 \\
 0 \\
 0 \\
\end{array}$ & $\begin{array}{c}
 C \text{D}_- \left(k_y+i k_x\right) \\
 -i C \text{D}_+ \left(k_x-i k_y\right) \\
 \text{D}_- C^* \left(k_y-i k_x\right) \\
 i \text{D}_+ C^* \left(k_x+i k_y\right) \\
\end{array}$\\ 
 \hline 
\end{tabular}
    \hspace*{\fill}
\end{table}

\begin{table}[h]
    \fontsize{8}{10}\selectfont
    \centering
    \hspace*{\fill}
            \begin{tabular}{|c|c|c|} 
\hline 
31: C3+C2x$ \times $C3&$\psi$ & $d$  \\ 
 \hline 
$\text{A}_{+}$ & $\begin{array}{c}
 1 \\
\end{array}$ & $\begin{array}{c}
 i \text{D}_{\text{z}} k_z \\
\end{array}$\\ 
 \hline 
$\text{A}_{-}$ & $\begin{array}{c}
 -k_x k_z \left(k_x^2-3 k_y^2\right) \\
\end{array}$ & $\begin{array}{c}
 -i \text{D}_{\text{z}} \left(k_x^3-3 k_x k_y^2\right) \\
\end{array}$\\ 
 \hline 
$\text{E}^1_{0}\text{E}^2_{0}$ & $\begin{array}{c}
 i r \left(k_x-i k_y\right){}^2+k_z \left(k_x+i k_y\right) \\
 i \left(r \left(k_x+i k_y\right){}^2+k_z \left(k_y+i k_x\right)\right) \\
\end{array}$ & $\begin{array}{c}
 \text{D}_{\text{z}} \left(k_y-i k_x\right) \\
 \text{D}_{\text{z}} \left(k_y+i k_x\right) \\
\end{array}$\\ 
 \hline 
$\text{A}_{\nu}\text{A}_{-\nu}$ & $\begin{array}{c}
 0 \\
 0 \\
\end{array}$ & $\begin{array}{c}
 \text{D}_- k_z \\
 -\text{D}_+ k_z \\
\end{array}$\\ 
 \hline 
$\text{E}^1_{\mu}\text{E}^2_{-\mu}$ & $\begin{array}{c}
 0 \\
 0 \\
 0 \\
 0 \\
\end{array}$ & $\begin{array}{c}
 C \text{D}_- \left(k_y-i k_x\right) \\
 -i C \text{D}_+ \left(k_x-i k_y\right) \\
 \text{D}_- C^* \left(k_y+i k_x\right) \\
 i \text{D}_+ C^* \left(k_x+i k_y\right) \\
\end{array}$\\ 
 \hline 
\end{tabular}
            \hspace*{\fill}
             \newline
    \vspace*{0.5 cm}
    \newline
    \hspace*{\fill}
    \begin{tabular}{|c|c|c|} 
\hline 
32: C3+IC2x$ \times $C3&$\psi$ & $d$  \\ 
 \hline 
$\text{A}_{+}$ & $\begin{array}{c}
 1 \\
\end{array}$ & $\begin{array}{c}
 -i \text{D}_{\text{z}} \left(k_x^3-3 k_x k_y^2\right) \\
\end{array}$\\ 
 \hline 
$\text{A}_{-}$ & $\begin{array}{c}
 -k_x k_z \left(k_x^2-3 k_y^2\right) \\
\end{array}$ & $\begin{array}{c}
 i \text{D}_{\text{z}} k_z \\
\end{array}$\\ 
 \hline 
$\text{E}^1_{0}\text{E}^2_{0}$ & $\begin{array}{c}
 i r \left(k_x-i k_y\right){}^2+k_z \left(k_x+i k_y\right) \\
 i \left(r \left(k_x+i k_y\right){}^2+k_z \left(k_y+i k_x\right)\right) \\
\end{array}$ & $\begin{array}{c}
 \text{D}_{\text{z}} \left(k_x+i k_y\right) \\
 \text{D}_{\text{z}} \left(k_x-i k_y\right) \\
\end{array}$\\ 
 \hline 
$\text{A}_{\nu}\text{A}_{-\nu}$ & $\begin{array}{c}
 0 \\
 0 \\
\end{array}$ & $\begin{array}{c}
 \text{D}_- k_z \\
 \text{D}_+ k_z \\
\end{array}$\\ 
 \hline 
$\text{E}^1_{\mu}\text{E}^2_{-\mu}$ & $\begin{array}{c}
 0 \\
 0 \\
 0 \\
 0 \\
\end{array}$ & $\begin{array}{c}
 C \text{D}_- \left(k_y-i k_x\right) \\
 C \text{D}_+ \left(k_y+i k_x\right) \\
 \text{D}_- C^* \left(k_y+i k_x\right) \\
 \text{D}_+ C^* \left(k_y-i k_x\right) \\
\end{array}$\\ 
 \hline 
\end{tabular}
    \hspace*{\fill}
    \newline
    \vspace*{0.5 cm}
    \newline
    \hspace*{\fill}
        \begin{tabular}{|c|c|c|} 
\hline 
33: C3i+IC2x$ \times $C3i&$\psi$ & $d$  \\ 
 \hline 
$\text{A}^{\text{g}}_{+}$ & $\begin{array}{c}
 1 \\
\end{array}$ & $\begin{array}{c}
 0 \\
\end{array}$\\ 
 \hline 
$\text{A}^{\text{g}}_{-}$ & $\begin{array}{c}
 -k_x k_z \left(k_x^2-3 k_y^2\right) \\
\end{array}$ & $\begin{array}{c}
 0 \\
\end{array}$\\ 
 \hline 
$\text{A}^{\text{u}}_{-}$ & $\begin{array}{c}
 0 \\
\end{array}$ & $\begin{array}{c}
 i \text{D}_{\text{z}} k_z \\
\end{array}$\\ 
 \hline 
$\text{A}^{\text{u}}_{+}$ & $\begin{array}{c}
 0 \\
\end{array}$ & $\begin{array}{c}
 -i \text{D}_{\text{z}} \left(k_x^3-3 k_x k_y^2\right) \\
\end{array}$\\ 
 \hline 
$\text{E}^{{\text{g}}^2}_{0}\text{E}^{{\text{g}}^1}_{0}$ & $\begin{array}{c}
 i r \left(k_x-i k_y\right){}^2+k_z \left(k_x+i k_y\right) \\
 i \left(r \left(k_x+i k_y\right){}^2+k_z \left(k_y+i k_x\right)\right) \\
\end{array}$ & $\begin{array}{c}
 0 \\
 0 \\
\end{array}$\\ 
 \hline 
$\text{E}^{{\text{u}}^1}_{0}\text{E}^{{\text{u}}^2}_{0}$ & $\begin{array}{c}
 0 \\
 0 \\
\end{array}$ & $\begin{array}{c}
 \text{D}_{\text{z}} \left(k_x+i k_y\right) \\
 \text{D}_{\text{z}} \left(k_x-i k_y\right) \\
\end{array}$\\ 
 \hline 
$\text{A}^{\text{g}}_{\nu}\text{A}^{\text{g}}_{-\nu}$ & $\begin{array}{c}
 0 \\
 0 \\
\end{array}$ & $\begin{array}{c}
 0 \\
 0 \\
\end{array}$\\
\hline 
\end{tabular}
        \hspace*{\fill}
\end{table}

\begin{table}[h]
    \fontsize{8}{11}\selectfont
    \centering
    \hspace*{\fill}
            \begin{tabular}{|c|c|c|} 
\hline 
34: C3v+IEe$ \times $C3v&$\psi$ & $d$  \\ 
 \hline 
$\text{A}^1_{+}$ & $\begin{array}{c}
 1 \\
\end{array}$ & $\begin{array}{c}
 i \text{D}_{\text{z}} k_z \\
\end{array}$\\ 
 \hline 
$\text{A}^1_{-}$ & $\begin{array}{c}
 0 \\
\end{array}$ & $\begin{array}{c}
 0 \\
\end{array}$\\ 
 \hline 
$\text{A}^2_{-}$ & $\begin{array}{c}
 0 \\
\end{array}$ & $\begin{array}{c}
 0 \\
\end{array}$\\ 
 \hline 
$\text{A}^2_{+}$ & $\begin{array}{c}
 -k_x k_z \left(k_x^2-3 k_y^2\right) \\
\end{array}$ & $\begin{array}{c}
 -i \text{D}_{\text{z}} \left(k_x^3-3 k_x k_y^2\right) \\
\end{array}$\\ 
 \hline 
$\text{E}_{-}$ & $\begin{array}{c}
 0 \\
 0 \\
\end{array}$ & $\begin{array}{c}
 0 \\
 0 \\
\end{array}$\\ 
 \hline 
$\text{E}_{+}$ & $\begin{array}{c}
 r \left(k_x^2-k_y^2\right)+k_y k_z \\
 -k_x \left(2 r k_y+k_z\right) \\
\end{array}$ & $\begin{array}{c}
 i \text{D}_{\text{z}} k_y \\
 -i \text{D}_{\text{z}} k_x \\
\end{array}$\\ 
 \hline 
$\text{A}^1_{\nu}\text{A}^1_{-\nu}$ & $\begin{array}{c}
 0 \\
 0 \\
\end{array}$ & $\begin{array}{c}
 \text{D}_- k_z \\
 -\text{D}_+ k_z \\
\end{array}$\\ 
 \hline 
$\text{A}^2_{\nu}\text{A}^2_{-\nu}$ & $\begin{array}{c}
 0 \\
 0 \\
\end{array}$ & $\begin{array}{c}
 -\text{D}_- \left(k_x^3-3 k_x k_y^2\right) \\
 \text{D}_+ \left(k_x^3-3 k_x k_y^2\right) \\
\end{array}$\\ 
 \hline 
$\text{E}_{\nu}\text{E}_{-\nu}$ & $\begin{array}{c}
 0 \\
 0 \\
 0 \\
 0 \\
\end{array}$ & $\begin{array}{c}
 \text{D}_- k_y \\
 -\text{D}_- k_x \\
 -\text{D}_+ k_y \\
 \text{D}_+ k_x \\
\end{array}$\\ 
 \hline 
\end{tabular}
            \hspace*{\fill}
             \newline
    \vspace*{0.5 cm}
    \newline
    \hspace*{\fill}
    \begin{tabular}{|c|c|c|} 
\hline 
35: D3+IEe$ \times $D3&$\psi$ & $d$  \\ 
 \hline 
$\text{A}^1_{+}$ & $\begin{array}{c}
 1 \\
\end{array}$ & $\begin{array}{c}
 -i \text{D}_{\text{z}} \left(k_x^3-3 k_x k_y^2\right) \\
\end{array}$\\ 
 \hline 
$\text{A}^1_{-}$ & $\begin{array}{c}
 0 \\
\end{array}$ & $\begin{array}{c}
 0 \\
\end{array}$\\ 
 \hline 
$\text{A}^2_{-}$ & $\begin{array}{c}
 0 \\
\end{array}$ & $\begin{array}{c}
 0 \\
\end{array}$\\ 
 \hline 
$\text{A}^2_{+}$ & $\begin{array}{c}
 -k_x k_z \left(k_x^2-3 k_y^2\right) \\
\end{array}$ & $\begin{array}{c}
 i \text{D}_{\text{z}} k_z \\
\end{array}$\\ 
 \hline 
$\text{E}_{-}$ & $\begin{array}{c}
 0 \\
 0 \\
\end{array}$ & $\begin{array}{c}
 0 \\
 0 \\
\end{array}$\\ 
 \hline 
$\text{E}_{+}$ & $\begin{array}{c}
 r \left(k_x^2-k_y^2\right)+k_y k_z \\
 -k_x \left(2 r k_y+k_z\right) \\
\end{array}$ & $\begin{array}{c}
 i \text{D}_{\text{z}} k_x \\
 i \text{D}_{\text{z}} k_y \\
\end{array}$\\ 
 \hline 
$\text{A}^1_{\nu}\text{A}^1_{-\nu}$ & $\begin{array}{c}
 0 \\
 0 \\
\end{array}$ & $\begin{array}{c}
 -\text{D}_- \left(k_x^3-3 k_x k_y^2\right) \\
 \text{D}_+ \left(k_x^3-3 k_x k_y^2\right) \\
\end{array}$\\ 
 \hline 
$\text{A}^2_{\nu}\text{A}^2_{-\nu}$ & $\begin{array}{c}
 0 \\
 0 \\
\end{array}$ & $\begin{array}{c}
 \text{D}_- k_z \\
 -\text{D}_+ k_z \\
\end{array}$\\ 
 \hline 
$\text{E}_{\nu}\text{E}_{-\nu}$ & $\begin{array}{c}
 0 \\
 0 \\
 0 \\
 0 \\
\end{array}$ & $\begin{array}{c}
 \text{D}_- k_x \\
 \text{D}_- k_y \\
 -\text{D}_+ k_x \\
 -\text{D}_+ k_y \\
\end{array}$\\ 
 \hline 
\end{tabular}
                \hspace*{\fill}
             \newline
    \vspace*{0.5 cm}
    \newline
    \hspace*{\fill}
    \begin{tabular}{|c|c|c|} 
\hline 
36: C3+IC2z$ \times $C3&$\psi$ & $d$  \\ 
 \hline 
$\text{A}_{+}$ & $\begin{array}{c}
 1 \\
\end{array}$ & $\begin{array}{c}
 i \text{D}_{\text{z}} k_z \\
\end{array}$\\ 
 \hline 
$\text{A}_{-}$ & $\begin{array}{c}
 k_z \left(r \left(k_x^3-3 k_x k_y^2\right)-3 k_x^2 k_y+k_y^3\right) \\
\end{array}$ & $\begin{array}{c}
 i \text{Dz} \left(r k_y \left(k_y^2-3 k_x^2\right)+k_x \left(k_x^2-3 k_y^2\right)\right) \\
\end{array}$\\ 
 \hline 
$\text{E}^2_{-}$ & $\begin{array}{c}
 C k_z \left(k_y+i k_x\right) \\
 C^* k_z \left(k_y-i k_x\right) \\
\end{array}$ & $\begin{array}{c}
 C \text{D}_{\text{z}} \left(k_y+i k_x\right) \\
 i C^* \text{D}_{\text{z}} \left(k_x+i k_y\right) \\
\end{array}$\\ 
 \hline 
$\text{E}^2_{+}$ & $\begin{array}{c}
 -C \left(k_x+i k_y\right){}^2 \\
 -C^* \left(k_x-i k_y\right){}^2 \\
\end{array}$ & $\begin{array}{c}
 -C \text{D}_{\text{z}} k_z \left(k_x+i k_y\right){}^2 \\
 C^* \text{D}_{\text{z}} k_z \left(k_x-i k_y\right){}^2 \\
\end{array}$\\ 
 \hline 
$\text{A}_{\nu}\text{A}_{-\nu}$ & $\begin{array}{c}
 0 \\
 0 \\
\end{array}$ & $\begin{array}{c}
 \text{D}_- k_z \\
 -\text{D}_+ k_z \\
\end{array}$\\ 
 \hline 
$\text{E}^2_{\nu}\text{E}^2_{-\nu}$ & $\begin{array}{c}
 0 \\
 0 \\
 0 \\
 0 \\
\end{array}$ & $\begin{array}{c}
 C \text{D}_- \left(k_y+i k_x\right) \\
 C \text{D}_+ \left(k_y+i k_x\right) \\
 \text{D}_- C^* \left(k_y-i k_x\right) \\
 \text{D}_+ C^* \left(k_y-i k_x\right) \\
\end{array}$\\ 
 \hline 
\end{tabular}
    \hspace*{\fill}
\end{table}

\begin{table}[h]
    \fontsize{8}{11}\selectfont
    \centering
    \hspace*{\fill}
            \begin{tabular}{|c|c|c|} 
\hline 
37: C3+C2z$ \times $C3&$\psi$ & $d$  \\ 
 \hline 
$\text{A}_{+}$ & $\begin{array}{c}
 1 \\
\end{array}$ & $\begin{array}{c}
 i \text{D}_{\text{z}} \left(-3 r k_x^2 k_y+r k_y^3-3 k_x k_y^2+k_x^3\right) \\
\end{array}$\\ 
 \hline 
$\text{A}_{-}$ & $\begin{array}{c}
 k_z \left(r \left(k_x^3-3 k_x k_y^2\right)-3 k_x^2 k_y+k_y^3\right) \\
\end{array}$ & $\begin{array}{c}
 i \text{D}_{\text{z}} k_z \\
\end{array}$\\ 
 \hline 
$\text{E}^2_{-}$ & $\begin{array}{c}
 C k_z \left(k_y+i k_x\right) \\
 C^* k_z \left(k_y-i k_x\right) \\
\end{array}$ & $\begin{array}{c}
 -C \text{D}_{\text{z}} k_z \left(k_x+i k_y\right){}^2 \\
 C^* \text{D}_{\text{z}} k_z \left(k_x-i k_y\right){}^2 \\
\end{array}$\\ 
 \hline 
$\text{E}^2_{+}$ & $\begin{array}{c}
 -C \left(k_x+i k_y\right){}^2 \\
 -C^* \left(k_x-i k_y\right){}^2 \\
\end{array}$ & $\begin{array}{c}
 C \text{D}_{\text{z}} \left(k_y+i k_x\right) \\
 i C^* \text{D}_{\text{z}} \left(k_x+i k_y\right) \\
\end{array}$\\ 
 \hline 
$\text{A}_{\nu}\text{A}_{-\nu}$ & $\begin{array}{c}
 0 \\
 0 \\
\end{array}$ & $\begin{array}{c}
 \text{D}_- k_z \\
 \text{D}_+ k_z \\
\end{array}$\\ 
 \hline 
$\text{E}^2_{\nu}\text{E}^2_{-\nu}$ & $\begin{array}{c}
 0 \\
 0 \\
 0 \\
 0 \\
\end{array}$ & $\begin{array}{c}
 C \text{D}_- \left(k_y+i k_x\right) \\
 -i C \text{D}_+ \left(k_x-i k_y\right) \\
 \text{D}_- C^* \left(k_y-i k_x\right) \\
 i \text{D}_+ C^* \left(k_x+i k_y\right) \\
\end{array}$\\ 
 \hline 
\end{tabular}
            \hspace*{\fill}
             \newline
    \vspace*{0.5 cm}
    \newline
    \hspace*{\fill}
    \input{FAST_basis/FAST_Basis_38_C6_C2x.txt}
                \hspace*{\fill}
             \newline
    \vspace*{0.5 cm}
    \newline
    \hspace*{\fill}
    \input{FAST_basis/FAST_Basis_39_D3_C2z.txt}
    \hspace*{\fill}
\end{table}

\begin{table}[h]
    \fontsize{7}{9}\selectfont
    \centering
    \hspace*{\fill}
            \input{FAST_basis/FAST_Basis_40_C3i_IC2z.txt}
            \hspace*{\fill}
             \newline
    \vspace*{0.5 cm}
    \newline
    \hspace*{\fill}
    \input{FAST_basis/FAST_Basis_41_C3h_IEe.txt}
    \hspace*{\fill}
\end{table}
\begin{table}[h]
    \fontsize{7}{9}\selectfont
    \centering
\input{FAST_basis/FAST_Basis_42_C6_IEe.txt}
\end{table}

\begin{table}[h]
    \fontsize{7}{9}\selectfont
    \centering
            \input{FAST_basis/FAST_Basis_43_C6_IC2y.txt}
\end{table}
\begin{table}[h]
    \fontsize{7}{9}\selectfont
    \centering
            \input{FAST_basis/FAST_Basis_44_C3v_C2z.txt}
\end{table}

\begin{table}[h]
    \fontsize{7}{9}\selectfont
    \centering
            \input{FAST_basis/FAST_Basis_45_C3h_IC2x.txt}
\end{table}

\begin{table}[h]
    \fontsize{7}{9}\selectfont
    \centering
            \input{FAST_basis/FAST_Basis_46_C3v_IC2z.txt}
\end{table}

\begin{table}[h]
    \fontsize{7}{9}\selectfont
    \centering
            \input{FAST_basis/FAST_Basis_47_D3_IC2z.txt}
\end{table}

\begin{table}[h]
    \fontsize{7}{8}\selectfont
    \centering
            \input{FAST_basis/FAST_Basis_48_D3d_IC2z.txt}
\end{table}

\begin{table}[h]
    \fontsize{7}{8}\selectfont
    \centering
            \input{FAST_basis/FAST_Basis_49_D3h_IEe.txt}
\end{table}

\begin{table}[h]
    \fontsize{7}{9}\selectfont
    \centering
            \input{FAST_basis/FAST_Basis_50_C3h_C2x.txt}
\end{table}

\begin{table}[h]
    \fontsize{7}{9}\selectfont
    \centering
            \input{FAST_basis/FAST_Basis_51_C6v_IEe.txt}
\end{table}

\begin{table}[h]
    \fontsize{7}{9}\selectfont
    \centering
            \input{FAST_basis/FAST_Basis_52_D6_IEe.txt}
\end{table}

\begin{table}[h]
    \fontsize{7}{9}\selectfont
    \centering
            \input{FAST_basis/FAST_Basis_53_T_IEe.txt}
\end{table}

\begin{table}[h]
    \fontsize{7}{8}\selectfont
    \centering
            \input{FAST_basis/FAST_Basis_54_Th_IC4z.txt}
\end{table}

\begin{table}[h]
    \fontsize{7}{8}\selectfont
    \centering
            \input{FAST_basis/FAST_Basis_55_T_C4z.txt}
\end{table}

\begin{table}[h]
    \fontsize{7}{9}\selectfont
    \centering
            \input{FAST_basis/FAST_Basis_56_Th_C4z.txt}
\end{table}

\begin{table}[h]
    \fontsize{8}{10}\selectfont
    \centering
            \input{FAST_basis/FAST_Basis_57_Td_IEe.txt}
\end{table}

\begin{table}[h]
    \fontsize{8}{10}\selectfont
    \centering
            \input{FAST_basis/FAST_Basis_58_O_IEe.txt}
\end{table}
\shipout\box255

\clearpage
\section{Tables of coreps}\label{tables_coreps}
We list irreducible co-representations for all spin-point groups describing symmetries of compensated collinear magnetic phases. In the tables $\nu$ takes positive integer values $\nu > 0$. Coreps with $\mu$ allow for the zero values $\mu \geq 0$. The coreps with $\mu = 0$ are written explicitly. For an example with described notations, see Table \ref{tab:EECoerp}. The data is available in the digital form at DOI: 10.5281/zenodo.16279047.

\begin{table}[h]
\noindent
    \fontsize{1.5}{7}\selectfont
    \centering
    \hspace*{\fill}
        \input{Coreps/Corep_1_C1_IEe.txt}
    \hspace*{\fill}
        \vspace*{0.5 cm}
    \newline
        \centering
        \hspace*{\fill}
        \input{Coreps/Corep_2_C1_C2z.txt}
    \hspace*{\fill}
        \centering
    \vspace*{0.5 cm}
    \newline
        \hspace*{\fill}
        \input{Coreps/Corep_3_C1_IC2z.txt}
    \hspace*{\fill}
    \vspace*{0.5 cm}
    \newline
        \hspace*{\fill}
        \input{Coreps/Corep_4_C2_IC2z.txt}
    \hspace*{\fill}
    \vspace*{0.5 cm}
    \newline
        \hspace*{\fill}
        \input{Coreps/Corep_5_Cs_C2z.txt}
    \hspace*{\fill}
    \vspace*{0.5 cm}
    \newline
        \hspace*{\fill}
        \input{Coreps/Corep_6_Ci_C2z.txt}
        \hspace*{\fill}
    \vspace*{0.5 cm}
    \newline
        \hspace*{\fill}
        \input{Coreps/Corep_7_C2_IC2x.txt}
            \hspace*{\fill}
        \vspace*{0.5 cm}
    \newline
        \hspace*{\fill}
        \input{Coreps/Corep_8_Cs_C2x.txt}
            \hspace*{\fill}    
\end{table}

\begin{table}[h]
\noindent
    \fontsize{1.5}{7}\selectfont
    \centering
    \hspace*{\fill}
    \input{Coreps/Corep_9_C2_C2x.txt}
            \hspace*{\fill}
        \vspace*{0.5 cm}
    \newline
        \centering
        \hspace*{\fill}
        \input{Coreps/Corep_10_C2h_IC2x.txt}
    \hspace*{\fill}
        \vspace*{0.5 cm}
    \newline
        \centering
        \hspace*{\fill}
        \input{Coreps/Corep_11_C2v_IC2z.txt}
    \hspace*{\fill}
        \centering
    \vspace*{0.5 cm}
    \newline
        \hspace*{\fill}
        \input{Coreps/Corep_12_D2_IEe.txt}
    \hspace*{\fill}
    \vspace*{0.5 cm}
    \newline
        \hspace*{\fill}
        \input{Coreps/Corep_13_C2_C4z.txt}
    \hspace*{\fill}
    \vspace*{0.5 cm}
    \newline
        \hspace*{\fill}
        \input{Coreps/Corep_14_C2_IC4z.txt}
    \hspace*{\fill}
\end{table}

\begin{table}[h]
\noindent
    \fontsize{1.5}{7}\selectfont
    \centering
    \hspace*{\fill}
        \input{Coreps/Corep_15_C2h_C4z.txt}
    \hspace*{\fill}
        \vspace*{0.5 cm}
    \newline
        \centering
        \hspace*{\fill}
        \input{Coreps/Corep_16_S4_C4z.txt}
    \hspace*{\fill}
        \centering
    \vspace*{0.5 cm}
    \newline
        \hspace*{\fill}
        \input{Coreps/Corep_17_C4_IC2z.txt}
    \hspace*{\fill}
    \vspace*{0.5 cm}
    \newline
        \hspace*{\fill}
        \input{Coreps/Corep_18_C4_C2x.txt}
    \hspace*{\fill}
    \vspace*{0.5 cm}
    \newline
        \hspace*{\fill}
        \input{Coreps/Corep_19_D2_C4z.txt}
    \hspace*{\fill}
\end{table}

\begin{table}[h]
\noindent
    \fontsize{1.5}{7}\selectfont
    \centering
    \hspace*{\fill}
        \input{Coreps/Corep_20_C2v_C4z.txt}
    \hspace*{\fill}
        \vspace*{0.5 cm}
    \newline
        \centering
        \hspace*{\fill}
        \input{Coreps/Corep_21_C2v_IC4z.txt}
    \hspace*{\fill}
        \centering
    \vspace*{0.5 cm}
    \newline
        \hspace*{\fill}
        \input{Coreps/Corep_22_D2_IC4z.txt}
    \hspace*{\fill}
    \vspace*{0.5 cm}
    \newline
        \hspace*{\fill}
        \input{Coreps/Corep_23_C4_IC2x.txt}
    \hspace*{\fill}
    \vspace*{0.5 cm}
    \newline
        \hspace*{\fill}
        \input{Coreps/Corep_24_S4_C2x.txt}
    \hspace*{\fill}
\end{table}

\begin{sidewaystable}
\noindent
    \fontsize{1.5}{7}\selectfont
    \centering
    \hspace*{\fill}
        \input{Coreps/Corep_25_D2d_IEe.txt}
    \hspace*{\fill}
        \vspace*{0.5 cm}
    \newline
        \centering
        \hspace*{\fill}
        \input{Coreps/Corep_26_C4v_IEe.txt}
    \hspace*{\fill}
\end{sidewaystable}

\begin{table}[h]
\noindent
    \fontsize{1.5}{7}\selectfont
    \centering
    \hspace*{\fill}
        \input{Coreps/Corep_27_D2h_C4z.txt}
    \hspace*{\fill}
\end{table}

\begin{sidewaystable}
\noindent
    \fontsize{1.5}{7}\selectfont
    \centering
        \hspace*{\fill}
        \input{Coreps/Corep_28_C4h_C2x.txt}
    \hspace*{\fill}
        \centering
    \vspace*{0.5 cm}
    \newline
        \hspace*{\fill}
        \input{Coreps/Corep_29_D4_IEe.txt}
    \hspace*{\fill}
\end{sidewaystable}

\begin{table}[h]
\noindent
    \fontsize{1.5}{7}\selectfont
    \centering
    \hspace*{\fill}
            \input{Coreps/Corep_30_C3_IEe.txt}
    \hspace*{\fill}
    \vspace*{0.5 cm}
    \newline
        \hspace*{\fill}
        \input{Coreps/Corep_31_C3_C2x.txt}
    \hspace*{\fill}
\end{table}

\begin{sidewaystable}
\noindent
    \fontsize{1.5}{7}\selectfont
    \centering
        \input{Coreps/Corep_32_C3_IC2x.txt}
            \hspace*{\fill}
    \vspace*{0.5 cm}
    \newline
        \hspace*{\fill}
        \input{Coreps/Corep_33_C3i_IC2x.txt}
    \hspace*{\fill}
\end{sidewaystable}

\begin{table}[h]
\noindent
    \fontsize{1.5}{7}\selectfont
    \centering
    \hspace*{\fill}
            \input{Coreps/Corep_34_C3v_IEe.txt}
    \hspace*{\fill}
    \vspace*{0.5 cm}
    \newline
        \hspace*{\fill}
        \input{Coreps/Corep_35_D3_IEe.txt}
    \hspace*{\fill}
        \vspace*{0.5 cm}
    \newline
        \hspace*{\fill}
        \input{Coreps/Corep_36_C3_IC2z.txt}
    \hspace*{\fill}
            \vspace*{0.5 cm}
    \newline
        \hspace*{\fill}
        \input{Coreps/Corep_37_C3_C2z.txt}
        \hspace*{\fill}
\end{table}
\begin{table}[h]
\noindent
    \fontsize{1.5}{7}\selectfont
    \centering
    \hspace*{\fill}
            \input{Coreps/Corep_38_C6_C2x.txt}
    \hspace*{\fill}
    \vspace*{0.5 cm}
    \newline
        \hspace*{\fill}
            \input{Coreps/Corep_39_D3_C2z.txt}
    \hspace*{\fill}
    \vspace*{0.5 cm}
    \newline
        \hspace*{\fill}
        \input{Coreps/Corep_40_C3i_IC2z.txt}
    \hspace*{\fill}
\end{table}
  \clearpage
\begin{sidewaystable}
\noindent
    \fontsize{1.5}{7}\selectfont
    \centering
        \hspace*{\fill}
        \input{Coreps/Corep_41_C3h_IEe.txt}
    \hspace*{\fill}
        \centering
    \vspace*{0.5 cm}
    \newline
        \hspace*{\fill}
        \input{Coreps/Corep_42_C6_IEe.txt}
    \hspace*{\fill}
\end{sidewaystable}
 \clearpage
\begin{sidewaystable}
\noindent
    \fontsize{1.5}{7}\selectfont
    \centering
        \hspace*{\fill}
        \input{Coreps/Corep_43_C6_IC2y.txt}
    \hspace*{\fill}
        \centering
    \vspace*{0.5 cm}
    \newline
        \hspace*{\fill}
        \input{Coreps/Corep_44_C3v_C2z.txt}
    \hspace*{\fill}
\end{sidewaystable}

\begin{sidewaystable}
\noindent
    \fontsize{1.5}{7}\selectfont
    \centering
        \hspace*{\fill}
        \input{Coreps/Corep_45_C3h_IC2x.txt}
    \hspace*{\fill}
        \centering
    \vspace*{0.5 cm}
    \newline
        \hspace*{\fill}
        \input{Coreps/Corep_46_C3v_IC2z.txt}
    \hspace*{\fill}
\end{sidewaystable}

\begin{sidewaystable}
\noindent
    \fontsize{1.5}{7}\selectfont
    \centering
        \hspace*{\fill}
        \input{Coreps/Corep_47_D3_IC2z.txt}
    \hspace*{\fill}
        \centering
    \vspace*{0.5 cm}
    \newline
        \hspace*{\fill}
        \input{Coreps/Corep_48_D3d_IC2z.txt}
    \hspace*{\fill}
\end{sidewaystable}

\begin{sidewaystable}
\noindent
    \fontsize{1.5}{7}\selectfont
    \centering
        \hspace*{\fill}
        \input{Coreps/Corep_49_D3h_IEe.txt}
    \hspace*{\fill}
        \centering
    \vspace*{0.5 cm}
    \newline
        \hspace*{\fill}
        \input{Coreps/Corep_50_C3h_C2x.txt}
    \hspace*{\fill}
\end{sidewaystable}

\begin{sidewaystable}
\noindent
    \fontsize{1.5}{7}\selectfont
    \centering
        \hspace*{\fill}
        \input{Coreps/Corep_51_C6v_IEe.txt}
    \hspace*{\fill}
        \centering
    \vspace*{0.5 cm}
    \newline
        \hspace*{\fill}
        \input{Coreps/Corep_52_D6_IEe.txt}
    \hspace*{\fill}
\end{sidewaystable}

\begin{sidewaystable}
\noindent
    \fontsize{1.5}{7}\selectfont
    \centering
        \hspace*{\fill}
        \input{Coreps/Corep_53_T_IEe.txt}
    \hspace*{\fill}
\end{sidewaystable}

\begin{sidewaystable}
\noindent
    \fontsize{1.5}{7}\selectfont
    \centering
        \hspace*{\fill}
        \input{Coreps/Corep_54_Th_IC4z.txt}
    \hspace*{\fill}
\end{sidewaystable}

\begin{sidewaystable}
\noindent
    \fontsize{1.5}{7}\selectfont
    \centering
        \hspace*{\fill}
        \input{Coreps/Corep_55_T_C4z.txt}
    \hspace*{\fill}
\end{sidewaystable}

\begin{sidewaystable}
\noindent
    \fontsize{1.5}{7}\selectfont
    \centering
        \hspace*{\fill}
        \input{Coreps/Corep_56_Th_C4z.txt}
    \hspace*{\fill}
\end{sidewaystable}
\begin{sidewaystable}
\noindent
    \fontsize{1.5}{7}\selectfont
    \centering
        \hspace*{\fill}
        \input{Coreps/Corep_57_Td_IEe.txt}
    \hspace*{\fill}
\end{sidewaystable}
\begin{sidewaystable}
\noindent
    \fontsize{1.5}{7}\selectfont
    \centering
        \hspace*{\fill}
        \input{Coreps/Corep_58_O_IEe.txt}
    \hspace*{\fill}
\end{sidewaystable}
\shipout\box255